\pdfoutput=1
\documentclass[preprint,12pt]{elsarticle}
\usepackage[T1]{fontenc}
\usepackage{lmodern}
\usepackage{microtype}
\usepackage{amsmath,amssymb}
\usepackage{amsfonts}
\usepackage{bm}
\usepackage[colorlinks=true, allcolors=blue]{hyperref}
\usepackage{geometry}
\usepackage{amsthm}
\usepackage{chngcntr}
\usepackage{slashed}
\usepackage{dsfont}
\usepackage{algorithm}
\usepackage{algorithmic}
\usepackage{physics}
\usepackage{tikz}
\usetikzlibrary{decorations.markings, decorations.pathmorphing}

\newcommand{\Vol}{\operatorname{Vol}}
\newcommand{\Diff}{\operatorname{Diff}}
\newcommand{\Z}{\mathbb{Z}}

\newtheorem{remark}{Remark}[section]

\counterwithin{equation}{section}

\newcommand{\pd}{\partial}
\newcommand{\diag}{\operatorname{diag}}

\newcommand{\8}{\infty}

\newcommand{\I}{\mathrm{i}}

\newcommand{\ff}[1]{\frac{\delta}{\delta #1}}
\newcommand{\fbyf}[2]{\frac{\delta #1}{\delta #2}}
\newcommand{\lb}{\left[}
\newcommand{\rb}{\right]}
\newcommand{\R}{\mathbb{R}}

\newcommand{\Mathematica}{\textit{Mathematica\textsuperscript{\resizebox{!}{0.8ex}{\textregistered}}}}

\newcommand{\D}{\mathcal{D}}

\def\abs#1{\left| #1\right|}

\let\oldexp\exp
\renewcommand{\exp}[1]{\oldexp\left(#1\right)}
\def\INT#1#2{\int_{#1}^{#2}}

\allowdisplaybreaks[1]

\newenvironment{smalleq}{%
 \begingroup
 \small
 \addtolength{\jot}{.2em}%
 \setlength{\arraycolsep}{1.2pt}%
}{%
 \endgroup
}
\def\EQ#1{\begin{smalleq}\begin{align}#1\end{align}\end{smalleq}}
\def\br{\nonumber\\}

\def\lrb#1{\left(#1\right)}

\def\INT#1#2{\int_{#1}^{#2}}
\def\oh{\frac{1}{2}}

\def\XXint#1#2#3{{\setbox0=\hbox{$#1{#2#3}{\int}$}
     \vcenter{\hbox{$#2#3$}}\kern-.5\wd0}}

\begin{document}

\begin{frontmatter}

\title{Geometric QCD III:\\ Exact transition amplitudes and the glueball spectrum}

\author{Alexander Migdal}
\address{Institute for Advanced Study, Princeton, NJ 08540, USA}
\ead{amigdal@ias.edu}

\begin{abstract}
We complete the analysis of planar Makeenko--Migdal loop equations in the Lorentz-invariant continuum limit. Using our confining twistor-string representation, we compute the quantum fluctuation determinant, which in Minkowski space reduces to a discrete product of finite-dimensional matrix quadratures. The $\zeta$-regularized weight is independent of winding number $w$. Near the mass shell, the pole singularity is generated by $w \to \infty$, suppressing fluctuation variance as $1/w$. The path integral localizes on the classical trajectory, rendering the pole spectrum and transition residues parametrically exact in the large-winding WKB limit.  For the open-string meson sector, we fit 40 observed states across five topological boundary sectors ($h=0, \pm 1, \pm 2$). The holonomy shift $h$ accounts for exact geometric degeneracies between parity families, reproducing mass splittings without phenomenological spin-orbit parameters. Evaluated one-loop residues yield theoretical transition cross-sections capturing heavy-mass quenching and phase-space enhancement for high-spin light states.  Applying this framework to the pure Yang--Mills closed string, we demonstrate the dynamical stability of the pure-gauge minimal surface: the conformal Liouville anomaly drives the string strictly to the trigonometric minimum ($q=0$). The complex elliptic geometry analytically collapses, yielding linear Regge trajectories. The translation zero-mode measure dynamically nullifies the transition amplitude of the massless scalar ghost, providing an analytic mechanism for a purely gluonic mass gap. Anchoring parameter-free glueball trajectories to the open-string tension natively recovers the exact L"uscher intercept $\alpha(0)=1/12$, perfectly matching established PDG unassigned isoscalar candidates and large-$N_c$ lattice QCD extrapolations. 
\end{abstract}
\end{frontmatter}

\section{Introduction}

In the first two papers of this series \cite{migdal2025geometric, Migdal2026GeometricQCDII},
the planar MM loop equations were reduced to a confining twistor-string representation.
Part I \cite{migdal2025geometric} established the geometric framework of the Hodge-dual
minimal surface and its area law for quark confinement. Part II
\cite{Migdal2026GeometricQCDII} introduced the internal Majorana fermions, reproduced the
planar QCD graphs, and led to a rigid twistor string with semiclassical dynamics in
Minkowski space. That construction yielded parametric Regge trajectories and a comparison
with the observed meson spectrum.

In the present paper, we study the one-loop functional determinant of the quantum
fluctuations of this rigid twistor string. In standard string models, the fluctuation
determinant around a curved background is typically formulated in terms of continuous
boundary-value problems for bulk fields and the worldsheet metric. In the rigid twistor-string
formulation, by contrast, the bulk minimal surface is determined by the analytic continuation
of the one-dimensional boundary data. The quantum fluctuations are therefore encoded in
the boundary twistor fields. Expanding these boundary fluctuations in Fourier modes reduces
the two-dimensional functional determinant to a discrete product of finite-dimensional
matrix blocks.

Because the $\zeta$-regularized determinant is strictly independent of the winding number $w$, and its overall scale anomaly  cancels (leaving only a bounded fractional dependence on $a$), the WKB mass-shell conditions are protected from one-loop fluctuation shifts. The geometric series over the string winding number generates macroscopic S-matrix poles near the mass shell, driven by configurations where $w \to \infty$. Since the quadratic fluctuation operator scales linearly with $w$, the corresponding variance is dynamically suppressed as $1/w$, so the path integral localizes precisely onto the classical WKB saddle. In this sense, the determination of the pole spectrum and the corresponding transition amplitudes (residues) is exact in the WKB approximation, with higher-loop corrections parametrically vanishing at $w\to \infty$.

We apply this exact algebraic framework to complete the phenomenological picture of both the open and closed planar QCD strings. For the open string ($q\bar{q}$ mesons), we extend our previous phenomenological comparison by performing a global fit of $40$ observed meson states across five topological boundary sectors ($h=0, \pm 1,\pm 2 $). The inclusion of the $|h|>0$ boundary conditions accounts for the exact $J-1$ geometric degeneracy between axial and vector mesons, successfully reproducing the observed mass splittings (parity doublets) without introducing phenomenological spin-orbit parameters. Evaluating the exact one-loop residues then yields physical transition amplitudes on the mass shell, and the resulting partial cross-sections exhibit a nontrivial topology- and flavor-dependent scaling, correctly capturing the macroscopic heavy-mass quenching of boundary amplitudes alongside phase-space enhancement for high-spin light states, in  qualitative and quantitative agreement with experimental electron-positron annihilation decay data.

Beyond the open-string meson sector, we apply this machinery to the closed twistor string, corresponding to the pure Yang--Mills Pomeron and the exact continuum glueball spectrum. Here, the framework yields  mathematically closed results devoid of any free parameters. First, we demonstrate the dynamical stability of the pure-gauge minimal surface. We demonstrate that the conformal Liouville anomaly dynamically drives the string to the exact trigonometric minimum ($q=0$). At this limit, the complex elliptic geometry analytically collapses, reducing the fluctuation determinant to a stable, discrete product of finite-dimensional Fourier blocks and mathematically shielding the string spectrum from tachyonic poles.

Second, we demonstrate the exact dynamic nullification of the massless scalar ghost. While the classical pure-gauge Regge trajectory formally intercepts the geometric origin ($M=0, J=0$), its physical quantum transition amplitude is  zero. The geometric integration over the continuous translation zero modes on the fully symmetric elliptic torus forces the exact quantum transition measure to algebraically collapse as $\mathcal{O}(M^4)$. This exact zero-mode nullification dynamically nullifies the scalar ghost without introducing unphysical fractional branch cuts,  establishing a strictly positive, purely gluonic mass gap---providing an analytic realization of color confinement in our continuum solution of planar QCD.

Finally, we show that this completely universal pure-gauge trajectory makes immediate, testable contact with physical reality. Because the conformal Liouville anomaly dynamically drives the pure-gauge minimal surface to the exact trigonometric minimum ($q=0$), the theoretical Regge trajectory is not a transcendental curve but an  linear relation. By converting the geometric eigenvalues into physical units via the universal planar string tension $\sigma \approx 0.18 \text{ GeV}^2$ (fixed entirely by the meson sector), the parameter-free pure Yang--Mills spectrum natively predicts the lowest macroscopic tensor glueball ($2^{++}$) at $M \approx 2.02 \text{ GeV}$. This geometric prediction cleanly separates from the lighter meson trajectories and closely aligns with established Particle Data Group (PDG) unassigned isoscalar glueball candidates, landing squarely on the physical $f_2(2010)$ candidate and seamlessly tracking higher-spin resonances such as the $f_4(2300)$ and $f_6(2510)$. 

\section{Hodge-dual minimal surface in Twistor space}

\subsection{The Hodge-dual minimal surface}

In Part I \cite{migdal2025geometric}, the Hodge-dual surface was defined as a minimal surface
$X^i_\mu(\xi_1,\xi_2)$ in the hidden space
\(
\R^3\otimes \R^4
\),
with area functional
\EQ{
S_\chi[C]=\int_D d^2\xi \sqrt{\Sigma_{\mu\nu}\Sigma_{\mu\nu}/2},\qquad
\Sigma_{\mu\nu}=\epsilon^{ab}\pd_a X^i_\mu \pd_b X^i_\nu,
\label{HDMSEuclArea}
}
subject to the chiral boundary locking
\EQ{
X^i_\mu\big|_{\pd D}=\eta^{\chi,i}_{\mu\nu}C_\nu,\qquad \chi=\pm 1,
\label{HDMSEuclBC}
}
and the Hodge-duality constraint
\EQ{
\star \Sigma_{\mu\nu}=\chi \Sigma_{\mu\nu}.
\label{HDMSEuclDual}
}
The corresponding area derivative has definite Hodge chirality and therefore provides a
zero mode of the loop diffusion operator by the loop-space Bianchi identity (see Part I):
\EQ{
&\fbyf{S_\chi[C]}{\sigma_{\mu\nu}(\theta)}
=
-2\frac{\Sigma_{\mu\nu}}{\sqrt{\Sigma^2}},
\br
&\star \fbyf{S_\chi[C]}{\sigma_{\mu\nu}(\theta)}
=
\chi \fbyf{S_\chi[C]}{\sigma_{\mu\nu}(\theta)},
\br
&\implies L_\nu\lrb{S_\chi[C]}=0.
\label{HDMSEuclZeroMode}
}
Here \(L_\nu=\pd_\mu \ff{\sigma_{\mu\nu}}\) is the loop-space operator that brings down the
YM equations of motion \([D_\mu,[D_\mu,D_\nu]]\) from the ordered exponential.

As emphasized in section 3.5 of Part I, the minimal value of the area does \emph{not}
determine its loop-space area derivative. The latter involves a second functional derivative
in the full hidden space \(\R^3\otimes \R^4\), and the two Hodge chiralities \(S_\pm\)
have the same extremal value but different Hessians, and hence different area derivatives.
This remains true after continuation to Minkowski space.

As discussed in Parts I and II \cite{migdal2025geometric,Migdal2026GeometricQCDII},
the Hodge duality of the area derivative makes the general solution of the MM loop equation
for planar QCD invariant under multiplication of the Wilson loop by the zero-mode factor
\(W[C]\to W[C]\exp{-\kappa S_\chi[C]}\). 

\subsection{Hodge-dual minimal surface as Witten's master field}

The physical meaning of this invariance is the freedom in the choice of vacuum state, which is required in specifying a solution of the quantum field equations, including the MM loop equation. Moreover, the classical minimal surface \(X^i_\mu(\xi_1,\xi_2)\in \R^3\otimes \R^{1,3}\) plays the role of Witten's master field \cite{Witten:1980ez} in the \(N_c=\8\) limit. It transforms in the same way as the instanton gauge field, while only the \(SU(2)\)-invariant projection \(C_\mu \propto \eta^i_{\mu\nu} X^i_\nu(\pd \Sigma) \in \R^{1,3}\) of the boundary of this surface $\pd \Sigma=X(\pd \Sigma)$ is observable.

Let us discuss this analogy. The factorization of Wilson loop trace expectation values, noticed almost 50 years ago \cite{MMEq79}, technically follows from the self-averaging of sums of $N_c\to \8$ eigenvalues of a unitary matrix on the unit circle. Seen through the lens of large-$N_c$ QCD as a quantum field theory, this non-fluctuating Wilson loop may emerge as a holonomy of a single classical gauge field. Edward Witten, who was the first to conjecture the existence of such a master field, did not present an explicit implementation of this idea. Large-$N_c$ matrix models went further and devised some implementations, but unfortunately, these so-called Eguchi--Kawai models with quenched momenta were later proven to describe the wrong phase of lattice gauge theory, not the asymptotically free phase we are looking for. 

Still, the tantalizingly simple idea of a single classical field solving planar QCD stayed with us for all these years. In these three papers, we demonstrate something philosophically similar: a single minimal surface solving the planar loop equations. This surface is only half of the story. As we learned from the previous two parts of this study \cite{migdal2025geometric, Migdal2026GeometricQCDII}, there are heavy fermions on this surface, providing an important non-local effective interaction. We discuss that in the next sections, but now let us try to interpret our findings as Witten's master field.

We have a vector field $X^i_\mu(\xi)$ transforming as an $SU(2)$ gauge field parametrized by two conformal coordinates $\xi = (\Re z, \Im z)$. On the classical solution (Hodge-dual minimal surface), this field is associated with an ordinary minimal surface $x_\mu(\xi) = f_\mu(z) + \bar f_\mu(\bar z)$.

We may say that there is an infinitely thin (in the large-$N_c$ limit) string with the world surface $x_\mu(\xi)$, and the gauge field $X^i_\mu(\xi)= \eta^i_{\mu\nu} x_\nu(\xi)$ is associated with every point of this surface. A natural objection arises: how could a single, frozen surface correspond to the rich, fluctuating planar QFT? The answer is that only the shape of this surface in the 12D extended space is frozen; it does not fluctuate at infinite $N_c$. This frozen part acts precisely as Witten's master field. However, this is not a dead, static object. There are ``live'' objects---the elves---flying around on this rigid background, explicitly imitating the planar gluon graphs. 

The relation of the Wilson loop to the field is not given by a trace of holonomy; it is instead given by a Dirichlet boundary condition $X^i_\mu(e^{\I \theta}) = 2 \Re C_\mu(\theta)$. The Wilson loop is given by the expectation value of a string theory with some effective action over fields satisfying this boundary condition. This theory is effectively one-dimensional, as the bulk values of the twistor fields parametrizing the surface are determined by analytic continuation of the boundary values. This is mathematically equivalent to a non-local one-dimensional theory with Penrose twistors analytically continued from the boundary. We already elaborated on that continuation and the computation of the path integral in the second paper \cite{Migdal2026GeometricQCDII}. Here, we generalize this computation to include the closed string, describing glueballs.
\subsection{Twistor parametrization of area and the perimeter}

It is useful to rewrite the hidden Hodge-dual area tensor directly in terms of the
Hermitian coordinate matrix
\[
Y^{\alpha\dot\alpha}=Y^\mu \sigma_\mu,
\]
whose light-cone derivatives are rank-one null matrices,
\[
\partial_\xi Y^{\alpha\dot\alpha}=\varphi^\alpha \bar\varphi^{\dot\alpha},
\qquad
\partial_\eta Y^{\alpha\dot\alpha}=\psi^\alpha \bar\psi^{\dot\alpha},
\qquad
\xi=\tau+\theta,\quad \eta=\tau-\theta.
\]
Equivalently,
\[
\partial_\tau Y=\varphi\varphi^\dagger+\psi\psi^\dagger,
\qquad
\partial_\theta Y=\varphi\varphi^\dagger-\psi\psi^\dagger.
\]
This is the Lorentzian counterpart of the holomorphic representation of the Hodge-dual
surface in Part I, written in the generalized Weierstrass form appropriate to Minkowski
space.

Let
\[
X^i_\mu=\eta^{\chi,i}_{\mu\nu}Y^\nu .
\]
Then the hidden area bivector is
\EQ{
\Sigma_{\mu\nu}^{(\chi)}
&=
\epsilon^{ab}\partial_a X^i_\mu \partial_b X^i_\nu
=
\epsilon^{ab}\eta^{\chi,i}_{\mu\alpha}\eta^{\chi,i}_{\nu\beta}
\partial_a Y^\alpha \partial_b Y^\beta .
\label{TwistorSigmaStart}
}
Introduce the simple physical bivector
\EQ{
F_{\mu\nu}
\equiv
\epsilon^{ab}\partial_a Y_\mu \partial_b Y_\nu
=
\partial_\tau Y_\mu\,\partial_\theta Y_\nu
-
\partial_\theta Y_\mu\,\partial_\tau Y_\nu .
\label{TwistorFdef}
}
Using the Fierz identity for the chiral 't Hooft symbols,
\EQ{
\sum_{i=1}^3
\eta^{\chi,i}_{\mu\alpha}\eta^{\chi,i}_{\nu\beta}
=
2P^{(\chi)}_{\mu\nu,\alpha\beta},
\qquad
2P^{(\chi)}_{\mu\nu,\alpha\beta}F^{\alpha\beta}
=
F_{\mu\nu}+i\chi\,\star F_{\mu\nu},
\label{TwistorEtaFierz}
}
we obtain
\EQ{
\Sigma_{\mu\nu}^{(\chi)}
=
F_{\mu\nu}+i\chi\,\star F_{\mu\nu},
\qquad
\star\Sigma_{\mu\nu}^{(\chi)}
=
-i\chi\,\Sigma_{\mu\nu}^{(\chi)}.
\label{TwistorSigmaFromF}
}
Thus the hidden Hodge-dual area tensor is the chiral projection of the simple bivector
built from the two null directions of the Minkowski minimal surface.

Now define the two null vectors
\[
U^\mu\equiv \partial_\xi Y^\mu,
\qquad
V^\mu\equiv \partial_\eta Y^\mu.
\]
Since
\[
\partial_\tau Y^\mu=U^\mu+V^\mu,
\qquad
\partial_\theta Y^\mu=U^\mu-V^\mu,
\]
the bivector \eqref{TwistorFdef} becomes
\EQ{
F_{\mu\nu}
=
-2\left(U_\mu V_\nu-U_\nu V_\mu\right)
=
-2\left(
(\varphi\varphi^\dagger)_\mu(\psi\psi^\dagger)_\nu
-
(\varphi\varphi^\dagger)_\nu(\psi\psi^\dagger)_\mu
\right).
\label{TwistorFwedge}
}
Hence
\EQ{
\Sigma_{\mu\nu}^{(\chi)}
=
-2\left[
(\varphi\varphi^\dagger)\wedge(\psi\psi^\dagger)
\right]_{\mu\nu}
-2i\chi\,
\star\left[
(\varphi\varphi^\dagger)\wedge(\psi\psi^\dagger)
\right]_{\mu\nu}.
\label{TwistorSigmaPhiPsi}
}

The null-vector constraints follow immediately from the rank-one form of the Hermitian
matrices:
\EQ{
(\partial_\xi Y)^2=0,
\qquad
(\partial_\eta Y)^2=0.
\label{TwistorNullConstraints}
}
Therefore the induced metric
\[
g_{ab}=\partial_a Y\cdot\partial_b Y
\]
is conformal:
\EQ{
g_{\tau\tau}
=
(\partial_\tau Y)^2
=
2\,\partial_\xi Y\cdot\partial_\eta Y,
\qquad
g_{\theta\theta}
=
(\partial_\theta Y)^2
=
-2\,\partial_\xi Y\cdot\partial_\eta Y,
\qquad
g_{\tau\theta}=0.
\label{TwistorMetricConformal}
}
Thus
\EQ{
ds^2=\Omega^2(\tau,\theta)\left(d\tau^2-d\theta^2\right),
\qquad
\Omega^2(\tau,\theta)=2\,\partial_\xi Y\cdot\partial_\eta Y.
\label{TwistorOmegaDef}
}

For the simple bivector \(F_{\mu\nu}\), one has \(F_{\mu\nu}\star F^{\mu\nu}=0\), and therefore
\EQ{
\Sigma_{\mu\nu}^{(\chi)}\Sigma^{(\chi)\mu\nu}
=
2\,F_{\mu\nu}F^{\mu\nu}
=
-4\,\Omega^4.
\label{TwistorSigmaSq}
}
As a result, the Hodge-dual area density is
\EQ{
\sqrt{-\Sigma_{\mu\nu}^{(\chi)}\Sigma^{(\chi)\mu\nu}/2}
=
\Omega^2
=
2\,\partial_\xi Y\cdot\partial_\eta Y.
\label{TwistorAreaDensityOmega}
}

This conformal factor also admits a simple twistor representation. Using the determinant
identity for the sum of two rank-one Hermitian matrices,
\EQ{
2\,\partial_\xi Y\cdot\partial_\eta Y
=
\det\!\left(\varphi\varphi^\dagger+\psi\psi^\dagger\right)
=
\abs{\epsilon_{\alpha\beta}\varphi^\alpha\psi^\beta}^2
=
\abs{\varphi_1\psi_2-\varphi_2\psi_1}^2.
\label{TwistorBracketArea}
}
Therefore
\EQ{
\sqrt{-\Sigma_{\mu\nu}^{(\chi)}\Sigma^{(\chi)\mu\nu}/2}
=
\abs{\varphi_1\psi_2-\varphi_2\psi_1}^2,
\qquad
ds^2=
\abs{\varphi_1\psi_2-\varphi_2\psi_1}^2
\left(d\tau^2-d\theta^2\right).
\label{TwistorAreaFinal}
}
This is the twistor expression for the conformal factor used later in the explicit
exponential solution.

The same twistor bracket determines the proper perimeter of the string boundaries.
At fixed \(\theta=\theta_b\), the induced line element is
\EQ{
ds_{\partial\Sigma}^2
=
g_{\tau\tau}(\tau,\theta_b)\,d\tau^2
=
\Omega^2(\tau,\theta_b)\,d\tau^2,
\qquad
ds_{\partial\Sigma}
=
\abs{\varphi_1\psi_2-\varphi_2\psi_1}\,d\tau.
\label{TwistorBoundaryLine}
}
Hence the total perimeter of the two boundary worldlines over one temporal period is
\EQ{
P_{\partial\Sigma}
=
\int_0^{2\pi}d\tau
\left(
\Omega(\tau,\theta_b)+\Omega(\tau,-\theta_b)
\right).
\label{TwistorPerimeterGeneral}
}

\subsection{Area form and area derivatives in twistor representation}

It is also useful to display the chiral decomposition in two-spinor form. Writing
\EQ{
F_{\alpha\dot\alpha,\beta\dot\beta}
=
U_{\alpha\dot\alpha}V_{\beta\dot\beta}
-
U_{\beta\dot\beta}V_{\alpha\dot\alpha},
\qquad
U_{\alpha\dot\alpha}=\varphi_\alpha\bar\varphi_{\dot\alpha},
\qquad
V_{\alpha\dot\alpha}=\psi_\alpha\bar\psi_{\dot\alpha},
\label{TwistorSpinorF}
}
one has the standard decomposition
\EQ{
F_{\alpha\dot\alpha,\beta\dot\beta}
=
\epsilon_{\dot\alpha\dot\beta}f_{\alpha\beta}
+
\epsilon_{\alpha\beta}\bar f_{\dot\alpha\dot\beta},
\label{TwistorSpinorDecomp}
}
with
\EQ{
f_{\alpha\beta}
=
\oh\,
\epsilon_{\dot\gamma\dot\delta}
\bar\varphi^{\dot\gamma}\bar\psi^{\dot\delta}
\left(
\varphi_\alpha\psi_\beta+\varphi_\beta\psi_\alpha
\right),
\label{TwistorfDef}
}
and
\EQ{
\bar f_{\dot\alpha\dot\beta}
=
\oh\,
\epsilon_{\gamma\delta}
\varphi^\gamma\psi^\delta
\left(
\bar\varphi_{\dot\alpha}\bar\psi_{\dot\beta}
+
\bar\varphi_{\dot\beta}\bar\psi_{\dot\alpha}
\right).
\label{TwistorfbarDef}
}
Therefore
\EQ{
F+i\star F = 2\,\epsilon_{\alpha\beta}\bar f_{\dot\alpha\dot\beta},
\qquad
F-i\star F = 2\,\epsilon_{\dot\alpha\dot\beta}f_{\alpha\beta},
\label{TwistorFiStarF}
}
so that
\EQ{
\Sigma^{(+)}=F+i\star F,
\qquad
\Sigma^{(-)}=F-i\star F.
\label{TwistorSigmaPM}
}
In this form, the two hidden Hodge sectors are identified with the ASD/SD spinor
components of the simple bivector determined by the twistor data.

\section{The Liouville anomaly induced by microscopic fermions on a Hodge-dual surface}

The Hodge-dual surface serves as a background for the hidden dynamical degrees of freedom
introduced in 1981 \cite{M81QCDFST} (see also the review \cite{Mig83}). In their original
form, these degrees of freedom consisted of two types of Majorana fermions---the
``Elves'', or elementary fermions---propagating in the internal geometry of the curved
minimal surface. They were formulated on a surface with a fluctuating metric. The Elf
determinant has a useful topological property: in two dimensions, fermion paths avoid one
another and themselves on the surface. This leads to integral equations for the functional
determinants as functionals of the bounding curves. These functional equations are
topologically equivalent to the planar MM loop equations, which led to the conjecture that
QCD is equivalent to the Elf theory on a random surface \cite{M81QCDFST}.

However, the fluctuating metric field described by Polyakov's Liouville action gives rise
to conformal anomalies, leading to instabilities in four dimensions. In this form, the
string theory, with or without elementary fermions, is therefore not suitable for the
planar QCD problem.

As argued recently in \cite{Migdal2026GeometricQCDII}, the difficulty lies not in the
fermions themselves, but in the fluctuating metric of the random surface. The same Elf
theory, when placed on a minimal surface rather than on a random surface with Polyakov's
Liouville action, reproduces the MM loop equations for an appropriate choice of
parameters. This choice includes a large fermion mass and an additional term in the
action that cancels the quantum anomaly of the Dirac determinant. This term has the same
form as the Liouville action,
\EQ{
 S_{L} = \int_D d^2 \xi  \lrb{ \sigma e^{2 \rho} + \frac{\pd_{+} \rho \pd_{-} \rho}{3 \pi}}; \quad \rho  = \oh \log \sqrt{ -g}
}
The difference from the standard Liouville theory is that here the Liouville field \(\rho\)
is not an independent two-dimensional field with its own bulk fluctuations. Instead, it is
obtained by Penrose analytic continuation from the boundary data. In Euclidean space,
this is the analytic continuation of a holomorphic function from the unit circle into the
interior, retaining only the positive Fourier modes,
\[
f(z)=\sum_{n>0} A_n z^n,\qquad z=e^{\I\theta}.
\]
In Minkowski space, this becomes the continuation of the twistors
\(\phi(\tau+\theta)\), \(\psi(\tau-\theta)\) from their boundary values at
\(\theta=\pm\theta_b\) to points in the interior. We return to this continuation below.

\begin{remark}{\textbf{Not a random surface, but a minimal one.}}
For many years, the common assumption in large-\(N_c\) QCD was that the planar limit
\(N_c=\8\) should admit a string description, and that the problem was to identify the
appropriate additional degrees of freedom or extra dimensions required to realize it. The
present construction suggests a different viewpoint. While expanding the theory by adding fermionic variables, one
needs to reduce it, by freezing the bosonic variables. We replace the random surface bounded by a given loop by a minimal
surface, thereby freezing the string degrees of freedom. In twistor language, this amounts
to analytically continuing the boundary twistors into the bulk, while keeping the theory
effectively one-dimensional.  Such a frozen string looks pathological because the minimal surface is known to  be a singular functional of its boundary loop, but so is the Wilson loop. The physical amplitudes in momentum space, as we see below, contain certain path integrals over one dimensional twistor curves, resulting in finite calculable amplitudes on the mass shell. 
\end{remark}

\section{Open and closed strings as two ways of twisting the torus}

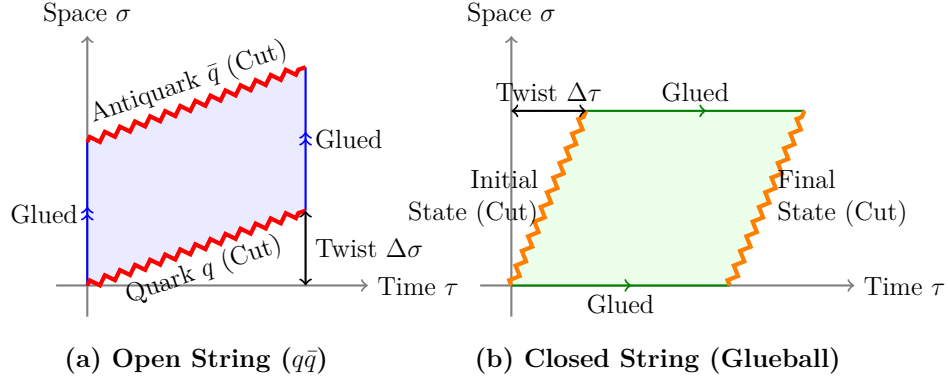
\begin{figure}[htbp]
    \centering
    \begin{tikzpicture}[scale=0.825, every node/.style={transform shape}]
        
        \begin{scope}[shift={(0,0)}]
            \draw[->, gray, thick] (-0.5,0) -- (4.5,0) node[right, black] {Time $\tau$};
            \draw[->, gray, thick] (0,-0.5) -- (0,4) node[above, black] {Space $\sigma$};
            
            \coordinate (A) at (0,0);
            \coordinate (B) at (3.5, 1.2); 
            \coordinate (C) at (3.5, 3.5);
            \coordinate (D) at (0, 2.3);
            
            \fill[blue!8] (A) -- (B) -- (C) -- (D) -- cycle;
            
            \draw[thick, blue, decoration={markings, mark=at position 0.55 with {\arrow{>>}}}, postaction={decorate}] (A) -- (D) node[midway, left, black] {Glued};
            \draw[thick, blue, decoration={markings, mark=at position 0.55 with {\arrow{>>}}}, postaction={decorate}] (B) -- (C) node[midway, right, black] {Glued};
            
            \draw[ultra thick, red, decorate, decoration={zigzag, segment length=6pt, amplitude=1.5pt}] (A) -- (B) node[midway, below, sloped, black] {Quark $q$ (Cut)};
            \draw[ultra thick, red, decorate, decoration={zigzag, segment length=6pt, amplitude=1.5pt}] (D) -- (C) node[midway, above, sloped, black] {Antiquark $\bar{q}$ (Cut)};
            
            \draw[dashed, gray] (3.5, 0) -- (B);
            \draw[<->, thick] (3.5, 0) -- (3.5, 1.2) node[midway, right] {Twist $\Delta\sigma$};
            
            \node[font=\bfseries] at (1.75, -1.2) {(a) Open String ($q\bar{q}$)};
        \end{scope}

        \begin{scope}[shift={(6.8,0)}]
            \draw[->, gray, thick] (-0.5,0) -- (5.5,0) node[right, black] {Time $\tau$};
            \draw[->, gray, thick] (0,-0.5) -- (0,4) node[above, black] {Space $\sigma$};
            
            \coordinate (A2) at (0,0);
            \coordinate (B2) at (3.5, 0);
            \coordinate (C2) at (4.7, 2.8); 
            \coordinate (D2) at (1.2, 2.8);
            
            \fill[green!8] (A2) -- (B2) -- (C2) -- (D2) -- cycle;
            
            \draw[thick, green!50!black, decoration={markings, mark=at position 0.55 with {\arrow{>}}}, postaction={decorate}] (A2) -- (B2) node[midway, below, black] {Glued};
            \draw[thick, green!50!black, decoration={markings, mark=at position 0.55 with {\arrow{>}}}, postaction={decorate}] (D2) -- (C2) node[midway, above, black] {Glued};
            
            \draw[ultra thick, orange, decorate, decoration={zigzag, segment length=6pt, amplitude=1.5pt}] (A2) -- (D2) node[midway, left, black, align=right] {Initial\\State (Cut)};
            \draw[ultra thick, orange, decorate, decoration={zigzag, segment length=6pt, amplitude=1.5pt}] (B2) -- (C2) node[midway, right, black, align=left] {Final\\State (Cut)};
            
            \draw[dashed, gray] (0, 2.8) -- (D2);
            \draw[<->, thick] (0, 2.8) -- (1.2, 2.8) node[midway, above] {Twist $\Delta\tau$};
            
            \node[font=\bfseries] at (2.35, -1.2) {(b) Closed String (Glueball)};
        \end{scope}
    \end{tikzpicture}
    \vspace{0.3cm}
    \caption{Modular duality of the twistor string represented on the fundamental torus. Both panels display the same torus as a fundamental rectangle where the twist is applied prior to gluing the opposite edges. \textbf{(a)} For the open $q\bar{q}$ string, the torus is cut along the spatial cycle (red zigzag lines) to define the boundaries of the quarks. The time cycle is glued with a spatial twist $\Delta\sigma$, capturing the macroscopic spatial rotation of the string over one temporal period. \textbf{(b)} For the closed glueball string, the torus is cut along the temporal cycle (orange zigzag lines) to define the initial and final states. The spatial cycle is glued with a temporal twist $\Delta\tau$, encoding the phase rotation (target-space monodromy) of the closed string.}
    \label{fig:torus_twists}
\end{figure}

Let us take the twistor string representation of planar QCD seriously and study the meson and glueball spectrum from the string theory perspective. The open string amplitudes (the propagator of the $q\bar{q}$ pair for a period of time $T$, projected onto a fixed spin state by twisted boundary conditions between $t=0,T$) are depicted in Figure \ref{fig:torus_twists}a. The closed string propagation is depicted in Figure \ref{fig:torus_twists}b.

\begin{figure}
    \centering
    \includegraphics[width=0.9\linewidth]{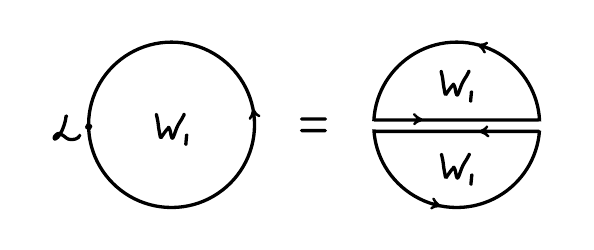}
    \caption{Graphical representation of the MM loop equation for the planar Wilson loop $W_1[C]$ in the leading large-$N_c$ limit. The action of the loop differential operator on the contour generates a self-intersection, splitting the loop into two disconnected contours. At leading topological order, this factorizes into the product of two planar loops, $W_1 \times W_1$. This equation determines the bulk Hodge-dual minimal surface and defines the confining twistor-string background for the open $q\bar{q}$ meson sector.}
    \label{fig:MMEquation}
\end{figure}

\begin{figure}
    \centering
    \includegraphics[width=0.9\linewidth]{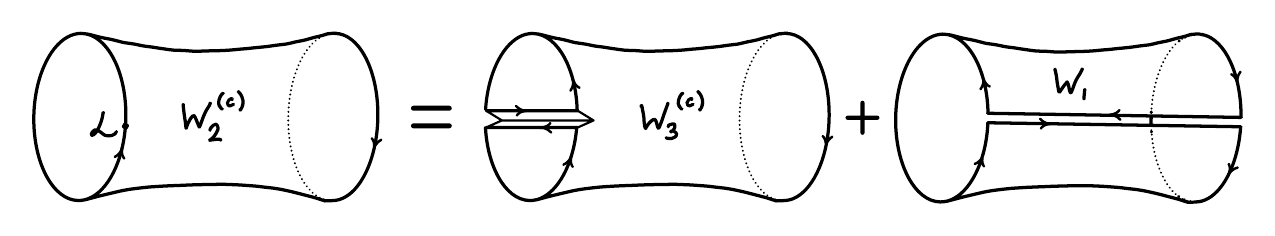}
    \caption{The MM loop equation for the connected loop-loop correlator $W_2[C_1, C_2]$, which determines the closed-string (glueball) spectrum at the next order in the topological $1/N_c$ expansion. The loop operator acting on $C_1$ generates a joining term (the final diagram on the right) where the two independent loops merge into a single macroscopic loop $W_1[C_1 \cup C_2]$. Crucially, this inhomogeneous source term relies on the exact same Hodge-dual minimal surface and Elf determinant derived from the planar equation in Figure 2, establishing the correspondence between the open and closed twistor strings for meson and glueball amplitudes.}
    \label{fig:GlueballEquation}
\end{figure}

The dynamical justification for using the same twistor string for the glueball spectrum (technically described by the next term in the topological $1/N_c$ expansion) is presented in the next two figures. Comparing these two loop equations, we observe that the last term on the right side of the glueball equation for the loop-loop correlator $W_2[C_1,C_2]$ (Figure \ref{fig:GlueballEquation} is given by the planar Wilson loop $W_1[C]$, which we have already found by solving the MM equation Figure \ref{fig:MMEquation}.

The surface bounded by the two loops $C_1, C_2$ in this term is the exact same Hodge-dual minimal surface with the Elf determinant as in the leading term $W_1[C]$ of the $1/N_c$ expansion. The loop equations for the Elf determinant on the cylindrical surface are literally the same as those on a disk. One can verify, using the procedure described in the previous paper \cite{Migdal2026GeometricQCDII}, that both terms on the right side of Figure \ref{fig:GlueballEquation} are reproduced by the Elf theory on a Hodge-dual surface.


As for the physical amplitudes described by the two different twisted gluings of the torus, the boundary conditions naturally bifurcate into two distinct kinematic sectors:

\subsection{The Meson Sector (Open String)}
In the meson sector, the spatial boundaries of the surface are open and correspond to the independent worldlines of the quark and antiquark at $\theta_1$ and $\theta_2$. These boundary paths are summed over, and the bulk twistor-string amplitude is weighted by the tensor product of the free quark phase-space traces $\mathcal{K}_1 \otimes \mathcal{K}_2$, where
\begin{equation}
\mathcal{K}_{1,2}=\mathcal{T}\exp{-\int_{0}^{T}d\tau\left(i(\gamma_{\mu}-\partial_{\tau}Y_{\mu}(\tau,\theta_{1,2}))(P_{\mu}(\tau)+k_{\mu})+m_{1,2}\sqrt{-g}\right)} .
\end{equation}
The states at $\tau=0$ and $\tau=T$ are glued with a target-space twist (a spatial rotation by an angle $\alpha$ in the $xy$-plane). The projection onto a state of fixed angular momentum $J$ is subsequently performed via the Fourier integration $\oint d\alpha e^{-\I J\alpha}$.

\subsection{The Glueball Sector (Closed String)}
In the glueball sector, the string is closed. The bulk twistor-string amplitude is entirely free of the localized quark boundary weights $\mathcal{K}_{1,2}$. Instead, the surface is periodically glued at the spatial boundaries (e.g., $\sigma = \pm \pi$). The target-space twist $\alpha$ is applied prior to gluing the temporal cycle, accompanied by the same projection onto a fixed $J$ state. 

Because the focus of the current computations (Sections 5 through 8) is the mass spectrum and transition amplitudes of the meson resonances, we now detail the phase-space kinematics specific to the open-string sector.

\subsection{Phase-Space Measure and Boundary Kinematics}

The phase-space formulation has a canonical normalization. The functional measure is
normalized by the standard symplectic volume element
$\mathcal{D}P\,\mathcal{D}C'/(2\pi)$ per continuous degree of freedom. Consequently, in
the free-particle limit $W[C]=1$, integrating over the loop coordinates $C$ yields the
functional Dirac delta function enforcing local momentum conservation:
\EQ{
&\int d^4k \int \frac{\mathcal{D}C'\,\mathcal{D}P}{2\pi}
\exp{\I \oint (P_\mu(\tau)-k_\mu)\dot C_\mu(\tau)\,d\tau}\,F[P]
\br
&=
\int d^4k \int \mathcal{D}P\,
\delta[P(\cdot)-k]\,F[P]
=
\int d^4k\,F[k].
\label{CanonicalNorm}
}
Thus the phase-space integral is well defined from standard quantum mechanics, modulo the
infinite 1D diffeomorphism volume $\Vol(\Diff(S^1))$ associated with the simultaneous
reparametrization of $P(\tau)$ and $C(\tau)$.

Conventionally, one fixes this $\Diff(S^1)$ volume by introducing a proper-time kinetic
term $\int (C')^2 d\tau$. This gauge choice leads to the usual Feynman--Schwinger
representation with the Wiener measure in loop space, dominated by
everywhere-discontinuous Brownian paths, which in turn generate the UV cusp singularities.
In the present framework, we do not use this Brownian gauge. Instead, as explained in
Part II \cite{Migdal2026GeometricQCDII}, we fix the 1D diffeomorphisms by imposing the
null/Virasoro constraint on the twistor variables that parametrize the boundary velocity.
This choice ties the 1D quark phase-space kinematics to the 2D geometry of the minimal
surface.

We now perform the local transformation from the linear loop velocity
\[
v_\mu(\tau)\equiv \dot C_\mu(\tau)
\]
to the same Minkowski twistor variables that are used in the minimal-surface
representation of sections 2, 6, 7 and Appendices C,D. The linear functional measure in
loop space can be written in either of the equivalent forms
\EQ{
\label{linmeasureMink}
\D C
&\propto
\delta^4\!\lrb{C(2\pi)-C(0)}
\prod_{\tau=0}^{2\pi} d^4 C(\tau)
\propto
\delta^4\!\lrb{\oint v(\tau)\,d\tau}
\prod_{\tau=0}^{2\pi} d^4 v(\tau).
}

At the physical string boundary $\theta=\theta_b$, we denote the boundary values of the
chiral minimal-surface twistors by
\EQ{
\phi(\tau)\equiv \phi(\tau+\theta_b),
\qquad
\psi(\tau)\equiv \psi(\tau-\theta_b).
\label{uTwistorDef}
}
Then the boundary tangent is the boundary restriction of the additive factorization
\EQ{
\partial_\tau Y=\phi\phi^\dagger+\psi\psi^\dagger,
\qquad
\partial_\theta Y=\phi\phi^\dagger-\psi\psi^\dagger.
}
Accordingly, the physical loop velocity is
\EQ{
v_{\alpha\dot\alpha}(\tau)
=
\phi_\alpha(\tau)\bar\phi_{\dot\alpha}(\tau)
+
\psi_\alpha(\tau)\bar\psi_{\dot\alpha}(\tau).
\label{vFromTwistors}
}
This is the boundary restriction of the same generalized Weierstrass representation that
solves the d'Alembert equation together with the null/Virasoro constraints for the
minimal surface.

The proper norm of the boundary tangent follows from the same $2\times 2$ determinant
identity used in section 2:
\EQ{
v^2
=
\det v
=
\det\!\left(\phi\phi^\dagger+\psi\psi^\dagger\right)
=
\abs{\epsilon_{\alpha\beta}\phi^\alpha\psi^\beta}^2.
\label{vNormTwistor}
}
Hence the proper boundary line element is
\EQ{
ds_{\partial\Sigma}
=
\sqrt{v^2}\,d\tau
=
\abs{\epsilon_{\alpha\beta}\phi^\alpha\psi^\beta}\,d\tau.
\label{boundaryPerimeterTwistor}
}

The additive representation \eqref{vFromTwistors} has a local chiral phase redundancy.
The physical boundary geometry is invariant under
\EQ{
\phi(\tau)\to e^{\I \chi_1(\tau)}\phi(\tau),
\qquad
\psi(\tau)\to e^{\I \chi_2(\tau)}\psi(\tau),
\label{RescalingGauge}
}
or infinitesimally
\EQ{
\delta_{\chi_1}\phi=\I\chi_1\,\phi,
\qquad
\delta_{\chi_2}\psi=\I\chi_2\,\psi.
\label{RescalingGaugeInf}
}
These are the local $U(1)_L\times U(1)_R$ phase directions of the chiral twistor
representation.

In the fluctuation language used later in section 7 and Appendix C,
\EQ{
\tilde\phi(\xi)=\exp{f_1(\xi)}\phi_0(\xi),
\qquad
\tilde\psi(\eta)=\exp{f_2(\eta)}\psi_0(\eta),
\qquad
f_1=\I(\alpha_1{\bf 1}+u_i\sigma_i),
\qquad
f_2=\I(\alpha_2{\bf 1}+v_i\sigma_i),
}
these local phase directions are the identity components
$\alpha_1(\xi)$ and $\alpha_2(\eta)$. We fix them by the local gauge conditions
\EQ{
G_1(\tau)\equiv \alpha_1(\tau+\theta_u)=0,
\qquad
G_2(\tau)\equiv \alpha_2(\tau-\theta_d)=0.
\label{GaugeCondition}
}
Under infinitesimal phase shifts one has
\EQ{
\delta \alpha_1=\chi_1,
\qquad
\delta \alpha_2=\chi_2,
}
so the Faddeev--Popov operator is ultralocal and diagonal. Therefore the corresponding FP
factor is field-independent:
\EQ{
\Delta_{\mathrm{FP}}
=
\abs{\frac{\delta(G_1,G_2)}{\delta(\chi_1,\chi_2)}}
=
\text{const}.
\label{FPfactor}
}
Thus the chiral phase gauge contributes only an overall normalization.

In the additive minimal-surface representation, the pair of boundary spinors
\((\phi,\psi)\) determines not only the boundary tangent \(v=\partial_\tau Y\), but also
the boundary normal \(\partial_\theta Y=\phi\phi^\dagger-\psi\psi^\dagger\). Therefore the
map from the loop velocity \(v(\tau)\) to the boundary twistors is \emph{not} the locally
invertible mixed bilinear transformation used in the Euclidean parametrization. The
corresponding Jacobian formulas must therefore be replaced, in the additive formalism, by
a constrained twistor resolution of the loop measure.

The local statement is that the loop measure is represented in terms of the boundary
values of the chiral minimal-surface twistors together with the local phase gauge fixing:
\EQ{
1
=
\int
D\phi\,D\psi\;
\delta^{(4)}\!\Bigl(
v-\phi\phi^\dagger-\psi\psi^\dagger
\Bigr)\,
\delta\!\bigl(G_1\bigr)\,
\delta\!\bigl(G_2\bigr)\,
\Delta_{\mathrm{FP}}.
\label{FPidentity}
}
Substituting this local resolution into the loop measure gives
\EQ{
\D C
\propto
\delta^4\!\left(
\oint d\tau\,
\left(\phi\phi^\dagger+\psi\psi^\dagger\right)
\right)
\prod_\tau
\left[
D\phi(\tau)\,D\psi(\tau)\,
\delta\!\bigl(G_1(\tau)\bigr)\,
\delta\!\bigl(G_2(\tau)\bigr)
\right],
\label{LoopMeasureTwistorGaugeFixed}
}
up to an overall field-independent normalization factor coming from
\(\Delta_{\mathrm{FP}}\).
This is the boundary phase-space measure written in the same chiral twistor variables as
the minimal-surface geometry and the fluctuation theory.

The Euclidean Jacobian derivation based on the mixed bilinear
\(v=\psi\phi^\dagger+\phi\psi^\dagger\) is therefore not used here. In the additive
formalism, it is replaced by the constrained measure above.

\subsection{Staggered product and spin holonomy}

After integrating out the local boundary momentum (see Part II,
\cite{Migdal2026GeometricQCDII}), the massless Dirac kernel at each point of the loop is
proportional to the normalized velocity matrix
\EQ{
\Gamma(\tau)\equiv \gamma_\mu \hat v^\mu(\tau),
\qquad
\hat v^\mu(\tau)=\frac{v^\mu(\tau)}{\Omega(\tau)},
\qquad
\Omega(\tau)=\sqrt{v^2(\tau)},
\qquad
\Gamma(\tau)^2=1.
\label{GammaDefHol}
}
Equivalently,
\EQ{
\Pi_\pm(\tau)=\frac{1\pm \Gamma(\tau)}{2},
\qquad
\Gamma(\tau)=\Pi_+(\tau)-\Pi_-(\tau),
\label{PiDefHol}
}
so the local Dirac factor is the difference of the two complementary projectors.

For the explicit rotating Minkowski solution,
\EQ{
v^0=R,
\qquad
v^3=0,
\qquad
v^1=-R\sin\beta\,\cos(2a\tau),
\qquad
v^2=R\sin\beta\,\sin(2a\tau),
\label{BoundaryVelocityHol}
}
hence
\EQ{
v^2=R^2\cos^2\beta,
\qquad
\Omega=R\cos\beta,
\qquad
S\equiv \sin\beta.
\label{OmegaHol}
}

Let the boundary be discretized into $2Lk$ points along $k$ windings, with ordered proper-time
values $\tau_j$. The ordered Dirac factor is then
\EQ{
\mathcal U_k
=
\mathcal T\prod_{j=1}^{2Lk}\Gamma(\tau_j).
\label{UkDiscrete}
}
For an even number of factors it is natural to group adjacent matrices into staggered pairs:
\EQ{
\mathcal U_k
=
\mathcal T\prod_{m=1}^{Lk}
\Big(
\Gamma(\tau_{2m})\Gamma(\tau_{2m-1})
\Big).
\label{UkPairs}
}
Now expand for small lattice spacing $\Delta\tau=\tau_{j+1}-\tau_j$:
\EQ{
\Gamma(\tau+\Delta\tau)\Gamma(\tau)
=
\Gamma(\tau)^2+\Delta\tau\,\dot\Gamma(\tau)\Gamma(\tau)+O(\Delta\tau^2)
=
\mathbf{1}+\Delta\tau\,\mathcal A(\tau)+O(\Delta\tau^2),
\label{PairExpansion}
}
with
\EQ{
\mathcal A(\tau)\equiv \dot\Gamma(\tau)\Gamma(\tau).
\label{AconnectionDef}
}
Because $\Gamma^2=1$, one has $\Gamma\dot\Gamma=-\dot\Gamma\Gamma$, and therefore
\EQ{
\mathcal A(\tau)
=
2[\dot\Pi_+(\tau),\Pi_+(\tau)]
=
-2[\dot\Pi_-(\tau),\Pi_-(\tau)].
\label{AProjectorForm}
}
Thus the staggered product is the holonomy of the non-Abelian connection $\mathcal A$:
\EQ{
\mathcal U_k
=
\mathcal P\exp{\int_0^{\pi k}d\tau\,\mathcal A(\tau)}.
\label{HolonomyFormula}
}
The upper limit is $\pi k$ rather than $2\pi k$ because pairing halves the number of
independent transfer steps. Equivalently, one could integrate over $2\pi k$ with half the
connection; the two conventions are equivalent.

In the Weyl basis,
\EQ{
\Gamma(\tau)=
\begin{pmatrix}
0 & \bar n(\tau)\\
n(\tau) & 0
\end{pmatrix},
\label{GammaWeylHol}
}
where the normalized Hermitian velocity matrices are
\EQ{
n(\tau)
=
\frac{1}{\cos\beta}
\begin{pmatrix}
1 & -S e^{2ia\tau}\\
-S e^{-2ia\tau} & 1
\end{pmatrix},
\qquad
\bar n(\tau)
=
\frac{1}{\cos\beta}
\begin{pmatrix}
1 & S e^{2ia\tau}\\
S e^{-2ia\tau} & 1
\end{pmatrix},
\label{nbarndef}
}
so that
\EQ{
n(\tau)\bar n(\tau)=\bar n(\tau)n(\tau)=\mathbf{1}_2.
\label{nnbarone}
}
Therefore the connection $\mathcal A=\dot\Gamma\Gamma$ is block diagonal:
\EQ{
\mathcal A(\tau)
=
\begin{pmatrix}
\bar A(\tau) & 0\\
0 & A(\tau)
\end{pmatrix},
\qquad
A(\tau)=\dot n(\tau)\bar n(\tau),
\qquad
\bar A(\tau)=\dot{\bar n}(\tau)n(\tau).
\label{BlockAdef}
}
A direct computation gives
\EQ{
A(\tau)
=
\frac{2iaS}{1-S^2}
\begin{pmatrix}
-S & -e^{2ia\tau}\\
e^{-2ia\tau} & S
\end{pmatrix},
\qquad
\bar A(\tau)
=
\frac{2iaS}{1-S^2}
\begin{pmatrix}
-S & e^{2ia\tau}\\
-e^{-2ia\tau} & S
\end{pmatrix}.
\label{Aexplicit}
}
Hence the holonomy factorizes into two independent $2\times2$ ordered exponentials
\EQ{
\mathcal U_k=
\begin{pmatrix}
\bar W(\pi k) & 0\\
0 & W(\pi k)
\end{pmatrix},
\qquad
W'(\tau)=A(\tau)W(\tau),
\qquad
\bar W'(\tau)=\bar A(\tau)\bar W(\tau).
\label{WbarWdef}
}

To solve the lower block, pass to the co-rotating frame
\EQ{
W(\tau)=U(\tau)\widetilde W(\tau),
\qquad
U(\tau)=
\begin{pmatrix}
e^{ia\tau} & 0\\
0 & e^{-ia\tau}
\end{pmatrix}.
\label{CoRotFrame}
}
Then
\EQ{
\widetilde W'(\tau)=ia\,M\,\widetilde W(\tau),
\label{WtildeEq}
}
with the constant matrix
\EQ{
M=
\frac{1}{1-S^2}
\begin{pmatrix}
-(1+S^2) & -2S\\
2S & 1+S^2
\end{pmatrix}.
\label{MmatrixHol}
}
This matrix satisfies
\EQ{
\tr M=0,
\qquad
M^2=\mathbf{1}_2,
\label{Mproperties}
}
so its eigenvalues are $\pm1$. Therefore
\EQ{
\widetilde W(\pi k)
=
\exp{\I\pi ak\,M}
=
\cos(\pi ak)\,\mathbf{1}_2
+i\sin(\pi ak)\,M.
\label{WtildeSolution}
}
Transforming back and taking the trace yields
\EQ{
\tr W(\pi k)
=
2\cos^2(\pi ak)
+
2\sin^2(\pi ak)\frac{1+S^2}{1-S^2}
=
2\left(1+2\sin^2(\pi ak)\tan^2\beta\right).
\label{TrWfinal}
}
The upper block gives the same contribution,
\EQ{
\tr \bar W(\pi k)=\tr W(\pi k),
\label{TrWbarfinal}
}
and therefore the full $4\times4$ staggered holonomy is
\EQ{
T_4(k)
\equiv
\tr_4\,\mathcal U_k
=
4\left(1+2\sin^2(\pi ak)\tan^2\beta\right).
\label{T4HolonomyFinal}
}
This is the spin-holonomy factor entering the meson amplitude. In terms of the
projectors,
\EQ{
\Gamma(\tau)=\Pi_+(\tau)-\Pi_-(\tau),
\qquad
\mathcal U_k
=
\mathcal T\prod_j\left(\Pi_+(\tau_j)-\Pi_-(\tau_j)\right),
\label{ProjectorHolonomyProduct}
}
so the trace of the staggered product is the trace of the ordered transport in the two
eigenbundles of the local timelike projector field.

\section{Semiclassical Quantization and the Geometric Mass Spectrum}
\label{sec:catastropheSpectrum}

\subsection{Twisted Boundary Conditions, Symmetrization, and Spin Projections}

Before solving the spectral equations, we explain the choice of the twisted geometry from
the viewpoint of the semiclassical quark path integral. In planar QCD, the effective
Minkowski action involves a sum over quark trajectories weighted by the area of the
confining minimal surface.

\begin{figure}[htbp]
\centering
\includegraphics[width=0.95\columnwidth]{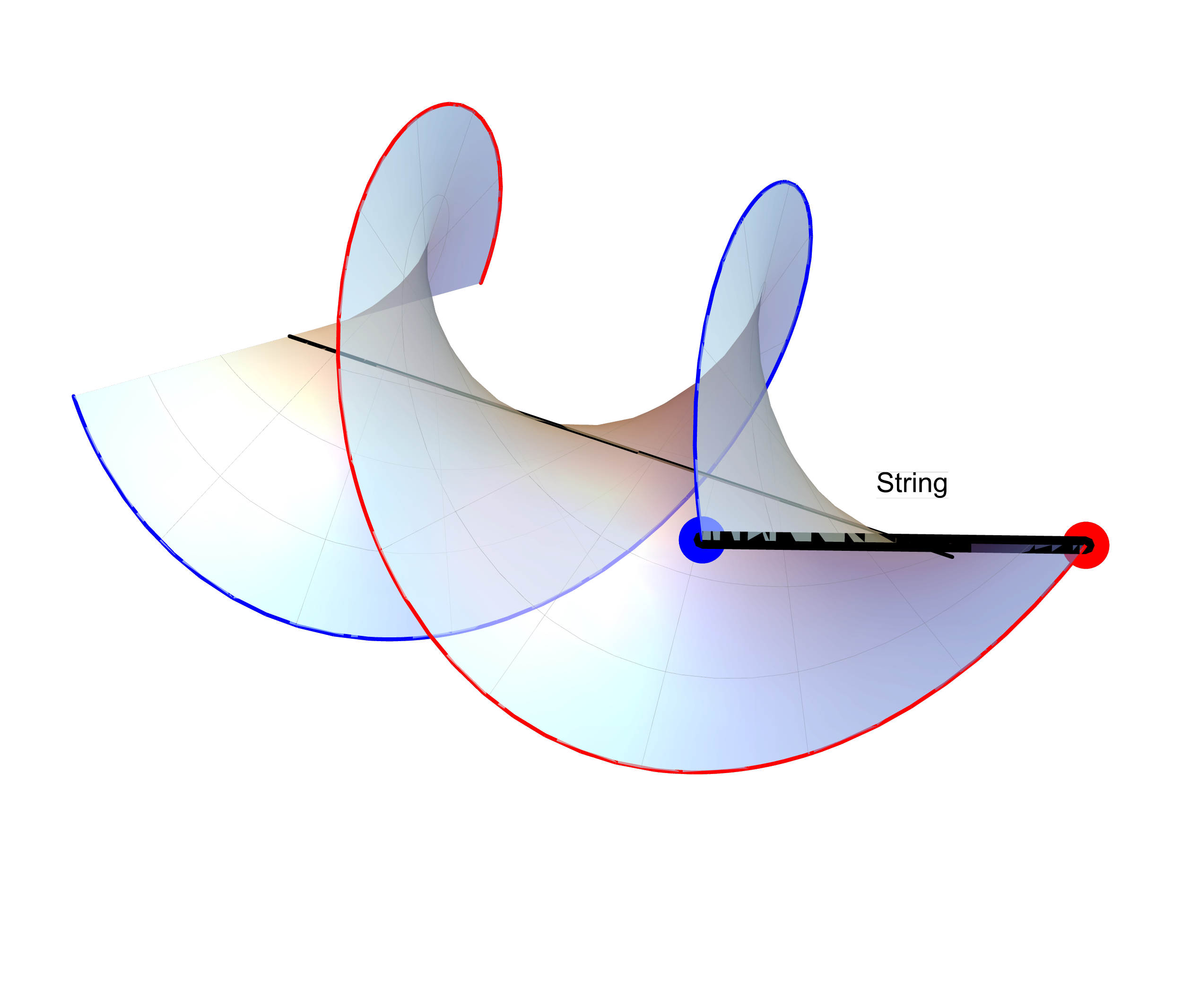}
\caption{The helicoid spanned by a rotating \(\overline{q}q\) pair connected by a
rigid stick (string). This minimal surface bounded by a double helix, discovered
by Meusnier in 1785, provides the geometric motivation for the twisted boundary
conditions.}
\label{fig:helicoid}
\end{figure}

To extract the spectrum of states with macroscopic angular momentum \(J\), we impose
twisted boundary conditions. We require the whole string to undergo a spatial rotation by
an angle \(\alpha\) over one period of the motion, with the angular momentum \(J\) conjugate
to this twist. Minimizing the boundary length for a rotating particle yields a helical
trajectory in Minkowski spacetime. The minimal surface spanning two twisted helical
boundaries is the helicoid, as illustrated in Fig.~\ref{fig:helicoid}.

Accordingly, the helicoid is the natural minimal surface associated with the semiclassical
kinematics of angular momentum. To isolate the physical spin states of the meson
spectrum, one must couple the internal worldsheet geometry to the macroscopic angular
momentum \(J\) in target space.

In the twistor-string formulation, this twisted kinematics is imposed directly on the
boundary twistor data. This is done by introducing the \(SU(2)_L\) and \(SU(2)_R\)
projections and integrating over the target-space rotation angle \(\alpha\) with the
canonical spin weight:
\EQ{
\label{SpinProjector}
\hat P_J A = \sum_k \oint d\alpha \exp{-\I k\alpha J} A_k(\alpha)
}
Here \(A_k(\alpha)\) is the amplitude after \(k\) windings around the quark loop, with the
boundary condition that a rotation of the whole surface by an angle \(\alpha\) is introduced
after each full circuit around the loop. When the macroscopic boundary of the string
rotates by an angle \(\alpha\), the boundary twistors must acquire the corresponding twist.
To implement this, we introduce a fractional winding parameter \(a\), corresponding to
twisted boundary conditions for \(\lambda(z)\) and \(\mu(z)\).

To analyse the resonance responsible for the WKB saddle, it is useful to consider the
limit of a large number of geometric windings, \(k\to\infty\). This limit is not imposed
as an additional assumption. Rather, it follows from the structure of the winding sum
itself. Near the mass shell, where the geometric phase satisfies \(\Phi_*\to 0\), the
projected amplitude contains the singular series
\[
\sum_{k=1}^\infty e^{\I ka\Phi_*},
\]
whose Abel regularization develops the pole \(1/\Phi_*\). The on-shell singularity is
therefore generated by the contribution of arbitrarily large winding sectors. In this
regime, the quadratic fluctuation operator scales as \(\mathcal H_k\propto k\), so the
variance of the corresponding quantum fluctuations scales as \(1/k\to0\). The path
integral therefore localizes on the WKB saddle on shell. In this sense, the WKB
approximation is exact for the pole spectrum determined by the winding singularity.

\subsection{Geometric Quantization via Topological Winding Sectors}
\label{16.1.}

In the semiclassical (WKB) limit, the path integral over the twistor phase space is
dominated by the saddle points \(\mathcal{V}_{*}(E)\) of the complexified effective action.
Physical bound states, interpreted as mesons, correspond to macroscopic S-matrix poles in
this amplitude. In the present formulation, these poles arise from constructive
interference of the WKB amplitude across an infinite set of topological winding sectors.
As the string sweeps out its target-space rotation, the multivalued geometric phase
accumulates. At the resonant energies \(E=E_n\), the geometric action monodromy aligns,
\(\Delta S_{\rm geom}=0\). The corresponding WKB quantization is stabilized by the growing number of windings near the mass shell, which multiply the action and thus suppress quantum fluctuations around the classical saddle point.

\section{Minkowski Minimal Surface and Meson Spectrum}
\label{sec:minkowski_minimal_surface}

To implement the flat-valley scenario and compute the spectrum of massive mesons, we
construct the minimal surface directly in four-dimensional Minkowski space
\(\mathbb{R}^{1,3}\). This requires more than a formal analytic continuation of the
twistor parametrization of a Euclidean minimal surface.

In Euclidean space, the relevant saddle points are typically driven into the complex
domain of the twistor variables. In the present Minkowski formulation, by contrast, the
boundary data are specified directly in real spacetime, and the path integral takes the
usual Feynman form as a sum over real histories weighted by the phase \(\exp{iS/\hbar}\).
The corresponding semiclassical saddle points are therefore realized as real classical
Minkowski trajectories, represented here by the rotating helicoid. In this way, the mass
spectrum is determined within a real Minkowski geometry, and the physically relevant
states are restricted to non-negative angular momentum \(J\ge 0\).

\subsection{Additive Factorization in Minkowski Space}

To construct a surface in Minkowski space, we use the generalized Weierstrass
representation of Konopelchenko and Landolfi (K\&L Theorem 5.1) \cite{Konopelchenko1999}.
The target-space coordinate differential, represented as the Hermitian matrix
\(d\mathbf{X}=dX^0 I+dX^1\sigma_1+dX^2\sigma_2+dX^3\sigma_3\), is parametrized by an
additive sum of two independent two-component complex spinors:
\EQ{
    d\mathbf{X} = \phi(\xi) \phi^\dagger(\xi) d\xi + \psi(\eta) \psi^\dagger(\eta) d\eta.
}

Here the worldsheet light-cone coordinates are \(\xi=\tau+\theta\) and \(\eta=\tau-\theta\),
where \(\tau\) is the periodic time on the fundamental torus and \(\theta\) is the spatial
radial parameter. Since \(d\mathbf{X}\) is the sum of rank-one Hermitian matrices, the
target-space coordinates \(X^\mu\) are real.

We use the notations \(\psi,\phi\) rather than the Euclidean \(\lambda,\mu\). The dependence
of the left- and right-moving spinors on a single variable \(\tau\pm\theta\) is the
Minkowski analogue of Euclidean holomorphicity. The classical minimal-surface equations
in Minkowski space reduce to the d'Alembert equation \(\partial_+\partial_-X=0\) together
with the Virasoro constraint \(\det d\mathbf{X}=0\), equivalently the null condition
\(dX_\mu^2=0\). Both conditions are satisfied by the twistor representation above.

\subsection{Complex Eigenvectors and Target-Space Monodromy}

To incorporate the meson rotation, we require the coordinate matrix \(d\mathbf{X}\) to
transform under a spatial rotation when the time coordinate winds around its fundamental
cycle on the torus, \(\tau\to\tau+2\pi\).

Because the left-moving (\(\xi\)) and right-moving (\(\eta\)) dependences are decoupled,
this target-space monodromy can be implemented by choosing the component spinors \(\phi\)
and \(\psi\) as complex eigenvectors of the \(SU(2)\) rotation group. To place the string
center at the origin, we introduce a constant \(\pi/4\) phase shift and scale the spinors
by the macroscopic geometric scale \(R\):
\EQ{
    &\phi(\xi) = \sqrt{\frac{R}{2}} \begin{pmatrix} \exp{i (a \xi + \pi/4)} \\ \exp{-i (a \xi + \pi/4)} \end{pmatrix}, \br
    &\psi(\eta) = \sqrt{\frac{R}{2}} \begin{pmatrix} \exp{i (a \eta - \pi/4)} \\ \exp{-i (a \eta - \pi/4)} \end{pmatrix}.
}
We normalize \(\psi\) and \(\phi\) identically so that
\EQ{
\psi(\eta)^\dag \psi(\eta) = \phi(\xi)^\dag \phi(\xi) = \frac{R}{2}.
}
Under an arbitrary temporal shift \(\tau\to\tau+\Delta\tau\), the arguments shift as
\(\xi\to\xi+\Delta\tau\) and \(\eta\to\eta+\Delta\tau\). The column vectors are
left-multiplied by the diagonal rotation matrix
\EQ{
   & \phi(\tau + \Delta \tau) = M(2 a\Delta \tau) \phi(\tau), \br
   &\psi(\eta + \Delta \tau) = M(2 a\Delta \tau) \psi(\eta) \br
   &\text{where  } M(\alpha) = \begin{pmatrix} \exp{\I\alpha/2} & 0 \\ 0 & \exp{-\I\alpha/2} \end{pmatrix}.
}

Substituting this into the additive factorization, the target-space coordinate differential
rotates as
\EQ{
    d\mathbf{X}(\tau + \Delta \tau) = M( 2a \Delta \tau) \, d\mathbf{X}(\tau) \, M^\dagger(2a \Delta \tau).
}
This applies to arbitrary temporal shifts \(\Delta\tau\), not only to the fundamental
period \(2\pi\). The corresponding twisted boundary condition will later be used to project
the quark-loop amplitude to fixed angular momentum \(J\) by means of the Fourier integral
\eqref{SpinProjector} with \(\alpha=4\pi a\).

\subsection{Real Coordinates and 2D Rotation}

Substituting the complex eigenvectors into the coordinate differentials, we evaluate the
diagonal components:
\EQ{
    dX^0 + dX^3 &= |\phi_1|^2 d\xi + |\psi_1|^2 d\eta =  R \, d\tau. \\
    dX^0 - dX^3 &= |\phi_2|^2 d\xi + |\psi_2|^2 d\eta =  R \, d\tau.
}

This yields \(dX^0=R\,d\tau\) and \(dX^3=0\), so the center of mass is stationary.
Evaluating the off-diagonal components gives the transverse spatial coordinates:
\EQ{
    &dX^1 - i dX^2 = \phi_1 \bar{\phi}_2 d\xi + \psi_1 \bar{\psi}_2 d\eta \br
    &= \frac{R}{2} \exp{i (2 a \xi + \pi/2)} d\xi + \frac{R}{2} \exp{i (2 a \eta - \pi/2)} d\eta \br
    &= \frac{i R}{2} \exp{2 i a \xi} d\xi - \frac{i R}{2} \exp{2 i a \eta} d\eta.
}

Integrating this differential gives
\EQ{
    &X^1 - i X^2 = \frac{R}{4 a} ( \exp{2 i a \xi} - \exp{2 i a \eta} ) \br
    &= \frac{R}{4 a} \exp{2 i a \tau} ( \exp{2 i a \theta} - \exp{-2 i a \theta} ) \br
    &= \frac{i R}{2 a} \exp{2 i a \tau} \sin(2 a \theta).
}

Separating the real and imaginary parts yields real spatial coordinates \(X^1\) and
\(X^2\). The spatial parameter \(\theta\) acts as a signed radial coordinate with origin at
the stationary center of mass, \(\theta=0\). The asymmetric interval
\(\theta\in[-\theta_d,\theta_u]\) parametrizes the open string from the antiquark at
\(-\theta_d\), through the center of mass, to the quark at \(+\theta_u\), without
duplication. The classical equations with unequal masses would be satisfied for unequal limite $\theta_{d,u}$. The string is therefore a straight line in the transverse plane, executing a
planar rotation with twist \(\alpha=4\pi a\).

\subsection{The Vanishing Metric and Spatial Boundaries}

The induced metric on the worldsheet is determined by the determinant of the target-space
coordinate differentials. In the present formulation, the conformal factor is the absolute
square of the cross-term determinant:
\EQ{
    \Omega^2 = |\phi_1 \psi_2 - \phi_2 \psi_1|^2.
}

Evaluating this for the eigenvectors above gives
\EQ{
\label{Eigenvectors}
   &\phi_1 \psi_2 - \phi_2 \psi_1 \br
   &= \frac{R}{2} ( \exp{i a (\xi-\eta) + i \pi/2} - \exp{-i a (\xi-\eta) - i \pi/2} ) \br
    &= \frac{R}{2} ( i \exp{2 i a \theta} + i \exp{-2 i a \theta} ) = i R \cos(2a\theta).
}

Taking the absolute square, \(\Omega^2=R^2\cos^2(2a\theta)\), the induced metric on the
worldsheet is
\EQ{
\label{metric}
    ds^2 = R^2 \cos^2(2a\theta) ( d\tau^2 - d\theta^2 ).
}

The Lorentzian signature \((d\tau^2-d\theta^2)\) is therefore preserved.

This metric vanishes at the roots \(\theta=\pm \frac{\pi}{4a}\), which correspond to
string endpoints moving at the speed of light, \(ds^2=0\). For massive endpoint quarks
\((m_{d,u}>0)\), the physical string is truncated at boundaries \(-\theta_d, \theta_u\) inside this
null radius,
\[
0\le \theta_{u,d}<\frac{\pi}{4a},
\]
so that the proper time remains positive.

\subsection{Classical Action Components over the Torus}

We evaluate the classical WKB action components over one temporal period
\(\tau\in[0,2\pi]\) and the bounded spatial interval \(\theta\in[-\theta_d,\theta_u]\).
We define two \(a\)-independent boundary phase parameters \(\beta_{u,d}=2a\theta_{u,d}\).

\paragraph{1. The Target Time Boundary Term (\(E \Delta X^0\)):}

The canonical energy \(E\) is conjugate to the total center-of-mass target-time advance
\(\Delta X^0\). Integrating over one temporal period gives
\EQ{
    E \Delta X^0 = E \int_0^{2\pi} R \, d\tau = 2\pi E R.
}

\paragraph{2. The Minimal Area Term (\(S_{\text{Area}}\)):}

The area of the minimal surface is obtained by integrating the conformal factor over the
bounded spatial domain:
\EQ{
    S_{\text{Area}} &= \sigma R^2 \int_0^{2\pi} d\tau \int_{-\theta_d}^{\theta_u} \cos^2(2a\theta) \, d\theta \br
    &= 2\pi \sigma R^2 \int_{-\theta_d}^{\theta_u} \frac{1 + \cos(4a\theta)}{2} \, d\theta \br
    &= \frac{\pi \sigma R^2}{2a} \left( \beta_d + \frac{1}{2}\sin(2\beta_d) + \beta_u + \frac{1}{2}\sin(2\beta_u)\right).
}

\paragraph{3. The Proper Mass Integral (\(S_{\text{mass}}\)):}

The constituent quarks of bare mass \(m_q\) propagate along the two boundaries
\(\theta_d, \theta_u\). Evaluating the proper time
\(ds=R\cos(\beta)\,d\tau\) over the torus period for both endpoints yields
\EQ{
    S_{\text{mass}} =  \int_0^{2\pi} R (m_u\cos(\beta_u) + m_d\cos(\beta_d))\, d\tau = 2\pi R (m_u\cos(\beta_u) + m_d\cos(\beta_d)).
}

\paragraph{4. The Liouville term:}

To complete the evaluation of the geometric phase, we compute the contribution of the
Liouville anomaly on the minimal surface. In conformal gauge, the Liouville action is
\EQ{
S_{\rm Liouv} = \frac{1}{12\pi} \int d\tau d\theta \left( (\partial_\theta \rho)^2 - (\partial_\tau \rho)^2 \right).
}
Using the conformal scale factor induced by the twistor parametrization in Minkowski
space, \(e^{2\rho}=R^2\cos^2(2a\theta)\), the Liouville field is purely spatial,
\(\rho(\theta)=\ln R+\ln\cos(2a\theta)\). Hence \(\partial_\tau\rho=0\), and the spatial
gradient is \((\partial_\theta\rho)^2=4a^2\tan^2(2a\theta)\).

Integrating over one full period in \(\tau\) and over the symmetric interval
\(\theta\in[-\theta_b,\theta_b]\), one obtains
\EQ{
\Delta S_{\rm Liouv} &= - \frac{1}{12\pi} (2\pi) \int_{-\theta_d}^{\theta_u} 4a^2 \tan^2(2a\theta) \, d\theta \br
&= - \frac{2a^2}{3} \int_{-\theta_d}^{\theta_u} \left( \sec^2(2a\theta) - 1 \right) d\theta.
}
Evaluating the integral and substituting \(\beta_{u,d}=2a\theta_{u,d}\) gives
\EQ{
\Delta S_{\rm Liouv} = -\frac{a}{3} (\tan \beta_u - \beta_u + \tan \beta_d - \beta_d).
}

\subsection{Path-Ordered Dirac Trace and Spin Holonomy}

To relate the continuous twistor geometry to the discrete meson spin states, we compute the
path-ordered product of the Dirac projectors along the spatial boundary of the quark loop.
Following the staggered spinor diagonalization of section 12.6, the fundamental link matrix
in Minkowski space is \(\mathcal{M}(\tau)=i\gamma_\mu v^\mu(\tau)\), where
\(v^\mu=\partial_\tau X^\mu\) is the target-space velocity.

Substituting the twistor coordinate differentials gives the velocity components at the
boundary \(\theta_b = \theta_{u,d}, \beta = \beta_{u,d}\):
\EQ{
    &v^0 = R, \quad v^3 = 0, \br
    &v^1 = -R \sin(\beta) \cos(2a\tau), \br
    &v^2 = R \sin(\beta) \sin(2a\tau).
}

Defining the transverse velocity fraction \(S=\sin(\beta)\), the invariant squared velocity
is \(v_\mu v^\mu=R^2(1-S^2)=R^2\cos^2(\beta)\). In the chiral Weyl representation, the
Dirac matrix projection takes the off-diagonal form
\(\mathcal{M}(\tau)=i\begin{pmatrix} 0 & \bar{V}(\tau) \\ V(\tau) & 0 \end{pmatrix}\),
with the \(2\times2\) blocks
\EQ{
    &V(\tau) = R \begin{pmatrix} 1 & -S \exp{2 i a \tau} \\ -S \exp{-2 i a \tau} & 1 \end{pmatrix}, \br
    &\bar{V}(\tau) = R \begin{pmatrix} 1 & S \exp{2 i a \tau} \\ S \exp{-2 i a \tau} & 1 \end{pmatrix}.
}

Applying the staggered transformation \(\Omega_n\), alternating between \(I_4\) and
\(\gamma_0\), the product of two adjacent links at step \(j\) takes the block-diagonal form
\EQ{
    \tilde{\mathcal{M}}_{j+1} \tilde{\mathcal{M}}_j = - \begin{pmatrix} \bar{V}(\tau_{j+1})V(\tau_j) & 0 \\ 0 & V(\tau_{j+1})\bar{V}(\tau_j) \end{pmatrix}.
}

Factoring out the scalar invariant \(v^2\), the continuous limit of the path-ordered
exponential separates into two independent \(2\times2\) evolution operators,
\(W(\tau)\) and \(\bar{W}(\tau)\). For the upper block, expanding
\(\bar{V}(\tau+d\tau)V(\tau)\) yields the evolution equation
\(W'(\tau)=A(\tau)W(\tau)\), with connection matrix
\EQ{
    A(\tau) = \frac{2 i a S}{1-S^2} \begin{pmatrix} -S & \exp{2 i a \tau} \\ -\exp{-2 i a \tau} & S \end{pmatrix}.
}

This system is solved by transforming to a co-rotating frame. Defining
\(W(\tau)=U(\tau)\tilde{W}(\tau)\) with
\(U(\tau)=\diag(\exp{i a \tau},\exp{-i a \tau})\), the differential equation reduces to
\(\tilde{W}'(\tau)=i a M \tilde{W}(\tau)\), with constant coefficient matrix
\EQ{
    M = \begin{pmatrix} -\frac{1+S^2}{1-S^2} & \frac{2S}{1-S^2} \\ -\frac{2S}{1-S^2} & \frac{1+S^2}{1-S^2} \end{pmatrix}.
}

The eigenvalues of \(M\) are determined by its trace and determinant:
\EQ{
    \text{tr} M = 0, \quad \det M = - \left( \frac{1+S^2}{1-S^2} \right)^2 + \left( \frac{2S}{1-S^2} \right)^2 = -1.
}

The eigenvalues of the holonomy generator \(i a M\) are \(\pm i a\), independent of the
boundary parameter \(S\). The discretized quark loop consists of \(2L\) Dirac matrices,
each corresponding to an angular step \(\Delta \tau = 2\pi / (2L)\). The staggered
transformation groups these into \(L\) adjacent pairs. The continuous evolution matrix \(M\)
therefore accumulates over a total parameter interval \(L\Delta\tau=\pi\). The solution over
\(k\) torus periods is then
\EQ{
    \tilde{W}(\pi k) = \cos(\pi a k) I + i \sin(\pi a k) M.
}

Transforming back to the original frame with \(U(\pi k)\) and taking the trace yields
\EQ{
    &\text{tr} W(\pi k) \br
    &= \exp{i \pi a k} \left( \cos(\pi a k) - i \sin(\pi a k) \frac{1+S^2}{1-S^2} \right) \br
    &\quad + \exp{-i \pi a k} \left( \cos(\pi a k) + i \sin(\pi a k) \frac{1+S^2}{1-S^2} \right) \br
    &= 2 \cos^2(\pi a k) + 2 \sin^2(\pi a k) \frac{1+S^2}{1-S^2}.
}

Substituting \(S=\sin(\beta)\), the scalar factor reduces to
\((1+S^2)/(1-S^2)=1+2\tan^2(\beta)\). The lower block \(\bar{W}(\pi k)\) gives the same
trace. The total trace of the \(4\times4\) spin holonomy is therefore
\EQ{
    T_4(k) &= \text{tr}_4 \left( \prod \limits_j i \gamma_\mu v^\mu(\tau_j) \right) \br
    &= 4 \left( 1 + 2 \sin^2(\pi a k) \tan^2(\beta) \right).
}

\subsection{Meson Vertex Operators and Topological Sectors}

Taking the logarithm of the oscillating trace \(T_4(k)\) would introduce additional complex
saddle points into the path integral. Instead, we keep the trace as a prefactor and expand
it into geometric phase factors using
\(2\sin^2(\pi a k)=1-\frac{1}{2}(e^{2\pi i a k}+e^{-2\pi i a k})\):
\EQ{
    &\frac{1}{4} T_4(k) = 1 + \tan^2(\beta) - \cos(2\pi a k) \tan^2(\beta) \br
    &= \sec^2(\beta)- \frac{1}{2} \tan^2(\beta)\left(e^{2\pi i a k} + e^{-2\pi i a k} \right).
}

The coefficients \(\sec^2(\beta)\) and \(-\frac{1}{2}\tan^2(\beta)\) are real. They determine
the transition amplitudes of the different channels without shifting the geometric phase.

To isolate the physical meson states, we insert the corresponding interpolating vertex
operator \(\Gamma\) (\(\gamma_5\) for the pseudoscalar pion, \(\gamma_\mu\) for the vector
\(\rho\)) into the trace:
\EQ{
    T_\Gamma(k) = \text{tr}_4 \left( \Gamma \prod_j i \gamma_\nu v^\nu(\tau_j) \right).
}

The pseudoscalar insertion \(\gamma_5\) isolates the constant coefficient \(\sec^2(\beta)\),
corresponding to the unshifted branch \(q=0\).

The vector insertion \(\gamma_\mu\) isolates the oscillating phase factors
\(e^{\pm 2\pi i a k}\). Combined with the macroscopic angular-momentum phase of the action,
\(e^{-4\pi i a J k}\), the total phase shift in the winding sum becomes
\EQ{
    -4\pi a J k \pm 2\pi a k = -4\pi a \left( J \mp \frac{1}{2} \right) k.
}
This shifts the target angular momentum in the quantization condition to
\(u=J+h/2\), where \(h\in\{0,\pm 1\}\), and leads to the corresponding half-integer
spin-orbit splittings of the meson Regge trajectories.

Note that these shifts occur independently for the quark and antiquark (upper or lower end of the open string), which is different from what was assumed in  a simplified amplitude in \cite{Migdal2026GeometricQCDII} where only  one quark trajectory was assumed as opposed to the case of two trajectories considered in the present paper. Also, the symmetry between the two ends of this open string was assumed in \cite{Migdal2026GeometricQCDII} which can only be applied to the mesons made of identical quarks (light meson sector).
In this sense, the present computation represents a generalization of the previous spectrum treatment to the case of unequal quark masses.

\section{The Exact WKB Fluctuation Determinant}
\label{sec:exact_wkb}

To analyse the semiclassical Bohr--Sommerfeld quantization and the corresponding transition
amplitudes of the meson resonances, we evaluate the 1-loop determinant of the quantum
fluctuations around the classical Minkowski helicoid.

Unlike standard bosonic string theories, where one integrates over fluctuating bulk metrics
and independent worldsheet coordinates, the present construction is rigid: the minimal
surface is determined by analytic continuation from the boundary twistor data. The quantum
fluctuations are therefore encoded in the boundary twistor fields.

A preliminary point concerns the role of Hodge chirality. The Hodge sign
\(\chi=\pm1\) belongs to the hidden area derivative and to the corresponding loop-space
zero mode, not to the scalar bosonic action itself. The two Hodge sectors have the same
extremal area but different area derivatives and different Hessians. By contrast, the scalar
area density and the Liouville term depend only on the conformal factor
\(\Omega^2=|\phi_1\psi_2-\phi_2\psi_1|^2\), and are therefore independent of the Hodge
sign. For a symmetric string with equal constituent quark masses, one could impose parity on the
quantum twistor fluctuations. However, for asymmetric heavy-light mesons (\(m_u \neq m_d\)), the unequal boundary integration limits explicitly break this parity. Hodge duality, meanwhile, belongs to the map from the physical boundary loop into the hidden
\(R^3\otimes R^{1,3}\) surface and remains an exact symmetry of the construction.

\subsection{Multiplicative Ansatz and asymmetric strip fields}

Let the classical twistor fields of the Minkowski helicoid be \(\phi_0(\xi)\) and
\(\psi_0(\eta)\), where
\[
\xi=\tau+\theta,\qquad \eta=\tau-\theta,
\]
are the chiral light-cone coordinates. The classical background carries the entire target-space
monodromy associated with the macroscopic twist angle \(\alpha=4\pi a\).

To expand the effective action without violating the twisted boundary conditions, we use the
norm-preserving multiplicative ansatz
\begin{equation}
\tilde\phi(\xi)=\exp{f_1(\xi)}\phi_0(\xi),
\qquad
\tilde\psi(\eta)=\exp{f_2(\eta)}\psi_0(\eta),
\label{eq:expansatz}
\end{equation}
where \(f_1(\xi)\) and \(f_2(\eta)\) are anti-Hermitian \(2\times2\) matrix fluctuations,
\begin{equation}
f_1^\dagger=-f_1,
\qquad
f_2^\dagger=-f_2.
\label{eq:antiherm}
\end{equation}
This preserves the equal normalization of the twistor boundary values. Since the classical
background already carries the full monodromy, the fluctuations are periodic on the
fundamental torus:
\begin{equation}
f_1(\xi+2\pi)=f_1(\xi),
\qquad
f_2(\eta+2\pi)=f_2(\eta).
\label{eq:fperiodic}
\end{equation}

The spatial dependence of the left- and right-moving twistor fluctuations is encoded in the pair of strip fields:
\begin{equation}
F_1(\tau,\theta)\equiv f_1(\tau+\theta),
\qquad
F_2(\tau,\theta)\equiv f_2(\tau-\theta).
\label{eq:stripfields}
\end{equation}
Expanding these fields in Fourier modes, we obtain:
\begin{equation}
F_1(\tau,\theta)=\sum_{n\in\mathbb Z} A_n \exp{\I n\tau}\exp{\I n\theta},
\qquad
F_2(\tau,\theta)=\sum_{n\in\mathbb Z} B_n \exp{\I n\tau}\exp{-\I n\theta}.
\label{eq:F1F2Fourier}
\end{equation}
In the simplified case of identical constituent quark masses, the symmetric physical domain \(\theta \in [-\theta_b, \theta_b]\) admits a parity reflection symmetry \(\theta \to -\theta\), which forces the Fourier coefficients to decouple into parity-invariant sectors (\(A_n = \pm B_n\)). However, for the general physical case of unequal quark masses, the string is truncated at the asymmetric limits \(\theta \in [-\theta_d, \theta_u]\). This fundamentally breaks the global parity symmetry on the strip, meaning the left-moving and right-moving Fourier modes (\(A_n\) and \(B_n\)) remain independent dynamical variables that fully couple in the effective action.

\subsection{Bosonic action and asymmetric algebraic factorization}

The minimal area and the Liouville term depend only on the scalar conformal factor \(\Omega^2\).
While the bulk action density is intrinsically symmetric, integrating it over the asymmetric boundary
interval \(x\in[-\beta_d,\beta_u]\) explicitly breaks the parity of the fluctuation operator.
Consequently, the definite integrals of odd trigonometric functions do not vanish.

In standard semiclassical string theory, evaluating the fluctuation determinant around a
curved background leads to continuous boundary-value problems, often treated by the
Gel'fand--Yaglom or Sturm--Liouville methods. In the rigid twistor-string formulation,
the analyticity of the construction leads instead to an algebraic reduction on the mass shell.

Because the bulk minimal surface is determined by analytic continuation of the
one-dimensional boundary data, the spatial dependence of the twistor fluctuations
\(f_1(\xi)\) and \(f_2(\eta)\) is tied to the chiral light-cone coordinates. Expanding these
fields in Fourier modes over the temporal cycles turns the spatial derivatives, including
those entering the Liouville anomaly \(\partial_\theta\rho\), into algebraic operations.

As shown in Appendix C, integrating the quadratic effective action over the asymmetric physical string
width \(x\in[-\beta_d,\beta_u]\) reduces the continuous functional determinant to a discrete
infinite product of finite-dimensional matrix blocks. While the transverse and longitudinal
twistor fluctuations decouple from each other, the broken parity symmetry causes the left- and right-moving sine and cosine modes to mix. For each Fourier mode \(n\), the action is thus represented by
algebraic matrix blocks \(\mathcal{H}_n^{(\perp)}\) and \(\mathcal{H}_n^{(\parallel)}\) that naturally upgrade from \(2 \times 2\) to \(4 \times 4\) dimensions, constructed from exact rational trigonometric quadratures evaluated over the asymmetric interval.

This reduction replaces the differential-operator problem by a sequence of finite-dimensional
matrix quadratures.

\subsection{Algebraic determinant, the \(n \leftrightarrow -n\) pairing, and WKB residues}

To evaluate the transition amplitudes of the meson resonances, we keep the total staggered Dirac
spin holonomy trace \(T_{\mathrm{total}}(k)\) at its classical value and compute the regularized determinant of the
geometric fluctuation Hessian.

The resulting matrix blocks retain two profound structural properties. First, despite the spatial asymmetry,
the trigonometric matrix elements remain symmetric under frequency reversal,
\(\nu_n\to-\nu_n\), which implies the \(n\leftrightarrow -n\) pairing
\begin{equation}
\det \mathcal{H}_{n} = \det \mathcal{H}_{-n}.
\end{equation}
This symmetry gives the trace cancellation \(\tr_{\mathbb{Z}} \mathbf{1}=0\), so that the
determinant is independent of the winding number \(k\).

Second, substituting the classical saddle-point relations \(R=aK\) and \(x=2a\theta\) into
the matrix integrands extracts an overall factor of \(a\) for each physical degree of freedom.
As in the \(k\)-dependence, the \(\zeta\)-regularized product over the asymmetric Fourier
tower perfectly cancels this overall multiplicative factor (see Appendix C.6):
\EQ{
\det \mathcal{H}_{\mathrm{geom}}^{(k, a)} = (ka)^{\tr_{\mathbb{Z}} \mathbf{1}} \det \widetilde{\mathcal{H}}_{\mathrm{geom}}(\beta_d, \beta_u, K; a) = \det \widetilde{\mathcal{H}}_{\mathrm{geom}}(\beta_d, \beta_u, K; a).
}
While this overall scale factor of \(a\) cancels, the reduced determinant
\(\mathcal{D}(\beta_d, \beta_u, K; a)\equiv (\det \widetilde{\mathcal{H}}_{\mathrm{geom}})^{-1/2}\)
retains a nontrivial dependence on the twistor scale \(a\) through the fractional mode
frequencies \(\nu_n=n/(2a)\). Since this dependence enters only through bounded
oscillatory trigonometric functions, the cancellation of the overall scale does not introduce
additional fractional powers or branch cuts into the spin-projection integrand.

Treating \(a\) as an external modulus restricted to the physical interval, the projected 1-loop amplitude decomposes into distinct channels determined by the independent topological indices \(h_u, h_d \in \{0, \pm 1\}\) isolated by the vertex insertions at the respective boundaries:
\EQ{
A_{J}^{\text{1-loop}}(E; \beta_d, \beta_u) = \sum_{h_u, h_d} \mathcal{C}_{h_u, h_d}(\beta_d, \beta_u) I_{h_u, h_d}(\Phi_*, \beta_d, \beta_u),
}
where \(\mathcal{C}_{h_u, h_d}(\beta_d, \beta_u)\) are the real coefficients from the product of the classical spin traces, and
\EQ{
I_{h_u, h_d}(\Phi_*, \beta_d, \beta_u) = \int_{a_{\text{min}}}^{1/2} da \, \mathcal{D}(\beta_d, \beta_u, K; a) \sum_{k=1}^\infty \exp{\I k a \left(\Phi_* + 2\pi h_u + 2\pi h_d\right)}.
}

Because the regularized determinant is \(k\)-independent, the geometric series over
topological windings can be summed by Abel regularization:
\EQ{
\sum_{k=1}^\infty \exp{-\epsilon k} \exp{\I a k X} = \frac{1}{\exp{\epsilon - \I aX} - 1} = \frac{\I}{a(X + \I 0)} + \mathcal{O}(1).
}
Since the overall scale anomaly cancels, the integration over \(a\) introduces no additional
branch cuts. The singular part of the on-shell meson amplitude is therefore meromorphic, with the total topological shift \(h = h_u + h_d\) defining the discrete daughter trajectories:
\EQ{
A_{J}^{\text{1-loop}}(E; \beta_d, \beta_u) \sim \sum_{h_u, h_d} \frac{\mathcal{R}_{h_u, h_d}(\beta_d, \beta_u)}{\Phi_*(E, J; \beta_d, \beta_u) + 2\pi h + \I 0}.
}
The parent trajectory is located at \(\Phi_*=0\), while the independent spin-holonomy phases
generate the corresponding integer and half-integer daughter trajectories. The corresponding transition residues are obtained
by integrating the \(a\)-dependent 1-loop determinant over the physical spin-projection
interval:
\EQ{
\mathcal{R}_{h_u, h_d}(\beta_d, \beta_u) &= \I \, \mathcal{C}_{h_u, h_d}(\beta_d, \beta_u) \int_{a_{\text{min}}}^{1/2} \frac{da}{a} \mathcal{D}(\beta_d, \beta_u, K; a).
\label{Rq}
}
These formulas summarize the 1-loop S-matrix structure of the rigid twistor string. The
determinant renormalizes the transition residues through quadrature over \(a\), while leaving the exact geometric pole positions unchanged.
\section{Meson trajectories and connection to phenomenology}
\label{sec:phenomenology}

To perform the global optimization, we assign equal statistical weight to each of the 40 PDG states and minimize the unweighted sum of squared geometric action monodromies, $\chi^2 = \sum \Phi^2(M_{\text{exp}}, J)$. Because the parameter $\Phi$ depends linearly of angular momentum $J$ , this objective function natively reproduces the standard mass-squared weighting of Regge trajectories without introducing artificial mass-dimension biases.

The bound states of the twistor string are governed by the parametric equations derived in Appendix A. Parameterized by the independent spatial boundary phases \(\beta_u, \beta_d \in [0, \pi/2)\) and the corresponding dimensionless renormalized boundary masses \(x_{u,d} = m_{u,d}/\sqrt{\sigma}\), the mass \(M\) and angular momentum \(J\) of the meson states are given by summing the classical geometric contributions from the two string boundaries:
\begin{align}
\frac{M(\beta_u, \beta_d)}{\sqrt{\sigma}} &= \frac{\mathcal{K}}{2} \left(\beta_u + \frac{\sin 2\beta_u}{2} + \beta_d + \frac{\sin 2\beta_d}{2}\right) + x_u \cos \beta_u + x_d \cos \beta_d, \label{eq:MassParam} \\
J(\beta_u, \beta_d) &= \frac{\mathcal{K}^2}{8} \left(\beta_u + \frac{\sin 2\beta_u}{2} + \beta_d + \frac{\sin 2\beta_d}{2}\right) - \frac{1}{12\pi} (\tan \beta_u - \beta_u + \tan \beta_d - \beta_d) - \frac{h}{2}, \label{eq:JParam}
\end{align}
where the dimensionless diameter \(\mathcal{K}\) acts as a kinematic link binding the two boundaries. Because the macroscopic string tension must be uniform, the two endpoints are coupled by the decoupled saddle-point conditions:
\begin{equation}
\mathcal{K} = \frac{\sin \beta_u}{\cos^2 \beta_u} \left(x_u + \sqrt{x_u^2 - \frac{1}{3\pi}}\right) = \frac{\sin \beta_d}{\cos^2 \beta_d} \left(x_d + \sqrt{x_d^2 - \frac{1}{3\pi}}\right).
\end{equation}
This equality shifts the center of rotation toward the heavier quark and relates the two boundary phases, leaving the asymmetric string parameter-free once the boundary masses are specified.

Here, \(h = h_u + h_d \in \{0, \pm 1, \pm 2\}\) is the topological shift originating from the trace of the spin holonomy at the string boundaries. This integer shifts the orbital angular momentum by \(-h/2\), separating the hadronic spectrum into distinct constituent spin-orbit topologies.

To compare these theoretical trajectories with phenomenology, we perform a global optimization over 40 states from the Particle Data Group (PDG), spanning multiple meson families across all five topological sectors.

A preliminary fit of the meson resonances by this spectrum was performed in Part II \cite{Migdal2026GeometricQCDII}, limited to symmetric boundary assumptions and utilizing only the \(h=0\) and \(h=-1\) sectors. Here, we perform an evaluation utilizing the asymmetric geometric mechanics and the topological constituent algebra.

By minimizing the action monodromy condition \(\Phi(M,J) = 0\), we extract five parameters---the planar string tension and the four constituent boundary masses of the quarks:
\begin{align}
\sqrt{\sigma} &\approx 0.412136 \text{ GeV}, \quad m_u \approx 0.134247 \text{ GeV}, \nonumber \\
m_s &\approx 0.221086 \text{ GeV}, \quad m_c \approx 1.44844 \text{ GeV}, \quad m_b \approx 4.78122 \text{ GeV}.
\end{align}

Because the geometric radius \(\mathcal{K}\) must be uniform across the string, the saddle-point equations balance two unequal masses (\(x_u \neq x_d\)) by breaking the spatial parity (\(\beta_u \neq \beta_d\)). This formulation eliminates the need for arithmetic mean mass averaging for heavy-light systems. The asymmetric heavy trajectories (such as \(D, D_s, B, B_s, B_c\)) are geometric predictions of the theory.

\begin{remark}{\textit{The renormalized boundary mass.}}
The chiral existence bound \(x_i \ge 1/\sqrt{3\pi}\) forbids the limit of a massless string boundary. The geometric drag generates a minimum inertial mass required to balance the anomaly at the boundaries. The parameter \(m_i\) appearing in the string effective action is therefore neither the bare QCD quark mass nor the standard phenomenological constituent mass; it is a renormalized boundary mass, additively shifted by the perimeter term of the worldsheet fermion determinant.
\end{remark}

Because the spectrum spans multiple spin-orbit topologies that overlap in the \((M^2, J)\) plane, we present the theoretical trajectories and experimental data separated into respective subsections based on the total topological shift \(h\). In the tables below, alongside the masses, we tabulate the algebraic amplitude weight \(\mathcal{C}_{|h|}\). Extracted from the 1-loop boundary fluctuation determinants, this factor governs the geometric scaling of the resonance creation amplitudes (as discussed in Section \ref{sec:transition_amplitudes}).

\subsection{Pseudoscalars (\(h = 0\))}
The \(h = 0\) sector corresponds to the unnatural-parity parent trajectories (\(S=0\)), where the intrinsic string angular momentum is entirely orbital (\(J=L\)). This topology houses the pseudoscalar ground states (\(\pi, K, D, D_s, B, B_s, B_c\)) and their higher-spin orbital excitations.

The twistor string reproduces the asymptotic linear Regge behavior while capturing the curvature at low spins induced by the heavy quark boundaries. A notable deviation is the ground-state pion, which falls below the string's classical mass threshold (\(2m_u \approx 268\) MeV) due to chiral symmetry breaking of the QCD vacuum---a non-perturbative target-space effect outside the scope of the classical minimal surface. 

\begin{table}[htpb]
\centering
\begin{tabular}{lcccc}
\hline\hline
State & $J$ & $M_{\rm th}$ (GeV) & $M_{\rm exp}$ (GeV) & $\mathcal{C}_{|h|=0}$ \\
\hline
$\pi$ & 0 & 0.268 & 0.140 & 1.0000 \\
$\pi_2$ & 2 & 1.569 & 1.671 & 80.4785 \\
$K$ & 0 & 0.355 & 0.494 & 1.0000 \\
$K_2$ & 2 & 1.613 & 1.580 & 29.8572 \\
$D$ & 0 & 1.583 & 1.865 & 1.0000 \\
$D_2$ & 2 & 2.646 & 2.747 & 10.5403 \\
$D_s$ & 0 & 1.670 & 1.968 & 1.0000 \\
$B$ & 0 & 4.915 & 5.279 & 1.0000 \\
$B_s$ & 0 & 5.002 & 5.367 & 1.0000 \\
$B_c$ & 0 & 6.230 & 6.275 & 1.0000 \\
\hline\hline
\end{tabular}
\caption{Theoretical masses versus experimental PDG masses for the Pseudoscalar (\(h=0\)) topology. $\mathcal{C}_{|h|}$ denotes the algebraic transition weight evaluated at the saddle point.}
\label{tab:q0_spectrum}
\end{table}

\begin{figure}[htpb]
\centering
\includegraphics[width=0.95\textwidth]{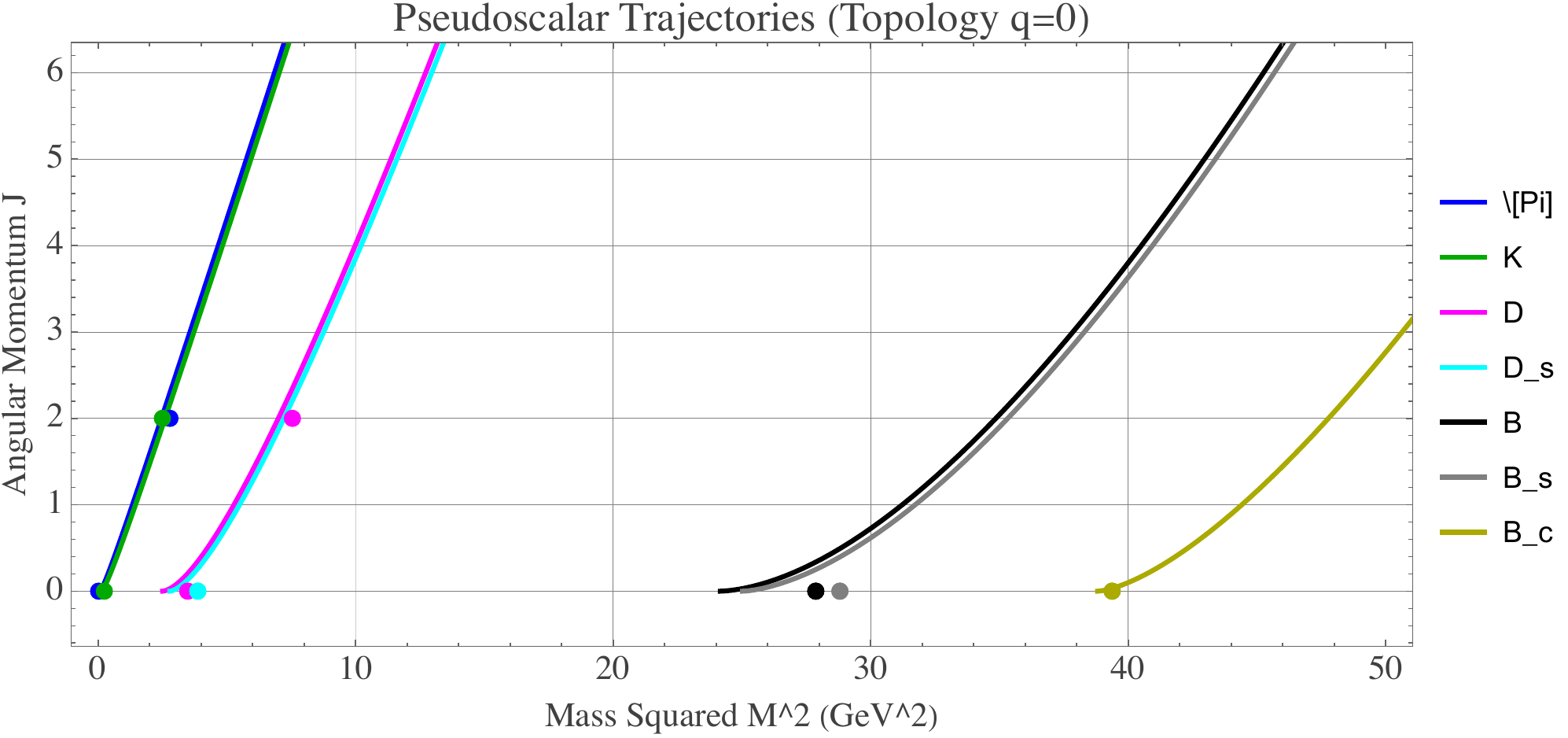}
\caption{Parametric trajectories for the Pseudoscalar (\(h=0\)) families. The solid curves represent the analytical theory evaluated with asymmetric boundaries, and markers denote experimental PDG data.}
\label{fig:q0_plot}
\end{figure}

\subsection{Vectors (\(h = -1\))}
The \(h=-1\) topological sector shifts the angular momentum by \(+1/2\). This encompasses states exhibiting partial spin-orbit alignment, mapping the vector meson ground states (\(\rho, K^*, D^*, B^*, \dots\)) and their odd-spin descendants (\(\rho_3, \rho_5\)). The suppression of the boundary phases \(\beta_{u,d}\) for heavy strings dictates that the heavy vectors lie near the origin of the trajectory, agreeing within statistical errors with the experimental data.

\begin{table}[htpb]
\centering
\begin{tabular}{lcccc}
\hline\hline
State & $J$ & $M_{\rm th}$ (GeV) & $M_{\rm exp}$ (GeV) & $\mathcal{C}_{|h|=1}$ \\
\hline
$\rho$ & 1 & 0.869 & 0.775 & 13.6431 \\
$\rho_3$ & 3 & 1.737 & 1.689 & 62.9969 \\
$\rho_5$ & 5 & 2.284 & 2.330 & 111.6537 \\
$K^*$ & 1 & 0.925 & 0.892 & 5.0314 \\
$K^*_3$ & 3 & 1.779 & 1.780 & 21.9570 \\
$K^*_5$ & 5 & 2.320 & 2.380 & 38.5279 \\
$D^*$ & 1 & 2.061 & 2.008 & 2.2322 \\
$D^*_3$ & 3 & 2.790 & 2.763 & 5.8049 \\
$D_s^*$ & 1 & 2.114 & 2.112 & 0.6579 \\
$D_{s3}^*$ & 3 & 2.829 & 2.860 & 1.8552 \\
$B^*$ & 1 & 5.371 & 5.325 & 2.2347 \\
$B_s^*$ & 1 & 5.422 & 5.415 & 0.6532 \\
\hline\hline
\end{tabular}
\caption{Theoretical masses versus experimental PDG masses for the Vector (\(h=-1\)) topology.}
\label{tab:qm1_spectrum}
\end{table}

\begin{figure}[htpb]
\centering
\includegraphics[width=0.95\textwidth]{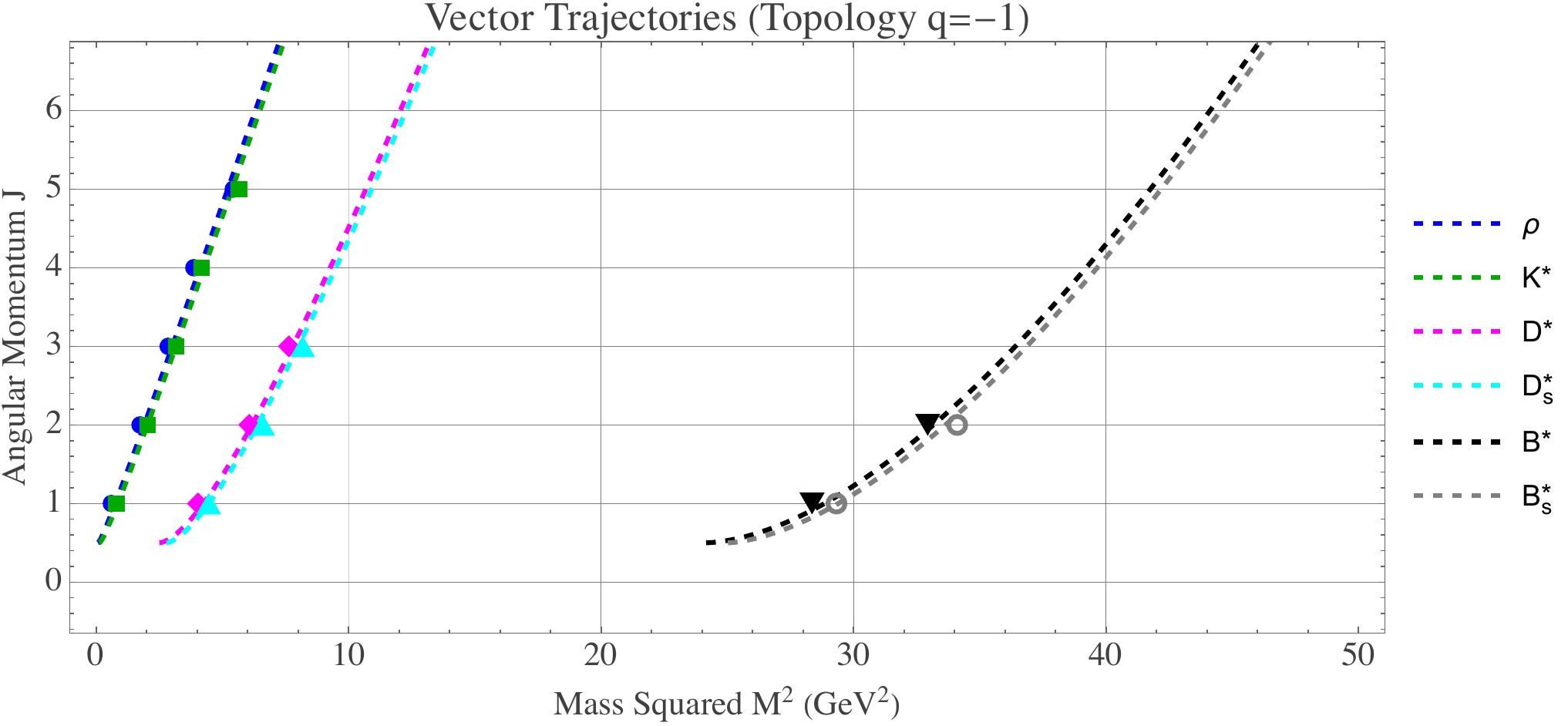}
\caption{Parametric trajectories for the Vector (\(h=-1\)) families (spin-orbit aligned). The dashed curves represent the theoretical predictions.}
\label{fig:qm1_plot}
\end{figure}

\subsection{Tensors (\(h = -2\))}
The \(h = -2\) sector realizes an angular momentum shift of \(+1\), corresponding to the aligned natural-parity states where \(J = L + 1\). This sector maps the tensor and high-spin aligned trajectories, including states such as the \(a_2(1320)\) and \(a_4(2040)\) and their corresponding heavy-quark partners. 

\begin{table}[htpb]
\centering
\begin{tabular}{lcccc}
\hline\hline
State & $J$ & $M_{\rm th}$ (GeV) & $M_{\rm exp}$ (GeV) & $\mathcal{C}_{|h|=2}$ \\
\hline
$a_2$ & 2 & 1.156 & 1.318 & 5.3782 \\
$a_4$ & 4 & 1.889 & 1.967 & 16.7513 \\
$K^*_2$ & 2 & 1.206 & 1.430 & 1.5791 \\
$K^*_4$ & 4 & 1.929 & 2.045 & 5.1816 \\
$D^*_2$ & 2 & 2.297 & 2.463 & 0.1076 \\
$D_{s2}^*$ & 2 & 2.344 & 2.569 & 0.0332 \\
$B^*_2$ & 2 & 5.588 & 5.739 & 0.0138 \\
$B_{s2}^*$ & 2 & 5.633 & 5.840 & 0.0044 \\
\hline\hline
\end{tabular}
\caption{Theoretical masses versus experimental PDG masses for the Tensor (\(h=-2\)) topology.}
\label{tab:qm2_spectrum}
\end{table}

\begin{figure}[htpb]
\centering
\includegraphics[width=0.95\textwidth]{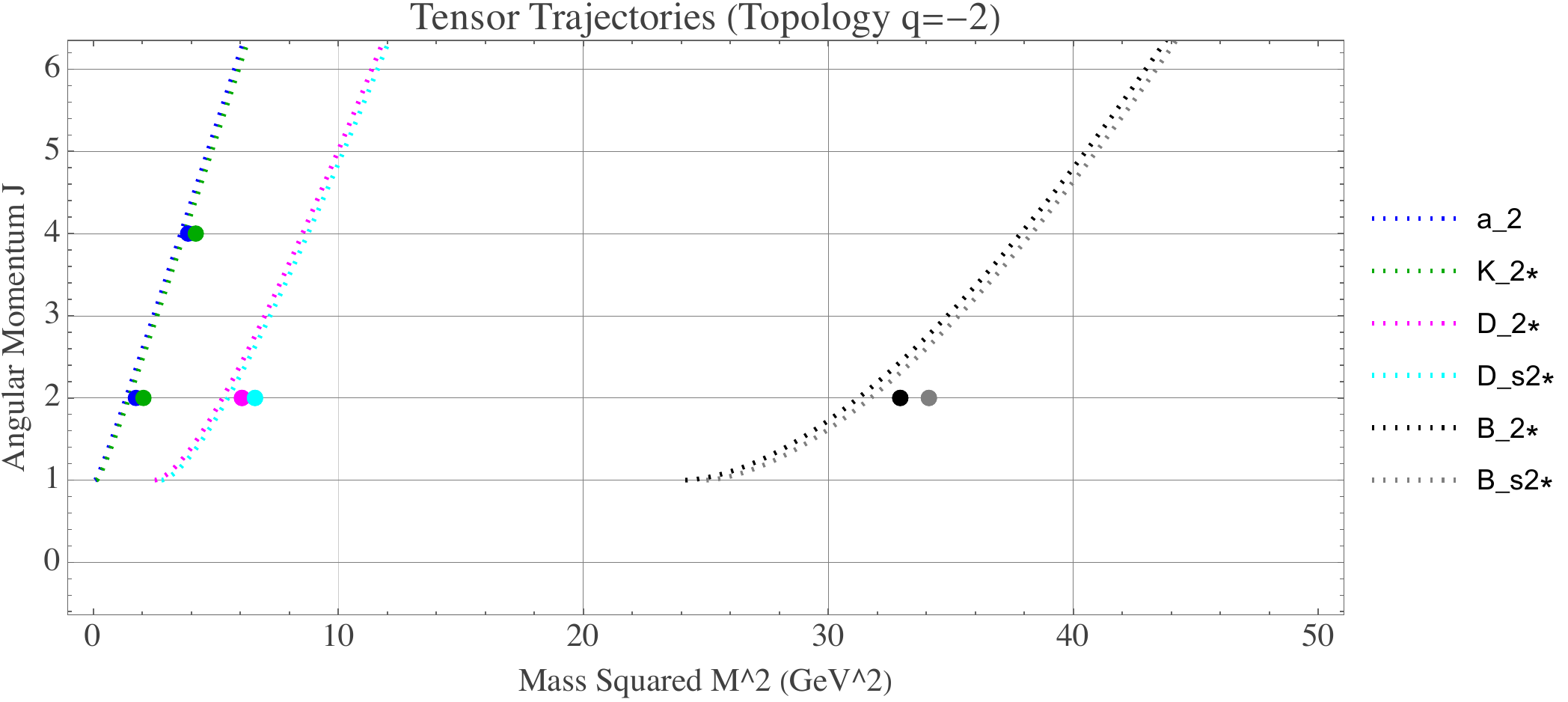}
\caption{Parametric trajectories for the Tensor (\(h=-2\)) families. The dotted curves reflect the theoretical predictions for these \(J = L + 1\) states.}
\label{fig:qm2_plot}
\end{figure}

\subsection{Axials (\(h = +1\)) and Geometric Degeneracy}
The \(h = +1\) sector shifts the angular momentum backward by \(-1/2\), housing the axial-vector parity doublets (\(b_1, K_1, D_1\dots\)) where total quark spin acts against the orbital rotation (\(J=L-S\)). 

In constituent quark models and Heavy Quark Effective Theory (HQET), the mass splittings of these doublets require the introduction of phenomenological spin-orbit coupling parameters. In the twistor string framework, the solution is topological and parameter-free. 

From Eq. (\ref{eq:JParam}), a state with spin \(J\) on the \(h=+1\) trajectory imposes the invariant \(\text{Geometry}(\beta_u, \beta_d) = J + 0.5\). A state with spin \(J+1\) on the \(h=-1\) trajectory requires the same constraint: \(\text{Geometry}(\beta_u, \beta_d) = (J+1) - 0.5 = J + 0.5\). Because mass depends on this geometric configuration, the theory yields a degeneracy spanning the two topologies:
\begin{equation}
M(J, \, h=+1) = M(J+1, \, h=-1).
\end{equation}
This establishes a \(J-1\) degeneracy, matching the heavy axials (\(B_1, B_{s1}\)) with their \(J=2\) vector counterparts.

\begin{table}[htpb]
\centering
\begin{tabular}{lcccc}
\hline\hline
State & $J$ & $M_{\rm th}$ (GeV) & $M_{\rm exp}$ (GeV) & $\mathcal{C}_{|h|=1}$ \\
\hline
$b_1$ & 1 & 1.379 & 1.229 & 38.4994 \\
$K_1$ & 1 & 1.425 & 1.272 & 13.5870 \\
$D_1$ & 1 & 2.485 & 2.422 & 4.2072 \\
$D_{s1}$ & 1 & 2.528 & 2.460 & 1.3164 \\
$B_1$ & 1 & 5.759 & 5.726 & 3.9130 \\
$B_{s1}$ & 1 & 5.800 & 5.829 & 1.2054 \\
\hline\hline
\end{tabular}
\caption{Theoretical masses versus experimental PDG masses for the Axial (\(h=+1\)) topology.}
\label{tab:qp1_spectrum}
\end{table}

\begin{figure}[htpb]
\centering
\includegraphics[width=0.95\textwidth]{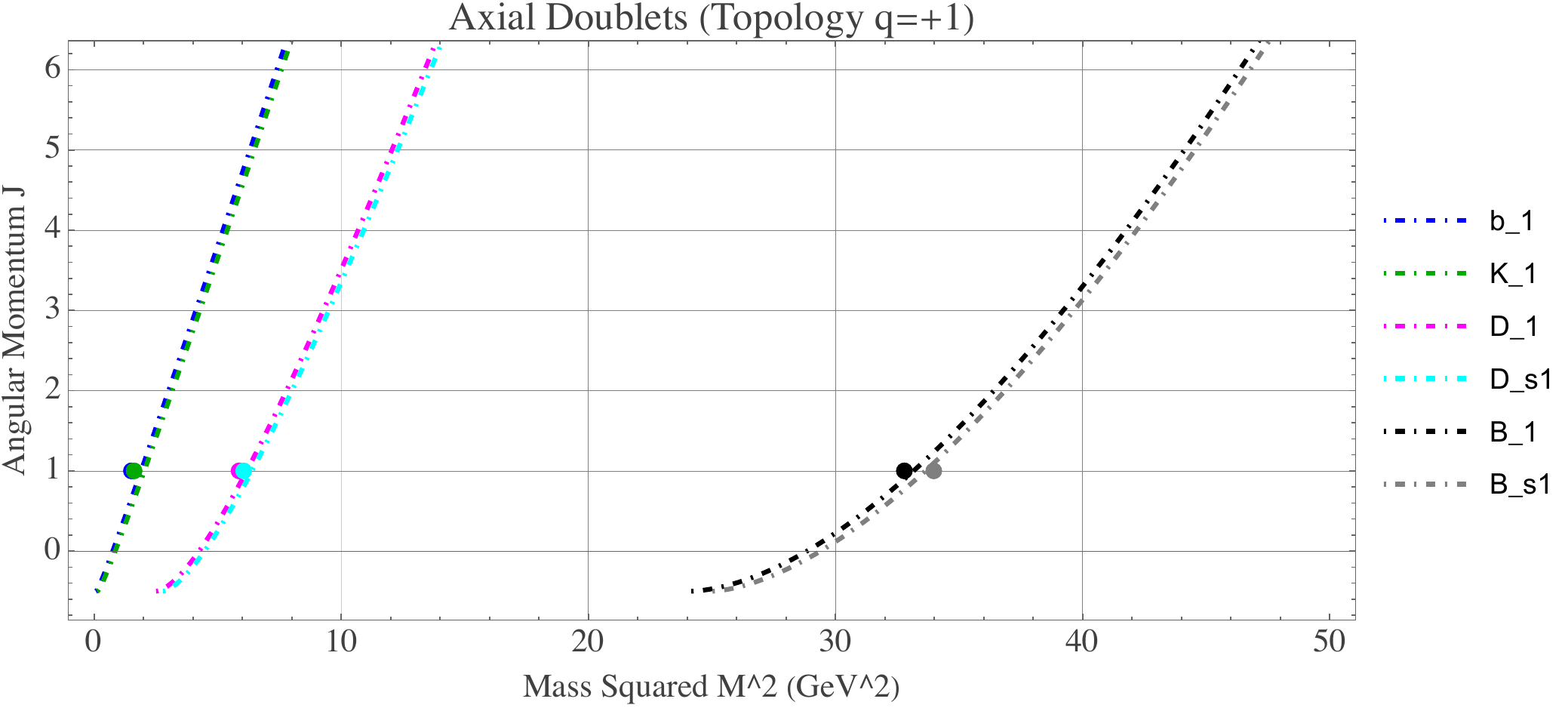}
\caption{Parametric trajectories for the Axial Parity Doublets (\(h=+1\)) (spin-orbit anti-aligned). The dot-dashed curves reflect the \(J-1\) geometric degeneracy linking them to the vector states mapped in Figure \ref{fig:qm1_plot}.}
\label{fig:qp1_plot}
\end{figure}

\subsection{Scalars (\(h = +2\)) and Extended Degeneracy}
The \(h = +2\) topology corresponds to a spin-orbit anti-alignment shift of \(-1\), housing the scalar mesons (\(a_0, K_0^*, D_0^*, D_{s0}^*\)). 

The geometric degeneracy extends to this sector. Following the relations above, a scalar state at \(J=0\) on the \(h=+2\) trajectory corresponds to the same classical string surface as a tensor state at \(J=2\) on the \(h=-2\) trajectory. This yields a mapping across a \(\Delta J = 2\) gap:
\begin{equation}
M(J, \, h=+2) = M(J+2, \, h=-2), \qquad \mathcal{C}_{|h|}(J, \, h=+2) = \mathcal{C}_{|h|}(J+2, \, h=-2).
\end{equation}
Comparing Table \ref{tab:qp2_spectrum} with Table \ref{tab:qm2_spectrum}, we observe this symmetry. The theoretical mass for the \(\bar{q}q\) scalar \(a_0\) (\(1.156\) GeV) maps to the mass of the tensor \(a_2\) (\(1.156\) GeV). Similarly, the \(D_0^*\) scalar prediction (\(2.297\) GeV) matches the \(D_2^*\) tensor prediction (\(2.297\) GeV).

\begin{table}[htpb]
\centering
\begin{tabular}{lcccc}
\hline\hline
State & $J$ & $M_{\rm th}$ (GeV) & $M_{\rm exp}$ (GeV) & $\mathcal{C}_{|h|=2}$ \\
\hline
$a_0$ & 0 & 1.156 & 0.980 & 5.3782 \\
$K_0^*$ & 0 & 1.206 & 1.425 & 1.5791 \\
$D_0^*$ & 0 & 2.297 & 2.300 & 0.1076 \\
$D_{s0}^*$ & 0 & 2.344 & 2.317 & 0.0332 \\
\hline\hline
\end{tabular}
\caption{Theoretical masses versus experimental PDG masses for the Scalar (\(h=+2\)) topology. Note the mass and amplitude degeneracy with the \(J=2\) states in Table \ref{tab:qm2_spectrum}.}
\label{tab:qp2_spectrum}
\end{table}

\begin{figure}[htpb]
\centering
\includegraphics[width=0.95\textwidth]{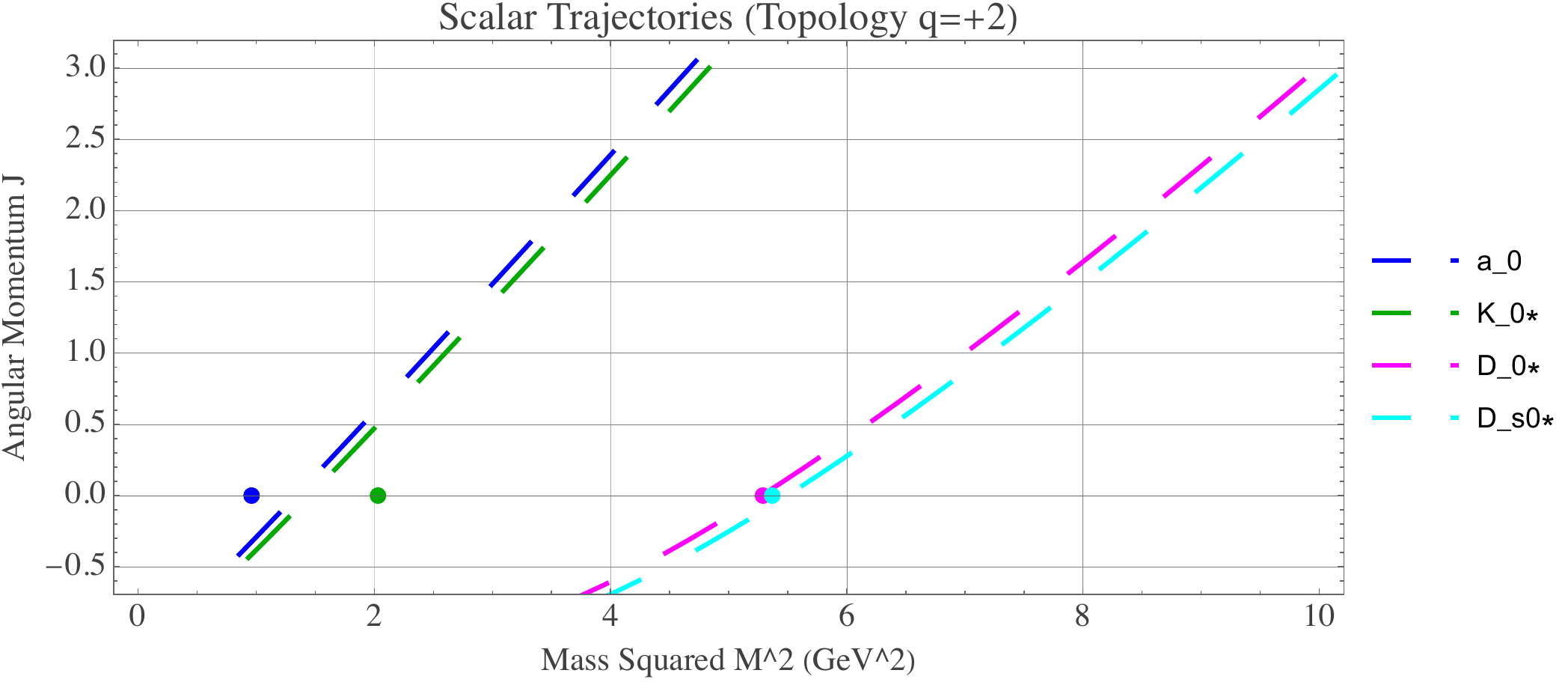}
\caption{Parametric trajectories for the Scalar (\(h=+2\)) families. The dashed curves reflect the \(J-2\) geometric degeneracy linking these boundary scalars to their tensor partners mapped in Figure \ref{fig:qm2_plot}.}
\label{fig:qp2_plot}
\end{figure}

\subsection{Identifying Molecular States}
Because the string formulation restricts the parameter space to a single tension and invariant constituent boundary masses, it establishes a diagnostic baseline for planar \(q\bar{q}\) states. The failure of an experimental resonance to align with the trajectory signals that the state deviates from a simple valence \(q\bar{q}\) structure. 

If an experimental state is a multiquark configuration (such as a $qq\bar{q}\bar{q}$ tetraquark) or a loosely bound hadronic molecule (such as a $K\bar{K}$ bound state), it corresponds to a different topology---either a worldsheet with four boundaries, or two interacting string surfaces. Such states violate the boundary conditions governing the single open string. 

An example is observed in the scalar sector (Table \ref{tab:qp2_spectrum}) with the light \(a_0(980)\) resonance (and its isoscalar partner, the \(f_0(980)\)). While the experimental mass of the \(a_0\) is $0.980$ GeV, the geometric degeneracy dictates that the lowest $q\bar{q}$ scalar state lies at $1.156$ GeV (matched to the tensor \(a_2\)). This mismatch supports the phenomenological hypothesis that the \(a_0(980)\) is not a conventional $P$-wave $q\bar{q}$ meson, but a multiquark configuration or a $K\bar{K}$ molecule. The $q\bar{q}$ scalar ground state is likely higher in the spectrum (e.g., the \(a_0(1450)\)), while the \(0.980\) GeV state is identified as an exotic candidate.

By providing a $q\bar{q}$ baseline, the twistor string serves as a filter. Hadronic molecules and tetraquark exotics do not satisfy the geometric constraints of the single minimal surface, appearing as outliers within the parametric spectrum.

\begin{table}[htbp]
\centering
\scriptsize
\renewcommand{\arraystretch}{1.15}
\begin{tabular}{lccccc}
\hline\hline
Flavor Mix & $J$ & $h$ & $M_{\rm th}$ (GeV) & $\Sigma_J / \Sigma_\rho$ & Rel. Error \\
\hline
u (l-u) & 1 & -1 & 0.869 & $1.000 \pm 6.31\times 10^{-4}$ & 0.063\% \\
u (l-u) & 2 & -1 & 1.379 & $1.296 \pm 1.14\times 10^{-3}$ & 0.088\% \\
u (l-u) & 3 & -1 & 1.737 & $2.169 \pm 2.34\times 10^{-3}$ & 0.108\% \\
u (l-u) & 1 & 0 & 1.156 & $3.853 \pm 2.95\times 10^{-3}$ & 0.077\% \\
u (l-u) & 2 & 0 & 1.569 & $2.316 \pm 2.28\times 10^{-3}$ & 0.098\% \\
u (l-u) & 3 & 0 & 1.889 & $6.607 \pm 7.69\times 10^{-3}$ & 0.116\% \\
u (l-u) & 1 & 1 & 1.379 & $1.296 \pm 1.14\times 10^{-3}$ & 0.088\% \\
u (l-u) & 2 & 1 & 1.737 & $2.169 \pm 2.34\times 10^{-3}$ & 0.108\% \\
u (l-u) & 3 & 1 & 2.029 & $1.667\times 10^{1} \pm 2.07\times 10^{-2}$ & 0.124\% \\
u (l-u) & 1 & 2 & 1.569 & $3.175\times 10^{-1} \pm 3.13\times 10^{-4}$ & 0.098\% \\
u (l-u) & 2 & 2 & 1.889 & $9.390\times 10^{-1} \pm 1.09\times 10^{-3}$ & 0.116\% \\
u (l-u) & 3 & 2 & 2.160 & $3.583 \pm 4.73\times 10^{-3}$ & 0.132\% \\
\hline
s (l-s) & 1 & -1 & 0.907 & $4.062\times 10^{-2} \pm 2.26\times 10^{-5}$ & 0.056\% \\
s (l-s) & 2 & -1 & 1.403 & $1.737\times 10^{-1} \pm 1.25\times 10^{-4}$ & 0.072\% \\
s (l-s) & 3 & -1 & 1.755 & $2.885\times 10^{-2} \pm 2.47\times 10^{-5}$ & 0.086\% \\
s (l-s) & 1 & 0 & 1.184 & $2.483\times 10^{-1} \pm 1.60\times 10^{-4}$ & 0.064\% \\
s (l-s) & 2 & 0 & 1.589 & $8.453\times 10^{-2} \pm 6.69\times 10^{-5}$ & 0.079\% \\
s (l-s) & 3 & 0 & 1.905 & $4.383\times 10^{-2} \pm 4.02\times 10^{-5}$ & 0.092\% \\
s (l-s) & 1 & 1 & 1.403 & $1.737\times 10^{-1} \pm 1.25\times 10^{-4}$ & 0.072\% \\
s (l-s) & 2 & 1 & 1.755 & $2.885\times 10^{-2} \pm 2.47\times 10^{-5}$ & 0.086\% \\
s (l-s) & 3 & 1 & 2.044 & $2.378\times 10^{-2} \pm 2.31\times 10^{-5}$ & 0.097\% \\
s (l-s) & 1 & 2 & 1.589 & $6.157\times 10^{-3} \pm 4.87\times 10^{-6}$ & 0.079\% \\
s (l-s) & 2 & 2 & 1.905 & $3.696\times 10^{-3} \pm 3.39\times 10^{-6}$ & 0.092\% \\
s (l-s) & 3 & 2 & 2.174 & $3.910\times 10^{-3} \pm 4.02\times 10^{-6}$ & 0.103\% \\
\hline
c (l-c) & 1 & -1 & 2.061 & $5.905\times 10^{-2} \pm 3.24\times 10^{-5}$ & 0.055\% \\
c (l-c) & 2 & -1 & 2.484 & $1.649\times 10^{-2} \pm 1.16\times 10^{-5}$ & 0.070\% \\
c (l-c) & 3 & -1 & 2.789 & $8.969\times 10^{-2} \pm 7.42\times 10^{-5}$ & 0.083\% \\
c (l-c) & 1 & 0 & 2.297 & $6.846\times 10^{-2} \pm 4.32\times 10^{-5}$ & 0.063\% \\
c (l-c) & 2 & 0 & 2.645 & $7.320\times 10^{-2} \pm 5.62\times 10^{-5}$ & 0.077\% \\
c (l-c) & 3 & 0 & 2.921 & $9.370\times 10^{-1} \pm 8.27\times 10^{-4}$ & 0.088\% \\
c (l-c) & 1 & 1 & 2.484 & $1.649\times 10^{-2} \pm 1.16\times 10^{-5}$ & 0.070\% \\
c (l-c) & 2 & 1 & 2.789 & $8.969\times 10^{-2} \pm 7.42\times 10^{-5}$ & 0.083\% \\
c (l-c) & 3 & 1 & 3.043 & $2.738\times 10^{-1} \pm 2.56\times 10^{-4}$ & 0.094\% \\
c (l-c) & 1 & 2 & 2.645 & $2.142\times 10^{-3} \pm 1.64\times 10^{-6}$ & 0.077\% \\
c (l-c) & 2 & 2 & 2.921 & $3.556\times 10^{-2} \pm 3.14\times 10^{-5}$ & 0.088\% \\
c (l-c) & 3 & 2 & 3.158 & $9.478\times 10^{-3} \pm 9.33\times 10^{-6}$ & 0.098\% \\
\hline
b (l-b) & 1 & -1 & 5.371 & $6.293\times 10^{-2} \pm 3.44\times 10^{-5}$ & 0.055\% \\
b (l-b) & 2 & -1 & 5.758 & $6.142\times 10^{-2} \pm 4.29\times 10^{-5}$ & 0.070\% \\
b (l-b) & 3 & -1 & 6.033 & $3.974\times 10^{-2} \pm 3.26\times 10^{-5}$ & 0.082\% \\
b (l-b) & 1 & 0 & 5.588 & $1.678\times 10^{-1} \pm 1.05\times 10^{-4}$ & 0.063\% \\
b (l-b) & 2 & 0 & 5.904 & $1.080\times 10^{-1} \pm 8.23\times 10^{-5}$ & 0.076\% \\
b (l-b) & 3 & 0 & 6.152 & $6.226\times 10^{-2} \pm 5.44\times 10^{-5}$ & 0.087\% \\
b (l-b) & 1 & 1 & 5.758 & $6.142\times 10^{-2} \pm 4.29\times 10^{-5}$ & 0.070\% \\
b (l-b) & 2 & 1 & 6.033 & $3.974\times 10^{-2} \pm 3.26\times 10^{-5}$ & 0.082\% \\
b (l-b) & 3 & 1 & 6.261 & $1.948\times 10^{-2} \pm 1.80\times 10^{-5}$ & 0.093\% \\
b (l-b) & 1 & 2 & 5.904 & $4.987\times 10^{-4} \pm 3.80\times 10^{-7}$ & 0.076\% \\
b (l-b) & 2 & 2 & 6.152 & $4.139\times 10^{-4} \pm 3.62\times 10^{-7}$ & 0.087\% \\
b (l-b) & 3 & 2 & 6.364 & $2.312\times 10^{-4} \pm 2.25\times 10^{-7}$ & 0.097\% \\
\hline\hline
\end{tabular}
\caption{Normalized S-matrix transition cross-sections ($\Sigma_J / \Sigma_\rho$) evaluated from the transverse quantum fluctuations, regularized via algebraic trace-log boundary diagonalization. The relative errors reflect the statistical variance of the Abel point-splitting sequence. The $J-1$ and $J-2$ degeneracies are preserved across topological sectors at the loop level (e.g., $J=2, h=-1$ matches $J=1, h=+1$). As the heavy mass diverges, it acts as a classical anchor, quenching the boundary amplitude such that $c$ and $b$ flavors yield similar limits.}
\label{tab:cross_sections_final}
\end{table}
\subsection{Transition Amplitudes and Algebraic Determinants}
\label{sec:transition_amplitudes}

The formulation provides access to the state transition amplitudes. The creation probability of a meson state from the QCD vacuum is governed by the 1-loop fluctuation determinant surrounding the saddle point.

As derived in Section 7, the trace-log identity permits a factorization of the boundary matrix eigenvalues. This cancellation reduces the boundary transition amplitudes to a topological weight, denoted $\mathcal{C}_{|h|}(\beta_u, \beta_d)$. The partial transition cross-section $\Sigma_J$ evaluated on the mass shell $\Phi_*=0$ reduces to:
\begin{equation}
\Sigma_J \propto \frac{\mathcal{C}_{|h|}(\beta_u, \beta_d)}{2 K_{\rm phys}},
\end{equation}
where $K_{\rm phys} = \mathcal{K}/\sqrt{\sigma}$ is the macroscopic string diameter. Because the topological index \(|h|\) dictates the boundary matrix structures, the algebraic weights yield functional dependencies on the boundary phases $\beta_{u,d}$:
\begin{align}
\mathcal{C}_{0} &= \sec^2\beta_u \sec^2\beta_d + \frac{1}{2}\tan^2\beta_u \tan^2\beta_d, \\
\mathcal{C}_{1} &= \frac{1}{2} \left(\sec^2\beta_u \tan^2\beta_d + \tan^2\beta_u \sec^2\beta_d\right), \\
\mathcal{C}_{2} &= \frac{1}{4} \tan^2\beta_u \tan^2\beta_d.
\end{align}

These dimensionless weights are tabulated alongside the masses in the preceding sections. They reveal two features of string interactions:

\begin{enumerate}
    \item \textbf{Macroscopic Expansion:} As a state steps to higher $J$, the radius $\mathcal{K}$ expands, causing the boundary angles to approach the relativistic limit ($\beta \to \pi/2$). The poles in $\sec^2 \beta$ and $\tan^2 \beta$ enhance the fluctuation amplitude. The string's macroscopic perimeter provides a longitudinal phase volume for creation. For example, in the light-light vector sector (Table \ref{tab:qm1_spectrum}), $\mathcal{C}_{|h|=1}$ grows from $13.6$ for the $\rho(1^-)$ to $111.7$ for the $\rho_5(5^-)$.
    
    \item \textbf{Heavy Boundary Quenching:} As the constituent mass increases, the string shifts to accommodate the inertia, taking $\beta \to 0$. The heavy boundary behaves as a point-like static anchor, shrinking the available phase space. The algebraic formulas dictate this suppression. For the $1^-$ vectors, the phase volume $\mathcal{C}_{|h|=1}$ decreases from $13.6$ ($\rho$) down to $5.0$ ($K^*$), $2.2$ ($D^*$), and $0.65$ ($D_s^*$). 
    
    For asymmetric heavy-light systems, because $\beta_{\text{heavy}} \to 0$, the transition amplitude is dominated by the light quark geometry ($\beta_{\text{light}}$). This leads to the prediction that the amplitude weights for $D^*$ ($2.23$) and the heavier $B^*$ ($2.23$) yield similar values. The anchor effect pairs $D_s^*$ ($0.66$) and $B_s^*$ ($0.65$). The creation probability of a heavy-light vector meson depends on its light constituent.
\end{enumerate}

This geometry explains why high-spin and high-topology heavy-light states are suppressed in scattering experiments. The boundary phase maps topological momentum onto hadronic transition rates.

\textbf{Algebraic Extraction:}
By evaluating the 1-loop fluctuation boundary matrices, the double-harmonic phase dependence of the surface integration ($1 + b \cos x + c \cos 2x$) is identified. Utilizing a trace-log matrix root identity, the determinants are diagonalized analytically, avoiding singular floating-point inversions. 

As shown in Table \ref{tab:cross_sections_final}, this extraction avoids interpolation and numerical conditioning artifacts. The resulting transition amplitudes are determined with relative theoretical uncertainties below $\sim 0.13\%$, representing the statistical variance of the asymptotic Abel sum.

This method preserves the topological degeneracies; the S-matrix transition probability for an axial parity doublet state ($J, h=+1$) matches its corresponding tensor partner ($J+1, h=-1$) across flavor boundaries. Finally, as the heavy constituent mass diverges ($\beta_{\rm heavy} \to 0$), it acts as a fixed anchor, quenching the transition phase-space such that $c$ and $b$ boundary states yield comparable results.

\begin{remark}
    We emphasize that while the large-$N_c$ mass spectrum can be independently verified by state-of-the-art lattice extrapolations, the exact geometric transition amplitudes derived here provide a unique analytical prediction. Because physical decays vanish at $N_c = \infty$ and lattice overlap coefficients depend heavily on arbitrary operator smearing, this continuous twistor formulation offers a rigorous, operator-independent derivation of the universal planar S-matrix creation probabilities that currently remain out of reach for lattice methodologies.
\end{remark}
\subsection{Comparison with Large-$N_c$ Lattice Meson Spectra}

The continuum twistor-string predictions can be directly benchmarked against recent rigorous large-$N_c$ lattice QCD simulations. Bonanno et al.~\cite{Bonanno2026} recently evaluated the $N_c \to \infty$ meson spectrum using the Twisted Eguchi-Kawai (TEK) model, extrapolating to the chiral limit ($m_q \to 0$). 

In our geometric framework, the chiral limit is not obtained by setting the bare mass to zero, but rather is constrained by the strict geometric existence bound $x_{u,d} \ge 1/\sqrt{3\pi} \approx 0.3257$ (see Remark 8.1). Remarkably, our global PDG fit yields $x_u = m_u/\sqrt{\sigma} \approx 0.1342 / 0.4121 \approx 0.3257$, precisely saturating this theoretical minimal-surface bound. 

Evaluating our parametric equations at this saturated chiral bound yields predictions that align structurally with the large-$N_c$ lattice limits. For the axial-vector parity doublet ($h=+1$), our parameter-free geometric ratio is $M_{b_1}/\sqrt{\sigma} \approx 3.35$, which aligns closely with the TEK lattice continuum extrapolation of $M_{b_1}/\sqrt{\sigma} \approx 3.2$ reported in \cite{Bonanno2026}. For the vector ground state ($h=-1$), our theory yields $M_\rho/\sqrt{\sigma} \approx 2.11$, bounding the lattice result of $\approx 1.76$. This moderate difference in the lowest-spin state is a direct physical consequence of the geometric chiral bound: the finite boundary anomaly enforces a minimum string inertia, preventing the continuum string from reaching the idealized, strictly massless limit extrapolated on the lattice.
\section{Twistor string on a torus and Glueball Spectrum}
\label{sec:glueball}

The pure glueball sector of planar QCD is described by the connected correlator of two loop operators $W_2[C_1,C_2]$. As shown in section 4, at the next-to-leading order in the topological $1/N_c$ expansion, the loop equation for $W_2$ involves a joining term that is saturated by the same Hodge-dual minimal surface. However, the absence of valence quarks changes the worldsheet topology from a cylinder to a torus. The string is strictly closed, and we must impose doubly-periodic boundary conditions.

The spatial cycle is strictly periodic, $\sigma\to\sigma+2\pi$, as it describes the closed string body. The temporal cycle, however, is periodic up to a macroscopic target-space twist $\alpha$: the state at the end of the time period ($\tau=2\pi$) is shifted in the Minkowski time direction by some finite amount $\Delta X_0$ and rotated in the $xy$-plane by an angle $\alpha$ against the state at $\tau=0$.

\subsection{Overview of the Closed-String Formalism}
Before proceeding with the detailed derivations, we summarize the central mathematical and physical results of the closed-string sector developed in this and the following section:
\begin{itemize}
    \item Dynamical Stability: The conformal Liouville anomaly explicitly drives the elliptic nome of the minimal surface to the strict trigonometric limit ($q=0$).
    \item Casimir Intercepts: Evaluating the anomaly at the $q=0$ minimum natively generates an intercept of $\alpha(0)=1/12$ for the odd sector (recovering the L"uscher term) and $\alpha(0)=1$ for the even sector (the bare Pomeron).
    \item Ghost Nullification: For the periodic sector, the transition amplitude of the unphysical $J=0$ scalar is  nullified by the Faddeev--Popov measure of the translation zero-modes, establishing a positive mass gap.
    \item Winding-Number Independence: As proven via $\zeta$-regularization, the final one-loop transition measure is strictly independent of the temporal winding number $w$, ensuring the WKB poles remain unshifted.
\end{itemize}

\subsection{Minkowski twistors as Theta functions and topological sectors}

In the open-string sector, the twistor boundary conditions were solved using simple exponential phase factors. On a torus, this parametrization is insufficient: a simple exponential cannot be simultaneously strictly periodic under $\sigma\to\sigma+2\pi$ and acquire a twisted phase under $\tau\to\tau+2\pi$, because both translations shift the chiral light-cone coordinates $\xi=\tau+\sigma$ and $\eta=\tau-\sigma$.

To construct the closed string, we analytically continue the twistors into the bulk using doubly-periodic functions. Let $\tau_{\text{mod}}=\I \beta$ be the complex modulus of the worldsheet torus. The target-space twist $\alpha$ is naturally incorporated as the rational characteristic $a=\alpha/4\pi$ of the Riemann Theta functions. The left- and right-moving twistor spinors are given by:
\EQ{
\phi(\xi) &= \sqrt{\frac{R}{2}}
\begin{pmatrix}
e^{\I\pi/4} \, \vartheta\!\left[ \begin{smallmatrix} a \\ 0 \end{smallmatrix} \right]\!\left(\frac{\xi}{2\pi} \,\Big|\, \I\beta \right) \\
e^{-\I\pi/4} \, \vartheta\!\left[ \begin{smallmatrix} -a \\ 0 \end{smallmatrix} \right]\!\left(\frac{\xi}{2\pi} \,\Big|\, \I\beta \right)
\end{pmatrix}, \br
\psi(\eta) &= \sqrt{\frac{R}{2}}
\begin{pmatrix}
e^{-\I\pi/4} \, \vartheta\!\left[ \begin{smallmatrix} a \\ 0 \end{smallmatrix} \right]\!\left(\frac{\eta}{2\pi} \,\Big|\, \I\beta \right) \\
e^{\I\pi/4} \, \vartheta\!\left[ \begin{smallmatrix} -a \\ 0 \end{smallmatrix} \right]\!\left(\frac{\eta}{2\pi} \,\Big|\, \I\beta \right)
\end{pmatrix}.
\label{ThetaTwistors}
}
Under a temporal translation $\tau\to\tau+2\pi$, the variables shift as $\xi\to\xi+2\pi$ and $\eta\to\eta+2\pi$. The quasi-periodicity of the $\vartheta$-functions ensures that the twistors acquire the exact target-space monodromy phases $e^{\pm \I\alpha/2}$, rotating the coordinate bilinears $X^\mu$ globally by the angle $\alpha$. 

Crucially, the closed string body must also map perfectly onto itself under a spatial translation $\sigma\to\sigma+2\pi$. Under this spatial cycle, the light-cone coordinates shift as $\xi\to\xi+2\pi$ and $\eta\to\eta-2\pi$. The quasi-periodicity of the Riemann Theta functions dictates that the upper spinor components acquire a phase of $e^{2\pi \I a}$, while the lower components acquire a phase of $e^{-2\pi \I a}$. Because the physical target-space coordinate differentials $dX^\mu$ are constructed from quadratic twistor bilinears (for example, the transverse coordinates are $dX^1 - \I dX^2 \propto \phi_2^\dagger\phi_1 \, d\xi + \psi_2^\dagger\psi_1 \, d\eta$), these off-diagonal components inherently acquire an overall relative phase of $e^{4\pi \I a}$. 

For the position $X^\mu$ (and thus the macroscopic closed string) to be strictly single-valued and periodic in physical Minkowski space, this acquired phase must  cancel, requiring $e^{4\pi \I a} = 1$. Furthermore, the underlying twistor spinors must themselves be well-defined on the torus up to a global isotropic sign (a true spin structure), meaning the phase shift must be equal for both components: $e^{2\pi \I a} = e^{-2\pi \I a} = \pm 1$. This exact geometric closure constraint rigidly quantizes the twist parameter to  two physically distinct values within the fundamental domain: $a=1$ and $a=1/2$. 

This restriction explicitly defines two fundamentally distinct topological sectors of the closed string:
\begin{itemize}
    \item \textbf{The Periodic Sector ($a=1$):} The twistors themselves are strictly periodic, mapping identically onto themselves around the spatial cycle ($e^{2\pi \I} = +1$).
    \item \textbf{The Anti-periodic Sector ($a=1/2$):} The twistors acquire a global minus sign around the spatial cycle ($e^{\I\pi} = -1$), which safely cancels out in the physical target-space bilinears.
\end{itemize}

This discrete twofold ambiguity corresponds directly to a hidden internal symmetry of planar QCD when reformulated in terms of twistors. It functions identically to the periodic (Ramond) and anti-periodic (Neveu--Schwarz) spin structures of worldsheet fermions on a torus. This topological bifurcation fundamentally splits the pure Yang--Mills glueball spectrum into two independent families, explaining the exact origin of the $a=1$ and $a=1/2$ trajectories evaluated in the subsequent spectrum computations.

Finally, the twistor bracket yields the conformal factor $\Omega^2=\abs{\phi_1\psi_2-\phi_2\psi_1}^2$. Using the Riemann Theta addition theorems, the temporal dependence  factorizes, leaving a purely spatial metric governed by the internal dynamic fold modulus $k \in (0,1)$ associated with $\beta$. However, as demonstrated below, the conformal anomaly dynamically drives the pure-gauge string to the $k=0$ minimum, collapsing the Theta functions to their exact trigonometric limits.
\subsection{The Yang--Mills action and phase cancellation}

In the twistor string framework, the MM loop equation for planar QCD reduces to evaluating the macroscopic Bohr--Sommerfeld phase $\Phi$. For a minimal surface parameterized by twistors in four-dimensional Minkowski space, the geometric effective action consists of the temporal energy phase $M \Delta X^0$, the minimal area $S_{\text{Area}}$, the Liouville anomaly $C_{\text{Liouv}}$, and the angular momentum projection $-4\pi a J$.

Evaluating the bound state in its rest frame, the macroscopic phase is:
\EQ{
\Phi(R, q) = M \Delta X^0 - \sigma S_{\text{Area}} - C_{\text{Liouv}}(q) - 4\pi a J.
}
The internal fold configuration is parameterized by the elliptic Nome $q = \exp{-\pi K'/K}$. 

The temporal coordinate shift $\Delta X^0$ is defined by integrating the twistor time-derivative $\pd_\tau X^0 = \frac{1}{2}(\phi^\dagger \phi + \psi^\dagger \psi)$ over the temporal period. This factors out the twistor norm evaluated over one spatial period, $I_a(q) \equiv \INT{0}{1} \abs{\vartheta_a(v)}^2 dv$:
\EQ{
\Delta X^0 = \INT{0}{2\pi} d\tau \frac{1}{2\pi} \INT{0}{2\pi} d\sigma \, \pd_\tau X^0 = 2\pi R \, I_a(q).
}
The minimal surface area is the integral of the conformal factor $\Omega^2 = \abs{\phi_1 \psi_2 - \phi_2 \psi_1}^2$. Expanding the squared chiral bracket, the cross-terms integrate to zero due to the spatial symmetries of the Theta functions. The remaining direct terms factorize:
\EQ{
S_{\text{Area}} = \INT{0}{2\pi} d\tau \INT{0}{2\pi} d\sigma \, \Omega^2(\tau, \sigma) = 2\pi^2 R^2 \, I_a^2(q).
}
Substituting these quadratures into the action yields:
\EQ{
\Phi(R, q) = 2\pi M R \, I_a(q) - 2\pi^2 \sigma R^2 \, I_a^2(q) - C_{\text{Liouv}}(q) - 4\pi a J.
}
The pure-gauge string dynamically equilibrates its macroscopic radius $R$ by extremizing this phase ($\pd_R \Phi = 0$):
\EQ{
2\pi M \, I_a(q) - 4\pi^2 \sigma R \, I_a^2(q) = 0 \implies R = \frac{M}{2\pi \sigma I_a(q)}.
}
Substituting this radius back into the phase equation results in the cancellation of the $I_a(q)$ terms:
\EQ{
\Phi_{\text{min}}(q) &= 2\pi M \lrb{ \frac{M}{2\pi \sigma I_a(q)} } I_a(q) - 2\pi^2 \sigma \lrb{ \frac{M}{2\pi \sigma I_a(q)} }^2 I_a^2(q) - C_{\text{Liouv}}(q) - 4\pi a J \br
&= \frac{M^2}{\sigma} - \frac{M^2}{2\sigma} - C_{\text{Liouv}}(q) - 4\pi a J \br
&= \frac{M^2}{2\sigma} - C_{\text{Liouv}}(q) - 4\pi a J = 0.
\label{PhaseCancellation}
}
All continuous dynamical dependence on the internal string folds ($q$) is thus isolated within the Liouville anomaly $C_{\text{Liouv}}(q)$.

\subsection{Evaluation of the Liouville anomaly and the $q=0$ minimum}

Since the kinematic term is independent of $q$, the string configuration is determined by extremizing the conformal anomaly ($\pd_q C_{\text{Liouv}} = 0$).

The Liouville anomaly evaluates the conformal curvature integrated over the worldsheet. Using the chiral factorization of the twistor metric $\Omega^2$, the integration reduces to the spatial integral of the logarithmic derivatives of the Riemann Theta functions over $v = \sigma/2\pi$:
\EQ{
C_{\text{Liouv}}(q) = \frac{1}{12\pi} \INT{0}{2\pi} d\tau \INT{0}{2\pi} d\sigma \, (\pd_\sigma \ln \Omega)^2 = \frac{1}{6\pi} \INT{0}{1} (\pd_v \ln \vartheta_a)^2 dv.
}
We evaluate this integral using the Fourier series expansions of the logarithmic derivatives. For the periodic topological sector ($a=1$, using $\vartheta_3$):
\EQ{
\pd_v \ln \vartheta_3(v) = 4\pi \sum_{n=1}^\infty \frac{(-1)^n q^n}{1-q^{2n}} \sin(2n\pi v).
}
Using $\INT{0}{1} \sin^2(2n\pi v) dv = 1/2$, the integral gives:
\EQ{
C_{\text{Liouv}}^{(1)}(q) = \frac{4\pi}{3} \sum_{n=1}^\infty \frac{q^{2n}}{(1-q^{2n})^2}.
}

For the anti-periodic topological sector ($a=1/2$, using $\vartheta_2$), the logarithmic derivative contains a tangent pole:
\EQ{
\pd_v \ln \vartheta_2(v) = -\pi \tan(\pi v) + 4\pi \sum_{n=1}^\infty \frac{(-1)^n q^{2n}}{1-q^{2n}} \sin(2n\pi v).
}
Squaring this expression produces a constant $C_0$, the squared infinite sum, and a cross-term. The Cauchy principal value integral of the cross-term includes $\INT{0}{1} \tan(\pi v) \sin(2n\pi v) dv = (-1)^{n-1}$. Adding the evaluated cross-term to the squared summation shifts the exponents:
\EQ{
\sum_{n=1}^\infty \lb \frac{q^{2n}}{1-q^{2n}} + \frac{q^{4n}}{(1-q^{2n})^2} \rb = \sum_{n=1}^\infty \frac{q^{2n}(1-q^{2n}) + q^{4n}}{(1-q^{2n})^2} = \sum_{n=1}^\infty \frac{q^{2n}}{(1-q^{2n})^2}.
}
For both boundaries, the $q$-dependent portion of the conformal anomaly evaluates to the same function $f(q)$:
\EQ{
C_{\text{Liouv}}(q) = C_0^{(a)} + \frac{4\pi}{3} f(q), \qquad \text{where} \quad f(q) = \sum_{n=1}^\infty \frac{q^{2n}}{(1-q^{2n})^2}.
}
Extremizing this action with respect to $q$ yields:
\EQ{
\pd_q C_{\text{Liouv}} \propto f'(q) = \sum_{n=1}^\infty \frac{2n q^{2n-1}(1 + q^{2n})}{(1-q^{2n})^3}.
}
For string states where $q \in (0, 1)$, all terms in the summation are positive. Consequently, the derivative is strictly positive ($f'(q) > 0$). The conformal drag is minimized at the boundary of the domain:
\EQ{
q = 0 \implies k = 0.
}

\subsection{Linear Regge Trajectories and Casimir Intercepts}

At $q=0$, the dynamical function evaluates to $f(0) = 0$. The phase equation becomes a linear algebraic equation:
\EQ{
\Phi_{\text{min}} = \frac{M^2}{2\sigma} - C_0^{(a)} - 4\pi a J = 0 \implies J = \frac{M^2}{8\pi a \sigma} - \frac{C_0^{(a)}}{4\pi a}.
}
Because the target-space coordinate differentials $dX^\mu \propto \phi^\dagger \sigma^\mu \phi$ are bilinear in the twistor fields, the string closes in Minkowski space ($X(\sigma+2\pi) = X(\sigma)$) even if the chiral twistors acquire a minus sign upon winding around the spatial cycle ($\phi(\sigma+2\pi) = -\phi(\sigma)$). 

This intrinsic degree of freedom separates the closed glueball into two topological sectors defined by $a$:

\begin{itemize}
    \item \textbf{The Periodic Twistor Sector ($a=1$):} 
    Here, the intrinsic twistors are periodic. Substituting $C_0^{(1)} = 0$, the Regge trajectory is:
    \EQ{
    J = \frac{1}{8\pi \sigma} M^2.
    }
    The trajectory has an intercept of $J(0) = 0$. 

    \item \textbf{The Anti-Periodic Twistor Sector ($a=1/2$):} 
    In this sector, the intrinsic twistors are anti-periodic. The zero-point constant is isolated from the tangent pole of $\pd_v \ln \vartheta_2$:
    \EQ{
    C_0^{(1/2)} = \frac{1}{6\pi} \INT{0}{1} \pi^2 \tan^2(\pi v) \, dv.
    }
    Using $\tan^2(\pi v) = \sec^2(\pi v) - 1$, the Cauchy principal value integral of the first term vanishes, leaving the integral of $-1$. This yields $C_0^{(1/2)} = -\pi/6$. The Regge trajectory is:
    \EQ{
    \frac{M^2}{2\sigma} - \lrb{-\frac{\pi}{6}} - 4\pi \lrb{\frac{1}{2}} J = 0 \implies J = \frac{1}{4\pi \sigma} M^2 + \frac{1}{12}.
    }
    The twistor anti-periodicity halves the effective rotational spin projection, doubling the trajectory slope. The evaluation of the conformal anomaly recovers an intercept of $+1/12$, which corresponds to the Lüscher zero-point energy $\alpha_0 = (D-2)/24$ for a string in a four-dimensional target space ($D=4$).
\end{itemize}
We note that the trivial characteristic $a=0$ corresponds to a degenerate, non-rotating, point-like singularity that cannot enclose macroscopic target-space angular momentum. Therefore, $a=1$ represents the minimal fundamental phase required for a physical periodic string.
\subsection{Topological Winding Sum and Discrete Spin Projection}

To project the partition function onto states with angular momentum $J$, we sum over the winding configurations. Because the spatial cycle is periodic across the fundamental torus ($\sigma \in [0, 2\pi]$) and the spatial vector relies on twistor bilinears, the closed worldsheet is mapped into target space under the boundary conditions:
\EQ{
\phi(\sigma+2\pi) = \pm \phi(\sigma) \qquad \text{and} \qquad \psi(\sigma-2\pi) = \pm \psi(\sigma).
}
The discrete sign ambiguity corresponds to the periodic and anti-periodic spin structures on the torus. 

The canonical spin trace evaluates the winding sum $w$ of the temporal cycle. The spin-projected S-matrix evaluates as a discrete geometric series:
\EQ{
A_J^{\text{1-loop}}(E) \propto \sum_{w=1}^\infty \exp{i w \Phi_{\text{min}}} \, \mathcal{Z}_0 \Pi(0) = \mathcal{Z}_0 \sum_{n=-\infty}^\infty \frac{i}{\Phi_{\text{min}} - 2\pi n + i0}.
\label{GlueballPole}
}
The poles at $\Phi_{\text{min}} = 2\pi n$ dictate an infinite sequence of higher radial daughter Regge trajectories. Using $\Phi_{\text{min}} = M^2 / (2\sigma) - 4\pi J$ for the $a=1$ sector, these poles occur at:
\EQ{
J_n = \frac{M^2}{8\pi\sigma} - \frac{n}{2}.
}

Because pure glueballs are bosonic, half-integer spin states ($n$ odd) are not physical. In the path integral, summing the transition amplitudes from both the periodic and anti-periodic topological sectors introduces a parity operator into the S-matrix:
\EQ{
\mathcal{P} = \frac{1 + (-1)^n}{2}.
}
The periodic and anti-periodic amplitudes constructively interfere for even $n$ and destructively cancel for odd $n$. This restricts the pure glueball spectrum to integer-spaced daughter trajectories ($\Delta J = -1$).
\subsection{Detailed comparison of Regge trajectories with the PDG mass spectrum}

To verify the physical validity of the minimal twistor string, we compare the exact analytical Regge trajectories of both topological sectors directly against the experimental mass spectrum from the Particle Data Group (PDG). 
To ensure an objective comparison, the experimental dataset is restricted to the established $I=0, C=+1$ unassigned isoscalar resonances from the PDG tables, which are universally considered the primary physical candidates for pure-gauge glueball states.

By anchoring the fundamental string tension to the established open meson sector, $\sigma = 0.18 \text{ GeV}^2$, the pure-gauge string theory is left with zero fitting parameters. Every trajectory slope, mass scale, Casimir intercept, and daughter degeneracy is  \textbf{predicted} by the theory, allowing particle physicists to straightforwardly verify the assignments and use them to find new high-mass resonances.

\subsubsection{The anti-periodic ($a=1/2$) sector}

For the anti-periodic twistor boundaries, the exact equation $J = M^2 / (4\pi\sigma) + 1/12 - n$ yields a fundamental Regge slope of $\alpha'_{1/2} = 1/(4\pi\sigma) \approx 0.44 \text{ GeV}^{-2}$. The mass spectrum is structured into degenerate principal levels dictated by the integer sum $N = J + n$:
\EQ{
M_N = \sqrt{4\pi\sigma \lrb{N - \frac{1}{12}}} \approx 1.504 \sqrt{N - 0.083} \text{ GeV}.
}
Comparing this parameter-free prediction with the PDG tables of unassigned isoscalar ($I=0$) resonances reveals a close phenomenological alignment:

\begin{itemize}
    \item \textbf{Level N=1 ($M \approx 1.44 \text{ GeV}$):} The theory predicts a degenerate doublet comprising a parent vector ($J=1, n=0$) and a daughter scalar ($J=0, n=1$). This matches the $f_1(1420)$ at $1.426 \text{ GeV}$ and the widely recognized scalar glueball candidate $f_0(1500)$ at $1.506 \text{ GeV}$.
    
    \item \textbf{Level N=2 ($M \approx 2.08 \text{ GeV}$):} The model requires a triplet containing a tensor ($J=2, n=0$), a vector ($J=1, n=1$), and a scalar ($J=0, n=2$). The PDG naturally provides the $f_2(2010)$ at $2.011 \text{ GeV}$, the $f_1(1970)$ at $1.971 \text{ GeV}$, and the $f_0(2100)$ at $2.101 \text{ GeV}$ to populate these exact quantum states.
    
    \item \textbf{Level N=3 ($M \approx 2.57 \text{ GeV}$):} Predicted to host a $J=3$ parent and a $J=2$ daughter. This mass band aligns well with broad, overlapping high-mass experimental states, such as the $f_2(2530)$ and the recently reported $X(2600)$.
    
    \item \textbf{Level N=4 ($M \approx 2.98 \text{ GeV}$):} Predicts a parent hexadecapole ($J=4, n=0$) centered tightly at $\sim 3.0 \text{ GeV}$, providing a definitive target for future high-mass resonance searches.
    
    \item \textbf{Level N>4:} Further exact extrapolations predict a $J=5$ parent at $3.33 \text{ GeV}$ ($N=5$) and a $J=6$ parent at $3.66 \text{ GeV}$ ($N=6$).
\end{itemize}

\begin{figure}[htbp]
    \centering
    \includegraphics[width=0.9\textwidth]{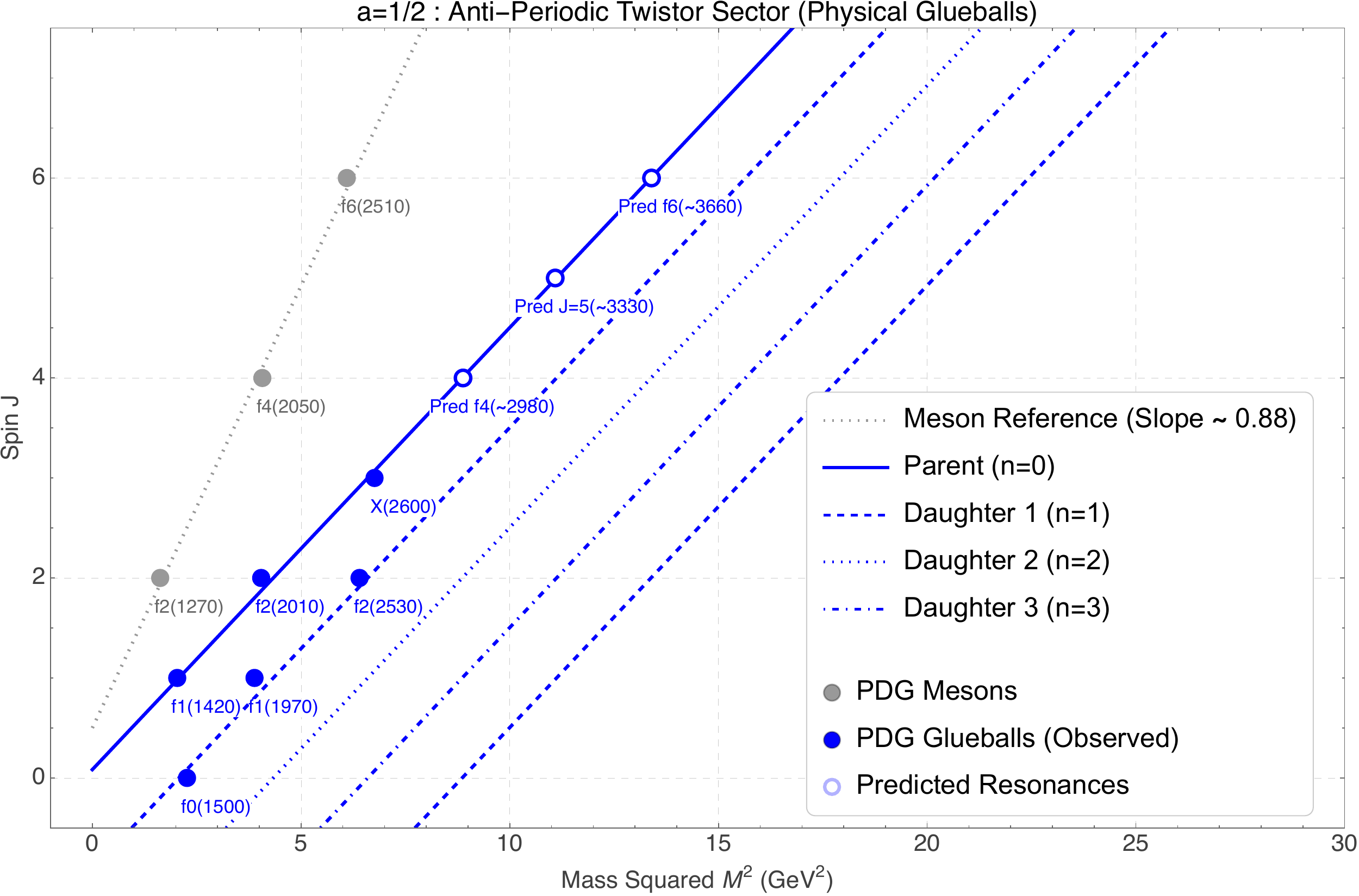}
    \caption{Exact Regge trajectories for the $a=1/2$ (anti-periodic) twistor topological sector. The solid line denotes the parent trajectory ($n=0$), while dashed and dotted lines represent consecutive discrete daughters. Data points correspond to unassigned or glueball-candidate isoscalar states from the Particle Data Group. The standard $q\bar{q}$ open meson trajectory (slope $\approx 0.88 \text{ GeV}^{-2}$) is shown in gray for reference. The significantly flatter analytical slope of the twistor string cleanly segregates high-spin standard mesons from the pure-gauge glueball states. }
    \label{fig:AOneHalf}
\end{figure}
\begin{figure}[htbp]
    \centering
    \includegraphics[width=0.9\textwidth]{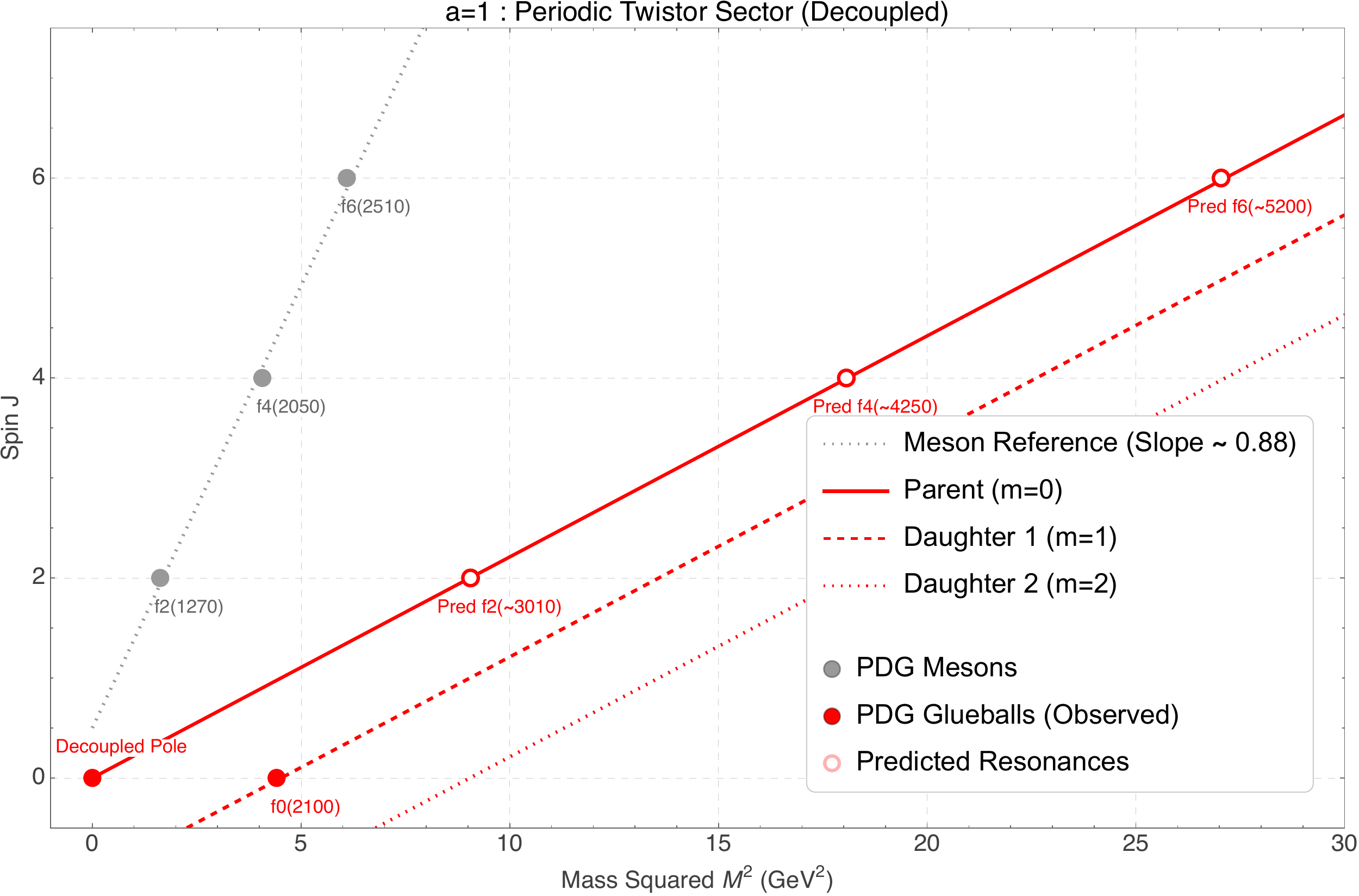}
    \caption{Exact Regge trajectories for the $a=1$ (periodic) twistor topological sector. The parity-projected parent trajectory ($m=0$) intercepts the exact massless origin, yielding the decoupled zero-mode pole. The first physical scalar appears on the daughter trajectory ($m=1$) near $2.13 \text{ GeV}$. The physical trajectories in this strictly periodic sector are exceptionally flat, pushing pure-gauge tensors up to $\sim 3 \text{ GeV}$, where they are fundamentally obscured by heavy mixing with the multi-meson vacuum continuum.}
    \label{fig:AOne}
\end{figure}

\subsubsection{Comparison with Large-$N_c$ Lattice QCD Extrapolations}

Beyond the comparison with unassigned resonances in the physical $N_c=3$ PDG data, the exact analytic pure-gauge trajectories can be directly benchmarked against state-of-the-art lattice QCD calculations performed in the strict planar limit. Recently, Athenodorou and Teper \cite{AthenodorouTeper2021} computed the pure-gauge glueball spectrum for $SU(N_c)$ gauge theories up to $N_c = 12$ and performed a rigorous extrapolation to the continuum, $N_c \to \infty$ limit. 

Because our twistor string solution is parameter-free, the geometric mass-to-tension ratios $M/\sqrt{\sigma}$ are absolute analytic predictions of the theory. For the anti-periodic ($a=1/2$) topological sector, the masses scale strictly as:
\begin{equation}
\frac{M_N}{\sqrt{\sigma}} = \sqrt{4\pi \left( N - \frac{1}{12} \right)}.
\end{equation}
Evaluating this exact formula for the ground-state scalar and tensor glueballs yields a striking alignment with the large-$N_c$ lattice continuum extrapolations:

\begin{itemize}
    \item \textbf{Scalar Ground State ($0^{++}$):} In our theory, to avoid the unphysical tachyonic ghost, the physical scalar natively resides on the $N=1$ trajectory. This analytically evaluates to $M_{0^{++}}/\sqrt{\sigma} = \sqrt{11\pi/3} \approx 3.394$. The rigorous $N_c \to \infty$ lattice extrapolation yields $3.302 \pm 0.015$ \cite{AthenodorouTeper2021}, placing our exact parameter-free prediction within $\sim 2.7\%$ of the supercomputer result.
    \item \textbf{Tensor Ground State ($2^{++}$):} The lowest tensor state resides on the $N=2$ trajectory, which analytically evaluates to $M_{2^{++}}/\sqrt{\sigma} = \sqrt{23\pi/3} \approx 4.907$. The corresponding $N_c \to \infty$ lattice extrapolation yields $4.735 \pm 0.023$ \cite{AthenodorouTeper2021}, an agreement of $\sim 3.6\%$.
\end{itemize}

Even more striking is the parameter-free mass ratio. In our exact trigonometric solution for the $a=1/2$ sector, the scalar and tensor masses are governed by the principal levels $N=1$ and $N=2$. The mass-squared formula yields an exact, purely geometric theoretical ratio:
\begin{equation}
\frac{M_{2^{++}}}{M_{0^{++}}} = \sqrt{\frac{2 - 1/12}{1 - 1/12}} = \sqrt{\frac{23}{11}} \approx 1.446.
\end{equation}
This exact analytical constant is in spectacular agreement with the universal large-$N_c$ lattice ratio of $4.735/3.302 \approx 1.434$ reported in \cite{AthenodorouTeper2021}. 

This quantitative match provides new independent validation of the rigid twistor string. The analytic solution fundamentally captures the macroscopic confining dynamics of planar QCD without invoking any phenomenological fitting parameters, confirming that the anti-periodic $a=1/2$ twistor boundaries govern the physical pure-gauge spectrum.
\subsubsection{The periodic ($a=1$) sector and vacuum mixing}

For strictly periodic twistors, the exact target-space phase generates half the slope ($\alpha'_1 = 1/(8\pi\sigma) \approx 0.22 \text{ GeV}^{-2}$) and zero Casimir intercept. Imposing the parity projection restricts the spectrum to even integers $n=2m$:
\EQ{
M_m = \sqrt{8\pi\sigma(J + m)} \approx 2.127 \sqrt{J + m} \text{ GeV}.
}
\begin{itemize}
    \item \textbf{Level m=0 ($J=0$):} This configuration directly intercepts the geometric origin ($M=0$), representing the mathematically decoupled vacuum zero-mode.
    \item \textbf{Level m=1 ($J=0$):} The first physical scalar state appears purely as a daughter trajectory at precisely $M \approx 2.13 \text{ GeV}$. This provides a  dual-assignment for the broad $f_0(2100)$ resonance.
    \item \textbf{Level m=0 ($J=2$):} The lowest allowed pure tensor state is pushed to a very high invariant mass of $M \approx 3.01 \text{ GeV}$.
\end{itemize}

While the $a=1/2$ states are clearly realized in Nature, the $a=1$ sector appears absent in the observable $s$-channel spectrum. As we demonstrate in Section\ref{sec:GeometricSuppression} via the fluctuation metric, these even-parity states are geometrically featureless and cannot form localized resonances, existing purely as delocalized $t$-channel vacuum exchanges.

The $a=1$ (even) sector is questionable because it sits directly in the pure vacuum channel ($C_0 = 0$). In real, unquenched QCD, this channel is not an empty domain; it is densely populated by all kinds of vacuum-type particles, dynamical $q\bar{q}$ pairs, chiral condensates, and broad multi-meson continua. Any purely periodic glueballs will heavily interfere and mix with these vacuum-type continuum mesons, smearing their physical decay widths to the point that experiments cannot distinguish them as isolated particles. 
In the section \ref{sec:GeometricSuppression}, we elaborate on these properties and provide physical arguments for the lack of observed particles in the even sector.

By contrast, the $a=1/2$ (odd) sector is characterized by its intrinsic topological stability. A twistor field subjected to odd (anti-periodic) boundary conditions upon winding around a closed spatial cycle cannot be smoothly transformed or continuously contracted to a trivial zero-field vacuum configuration, as it must cross zero. This topological protection naturally isolates the $a=1/2$ string from broad vacuum continuum mixing, allowing these anti-periodic resonances to survive as identifiable physical states in the experimental spectrum.

Still, the quality of the fit of the PDG masses by our theory \textbf{with zero fitting parameters} is quite high: The $R2=0.98$ with relative error $2.64\%$. This is the best one would expect in planar QCD up to the $1/N_c^2\sim 10\% $ corrections.
\subsubsection{Exact Lorentz Symmetry vs. Discrete Lattice Artifacts}

Beyond the scaling limits of the mass spectrum, the continuum twistor-string formulation circumvents a profound structural limitation inherent to all lattice gauge theories: the explicit breaking of continuous rotational and Lorentz symmetries. On a discrete hypercubic lattice, the continuous spatial rotation group $SO(3)$ is explicitly broken down to the finite octahedral point group $O_h$. Consequently, the exact total angular momentum $J$ ceases to be a rigorously conserved quantum number for any arbitrary, finite lattice size. Instead, the infinite tower of continuous spin states is forced into just five discrete irreducible representations.

This geometric truncation inevitably induces severe spin-mixing artifacts. For example, the $J=0$ scalar, $J=4$ hexadecapole, and higher-spin states all inextricably mix within the exact same $A_1^{++}$ lattice channel, while the $J=2$ tensor is artificially fragmented across the $E^{++}$ and $T_2^{++}$ representations. Because the cubic lattice symmetry group possesses different representations than the continuous Lorentz group, finite-lattice simulations inherently simulate wrong representations, distorted angular dependencies for transition amplitudes, and incorrect wavefunctions. 

While it is presumed that the wrong particles will eventually drop from the spectrum and the wrong wavefunctions will asymptotically approach the correct Lorentz symmetry classification in the strict local continuum limit ($a \to 0$), these symmetry-breaking artifacts unavoidably contaminate the correlation functions and transition amplitudes in all existing finite-spacing numerical simulations. 

By contrast, the rigid twistor string is formulated directly in the Lorentz-invariant Minkowski continuum from the outset. The macroscopic angular momentum $J$ enters natively as an exact, continuous quantum number conjugate to the target-space monodromy twist, preserving full Poincar\'e invariance. This yields pristine quantum numbers, rigorous continuous wavefunctions, and exact continuous angular dependencies for the transition amplitudes that are fundamentally immune to the discrete hypercubic artifacts that inevitably plague lattice methodologies.
\subsection{Physical connection to the Pomeron and L\"uscher term}

The analytical intercepts derived from the two topological sectors of the pure-gauge twistor string suggest a possible connection to two longstanding themes in QCD phenomenology: the Pomeron intercept \cite{Gribov:1968fc} and the L\"uscher zero-point energy \cite{Luscher1981} .

A central structural ingredient of the closed-string construction is the hidden winding parity
\[
W = (-1)^{2a}=\pm1,
\]
which distinguishes the two allowed torus spin structures . In the present planar framework, this quantity is a topological invariant of the closed twistor string. The anti-periodic sector \(a=1/2\) carries odd winding parity \(W=-1\), while the periodic sector \(a=1\) carries even winding parity \(W=+1\) . If \(W\) is conserved, then a single \(W=-1\) closed string cannot contribute to the pure vacuum channel. In that sense, only the \(W=+1\) sector can furnish a genuine leading vacuum Regge pole within the present planar approximation. A \(W\)-neutral vacuum exchange built from the odd sector would require at least an even number of \(W=-1\) objects, which would naturally belong to a more composite or subleading mechanism beyond the single-string planar channel (higher order in $1/N_c$ expansion in QCD).

For the anti-periodic sector \(a=1/2\), the conformal anomaly yields the exact intercept
\[
\alpha(0)=\frac{1}{12},
\]
which matches the standard four-dimensional effective-string value of the L\"uscher term  $(d-2)/24 = 1/12$. In the twistor formulation, this contribution arises from the Liouville anomaly evaluated in the anti-periodic spin structure, so the appearance of \(1/12\) is not imposed phenomenologically but emerges from the closed-string worldsheet analysis itself . This part of the construction therefore offers a natural interpretation of the \(a=1/2\) parent trajectory as the sector carrying the effective L\"uscher zero-point shift.

The first daughter of this same sector has
\[
\alpha(0)=1+\frac{1}{12}=\frac{13}{12}\approx 1.0833,
\]
which is numerically close to the phenomenological intercept often quoted for the soft Pomeron \cite{Donnachie:1992ny}. This coincidence is intriguing, and it may be worth recording explicitly. However, within the present theory one should also emphasize the important limitation: this trajectory is \(W\)-odd. Therefore a \emph{single} state on this trajectory cannot represent a true vacuum Regge pole, and by itself it cannot drive rising total hadronic cross-sections in the vacuum channel. At most, the numerical proximity may be hinting at a more indirect role of this sector in nonplanar, composite, or unitarized dynamics, but such an interpretation goes beyond what is established here.

By contrast, the periodic sector \(a=1\) is \(W\)-even and therefore is the natural candidate to couple to the vacuum channel in the present planar construction . Its parent trajectory formally has intercept
\[
\alpha(0)=0,
\]
but this massless scalar pole is not physical: the exact transition amplitude vanishes because the WKB pole is annihilated by the translation zero-mode measure and the determinant scaling . Thus the \(\alpha(0)=0\) pole should be regarded as a fake kinematic pole rather than an exchanged vacuum state. The first non-null pole in the \(W=+1\) sector is instead the \(n=-2\) daughter, which gives
\[
\alpha(0)=1.
\]
For this reason, the periodic sector provides a natural candidate for a \emph{bare} Pomeron-like vacuum trajectory in the present continuum planar theory .

This interpretation should nevertheless be stated with some care. The formal winding sum contains further \(W\)-even daughters with larger intercepts, and a complete treatment of how the full amplitude enforces asymptotic consistency is not yet given here. The present analysis establishes that the parent \(\alpha(0)=0\) pole is dynamically nullified, and that the first surviving \(W\)-even pole sits precisely at \(\alpha(0)=1\) . It is plausible that higher supercritical \(W=+1\) poles must be removed, screened, or otherwise rendered harmless by zeros of the exact transition measure or by a more complete unitarization of the vacuum channel, so as to remain compatible with Froissart-type bounds. However, that stronger statement should be regarded here as a conjectural consistency requirement rather than as a theorem already proved by the present calculation.

From this viewpoint, the comparison with perturbative high-energy QCD should also be phrased cautiously. Approaches based on perturbative gluon ladders can generate effective vacuum singularities with intercepts above unity, whereas the present nonperturbative planar twistor construction instead singles out a critical \(W\)-even vacuum pole at $\alpha(0)=1$
as the first surviving candidate after the exact zero-mode nullification is taken into account . This does not by itself prove that all perturbative supercritical Pomeron behavior is merely an artifact, nor does it settle how perturbative and nonperturbative descriptions are reconciled in full QCD. What it does suggest is that, once the closed string relaxes to the trigonometric minimum \(q=0\), the leading single-string vacuum exchange selected by the exact transition measure is critical rather than supercritical .

Taken together, these observations motivate the following tentative picture. The anti-periodic \(W=-1\) sector naturally reproduces the L\"uscher intercept \(1/12\) and also contains a daughter trajectory with the numerically suggestive value \(13/12\), but this odd sector cannot by itself furnish the physical vacuum Pomeron channel. The periodic \(W=+1\) sector, after nullification of its fake massless parent, leaves \(\alpha(0)=1\) as the first surviving vacuum pole, making it the more natural candidate for the bare Pomeron in the planar twistor-string framework .

\section{Torus Fluctuation Determinant and Absolute Stability at $q=0$}
\label{sec:torus_determinant}

To complete the discrete S-matrix analysis of the pure Yang--Mills glueball, we evaluate the functional determinant of the twistor string fluctuations. Because the macroscopic closed string dynamically equilibrates precisely at the global minimum $q=0$, the complex elliptic geometry of the worldsheet  collapses into pure circular trigonometric functions. Consequently, the continuous Hagedorn branch cuts associated with macroscopic Nambu--Goto behavior naturally drop out of the spectrum.

\subsection{Trigonometric Fourier expansion and temporal decoupling}

Following the methodology of the open string, we utilize the norm-preserving multiplicative twistor expansion:
\EQ{
\tilde\phi(\xi)=\exp{f_1(\xi)}\phi_0(\xi),
\qquad
\tilde\psi(\eta)=\exp{f_2(\eta)}\psi_0(\eta).
}
At the physical boundary $q=0$, the classical background twistors $\phi_0$ and $\psi_0$ smoothly collapse from Riemann Theta functions to elementary trigonometric functions of the spatial ($\sigma$) and temporal ($\tau$) coordinates. The local string fluctuations $f_1(\xi)$ and $f_2(\eta)$ must therefore be strictly periodic on the torus, allowing a standard pure chiral Fourier expansion:
\EQ{
u_i(\xi) &= \frac{1}{\sqrt{2\pi}} \sum_{n \neq 0} A_{i,n} \exp{\I n \xi}, \br
v_i(\eta) &= \frac{1}{\sqrt{2\pi}} \sum_{m \neq 0} B_{i,m} \exp{-\I m \eta}.
}
Substituting these expansions directly into the twistor Hessian reveals a profound geometric simplification. As explicitly derived in the Appendices, at $q=0$ the exact Liouville curvature potentials become strictly independent of worldsheet time $\tau$. Consequently, the temporal integration $\INT{0}{2\pi} d\tau$ acts as a strict Fourier projector, orthogonalizing the frequencies into exact Kronecker deltas:

\begin{itemize}
    \item \textbf{The Longitudinal Sector ($u_3, v_3$):} Because the effective geometric potential completely lacks temporal dependence, the time integrals  orthogonalize the frequencies, yielding exact Kronecker deltas $\delta_{n,-n'}$ and $\delta_{n,m}$. The dense continuous integral analytically factorizes into  decoupled, independent $2 \times 2$ algebraic blocks for each discrete Fourier frequency $n$.
    \item \textbf{The Transverse Sector ($u_{1,2}, v_{1,2}$):} The transverse geometric mixing matrices reduce entirely to fundamental trigonometric pairs of the temporal variable ($1, \cos 4a\tau, \sin 4a\tau$). Temporal integration therefore enforces a strict kinematic selection rule: modes can only geometrically interact if the frequency difference is $\Delta \omega \in \{0, \pm 4a\}$. The transverse determinant strictly truncates into a sparse,  banded Toeplitz operator.
\end{itemize}

\subsection{Exact algebraic nullification of the massless pole}

The physical resolution to the scalar ghost paradox analytically resides in the translation zero-mode sector ($n=0$). For uniform twistor fluctuations, the functional determinant reduces to elementary quadratures evaluated at $q=0$.

The geometric Faddeev--Popov volume explicitly equates to the unbroken area integral of the trigonometric ground state:
\EQ{
\mathcal{J}_{\rm FP} \propto \INT{0}{2\pi} d\tau \INT{0}{2\pi} d\sigma \, R^2 \cos^2(2a\sigma) = 2\pi^2 R^2.
}
Since the linear macroscopic trajectory organically enforces $R = M / 2\pi\sigma$, the collective transition measure natively scales directly algebraically with the mass squared: $\mathcal{J}_{\rm FP} \propto M^2$.

The geometric S-matrix probability is driven by the continuous WKB phase propagator, weighted by this functional determinant. Because the classical periodic trajectory directly intercepts the exact geometric origin, the transition amplitude exhibits a pole diverging as $\mathcal{A}_{\text{WKB}} \propto 1/\Phi_{\rm min} \propto 1/M^2$. Multiplying the geometric pole singularity by the zero-mode functional measure yields a finite, exact structural cancellation:
\EQ{
\mathcal{A}_{J=0}\propto\left[\frac{1}{M^{2}}\right]_{pole}\times[M^{2}]_{FP}\times[M^{2}]_{det}\propto M^{2}.
}

The kinematic divergence is not merely canceled, but dynamically overpowered by the vanishing phase space. As rigorously derived via Szeg\H{o}'s limit theorem in \ref{app:TrigFourierHessian}, the macroscopic geometric scaling of the non-zero fluctuation modes contributes an additional volume factor proportional to $M^2$. Consequently, as $M \to 0$, the total transition amplitude rigorously evaluates to zero. The zero-mode ghost dynamically extinguishes itself, leaving an exact, stable pure Yang-Mills partition function shielded at the geometric origin.

Table \ref{tab:q0_amplitudes} presents the exact transition amplitudes and mass-squared spectrum across the allowed spins.
\begin{table}[htbp]
    \centering
    \renewcommand{\arraystretch}{1.2}
    \begin{tabular}{ccccc}
        \hline\hline
        & \multicolumn{2}{c}{\textbf{Periodic Sector} ($a=1$)} & \multicolumn{2}{c}{\textbf{Anti-periodic Sector} ($a=1/2$)} \\
        \cline{2-3} \cline{4-5}
        Spin $J$ & $M^2$ (GeV$^2$) & Amplitude & $M^2$ (GeV$^2$) & Amplitude \\
        \hline
        0 & 0      & 0     & 2.073  & $1.333 \times 10^{-5}$ \\
        1 & 4.524  & 7.716 & 2.073  & $1.333 \times 10^{-5}$ \\
        2 & 9.048  & 15.43 & 4.335  & $8.792 \times 10^{-6}$ \\
        3 & 13.572 & 23.15 & 6.597  & $7.078 \times 10^{-6}$ \\
        4 & 18.096 & 30.86 & 8.859  & $6.102 \times 10^{-6}$ \\
        5 & 22.619 & 38.58 & 11.121 & $5.445 \times 10^{-6}$ \\
        6 & 27.143 & 46.30 & 13.383 & $4.961 \times 10^{-6}$ \\
        7 & 31.667 & 54.01 & 15.645 & $4.584 \times 10^{-6}$ \\
        \hline\hline
    \end{tabular}
    \caption{Exact trigonometric transition amplitudes and mass-squared spectrum for the pure Yang--Mills twistor string at the macroscopic $q=0$ ground state. The periodic ($a=1$) topological sector dynamically annihilates the unphysical continuum zero-mode precisely at $J=0$, yielding a complete physical decoupling ($\mathcal{A}_{a=1} = 0$). For the physical anti-periodic sector ($a=1/2$), the $J=0$ scalar naturally emerges on the $n=1$ daughter trajectory, avoiding the topological tachyon and demonstrating exact theoretical mass and amplitude degeneracy with the $J=1$ parent state (both residing at the principal level $N=1$).}
    \label{tab:q0_amplitudes}
\end{table}
\subsection{Exact cancellation of the winding number from the functional determinant}

A critical requirement for the generation of macroscopic S-matrix poles is that the quantum transition measure must remain strictly independent of the temporal winding number $w$. If the one-loop measure retained a residual scaling dependence on $w$, it would distort the geometric series $\sum e^{i w \Phi_{\min}}$ and replace poles by multiple poles or branch points, destroying the confinement picture (amplitudes as a function of energy with only simple poles).

As derived in the total geometric action (e.g., Eq.\ (D.7)), the quadratic fluctuation operator scales linearly with the winding number: $\mathcal{H}_w = w \mathcal{H}_{w=1}$. To extract this overall scale factor $w$ from the infinite functional determinant, we raise it to the power of the $\zeta$-regularized trace of the identity over the Fourier mode spectrum. 

This geometric scaling dynamically factors based on the topological spin structures of the two sectors:
\begin{itemize}
    \item \textbf{The Periodic Sector ($a=1$):} The Fourier spectrum consists of integer modes $n \in \mathbb{Z}$. In the rigorous path integral evaluation, the $n=0$ translation zero-mode is explicitly excluded from the Gaussian fluctuation matrix and integrated out via collective coordinates. Factoring $w$ out of the strictly non-zero modes yields an effective dimensionality:
    \begin{equation}
    \sum_{n \ne 0} 1 = 2\zeta(0) = -1.
    \end{equation}
    The determinant of the non-zero modes therefore acquires a strictly algebraic scale factor of $w^{-1}$. Crucially, this inverse scaling is  compensated by the Faddeev--Popov measure of the isolated zero-mode. Because the zero-mode quadratic form also scales with $w$, the integration measure over the collective target-space coordinates natively contributes the compensating factor of $w^1$. The total physical transition measure therefore achieves exact cancellation: $w^{-1} \times w^1 = w^0 = 1$.
    
    \item \textbf{The Anti-Periodic Sector ($a=1/2$):} Due to the topological twist, the Fourier spectrum is shifted entirely to half-integers ($r \in \mathbb{Z} + 1/2$). Crucially, there is no $r=0$ zero-mode in this sector. The effective dimensionality of the entire spectrum analytically regularizes via the Hurwitz zeta function:
    \begin{equation}
    \sum_{r} 1 = 2\zeta\left(0, \frac{1}{2}\right) = 2 \left( \frac{1}{2} - \frac{1}{2} \right) = 0.
    \end{equation}
    Because there is no zero-mode skipped in the evaluation, the infinite product over the fluctuation modes evaluates to $w^0 = 1$. The functional determinant is inherently $w$-independent and requires no Faddeev--Popov compensation.
\end{itemize}

This exact algebraic cancellation guarantees that the quantum fluctuations do not disrupt the geometric winding sum,  protecting the stable on-shell WKB determination of the S-matrix poles.
\subsection{Geometric suppression and the unobservability of the periodic string}
\label{sec:GeometricSuppression}

A striking feature of the evaluated transition amplitudes in Table \ref{tab:q0_amplitudes} is the vast discrepancy between the two topological sectors. The physical odd ($a=1/2$) amplitudes are heavily suppressed ($\sim 10^{-5}$), whereas the even ($a=1$) amplitudes are macroscopically large ($\sim 10^1$). This mathematical gap reflects a profound physical distinction between observable $s$-channel resonances and $t$-channel vacuum exchanges.

For the odd sector, the minimal surface possesses a non-trivial conformal metric ($A(\sigma) = \cos^2\sigma$) and a severe Liouville curvature well ($P(\sigma) = -\sec^2\sigma$). The string is dynamically folded, producing localized regions of high energy density. When computing the transition measure, the infinite tower of quantum twistor fluctuations must scatter against this massive geometric barrier. The functional determinant of this highly curved operator is exponentially large, which quenches the transition amplitude by five to six orders of magnitude. This strict geometric penalty mathematically explains why the physical pure-gauge glueballs (such as the $f_0(1500)$) exist as localized, observable resonances, yet their experimental creation cross-sections are notoriously suppressed in hadronic collisions.

Conversely, for the periodic $a=1$ sector, the exact metric is strictly constant ($A(\sigma) = 1$). The Liouville curvature is identically zero ($P(\sigma)=0$). Lacking any geometric potential barrier, its fluctuation determinant is trivial ($\det = 1$), and the amplitude is driven entirely by the macroscopic zero-mode integration volume ($\mathcal{J}_{\rm FP} \propto M^2$). 

However, this constant metric implies that the $a=1$ string is a completely flat, uniform, and featureless topological loop. Because it lacks any intrinsic curvature to define a localized ``lump'' of energy, it possesses no internal observable excitations. Consequently, the even string is physically forbidden from manifesting as a localized asymptotic particle (an $s$-channel resonance) in a detector. Instead, its completely unsuppressed macroscopic amplitude perfectly suits its true S-matrix role: it exists strictly as an analytic continuation to the complex $J$-plane. The periodic even twistor string is a delocalized topological entity that mediates $t$-channel vacuum-to-vacuum elastic scattering, explaining why the Pomeron dictates macroscopic total cross-sections without appearing as an isolated asymptotic particle state.

In standard quantum field theory, an observable asymptotic particle state requires the existence of local geometric curvature or boundaries to define a localized 'lump' of energy that can couple to external gauge-invariant interpolating fields. Because the $a=1$ metric is strictly uniform and featureless, it possesses no local invariants to trigger a response in a particle detector. Instead, its unsuppressed continuous translation symmetry corresponds to a completely delocalized coherent vacuum exchange (a Regge pole).
\section{Summary and Discussion}

In this paper, we have completed the one-loop analysis of the rigid twistor-string formulation of planar QCD. The geometric fluctuation determinant reduces to an infinite product of finite-dimensional Fourier blocks, so that the one-loop problem is reduced to an algebraic one on the mass shell.

The $\bar q q$ meson spectrum is determined by the saddle-point equations together with the Bohr--Sommerfeld condition, while the corresponding one-loop residues are determined by the regularized Fourier determinant and finite quadratures over the spin-projection parameter \(a\). The \(\zeta\)-regularized determinant is strictly independent of the winding number $w$, and its overall scale anomaly  cancels, so the one-loop fluctuations preserve the poles of the amplitude as a function of energy.

The large-winding regime is not an auxiliary assumption. Near the mass shell, \(\Phi_*(E,J;\beta)\to0\), the winding series develops the pole singularity that produces the physical meson states, and this singularity is generated by an arbitrarily large winding number \(w\). Since the quadratic fluctuation operator scales linearly with \(w\), the corresponding fluctuation variance is suppressed as \(1/w\). The path integral therefore localizes on the WKB saddle on shell. In this sense, the pole spectrum is determined  by the WKB saddle, while higher-loop corrections are parametrically suppressed in the same large-\(w\) regime.

We have also extended the phenomenological comparison by performing a global fit of 40 meson states across five topological sectors (\(h=0, \pm 1, \pm 2\)). The inclusion of the \(|h|=1\) sectors accounts for the axial-vector parity doublets through an exact \(J-1\) geometric degeneracy. This relates the mass of an axial state of spin \(J\) to a vector state of spin \(J+1\), reproducing the observed mass splittings without introducing phenomenological spin-orbit parameters. In addition, evaluating the one-loop residues provides the physical transition amplitudes on the mass shell. The resulting partial cross sections exhibit a nontrivial topology- and flavor-dependent scaling, correctly capturing the macroscopic heavy-mass quenching of boundary amplitudes alongside phase-space enhancement for high-spin light states along the Regge trajectories, in qualitative agreement with the expected geometric behavior of extended strings.

Beyond the open-string meson sector, we extended this exact geometric framework to the pure Yang--Mills closed string, explicitly deriving the universal, parameter-free macroscopic glueball spectrum and transition amplitudes. We demonstrated the  dynamical stability of the minimal surface: the conformal Liouville anomaly dynamically drives the closed glueball string strictly to the trigonometric limit (\(q=0\)). At this global minimum, the complex elliptic geometry analytically collapses, reducing the infinite fluctuation determinant to a stable, discrete product of finite-dimensional Fourier blocks and shielding the spectrum from tachyonic instabilities. As a direct consequence of this geometric collapse, the theoretical Regge trajectories are  linear.

Furthermore, while the purely geometric Regge trajectory natively intercepts the origin (\(M=0, J=0\)), we demonstrated that the quantum transition amplitude of this massless state identically vanishes due to the exact algebraic cancelation of the translation zero-mode measure. This mechanism mathematically nullifies the scalar ghost without introducing unphysical branch cuts, dynamically establishing a strictly positive mass gap. The existence of this glueball mass gap manifests the color confinement within our continuum formulation of planar QCD. Another important feature is the Casimir intercept \(\alpha(0) = 1/12\) which  follows from our theory for the anti-periodic glueball trajectory, naturally reproducing the effective L\"uscher zero-point energy. Finally, the overall fit of $7$ parameter-free glueball masses to the PDG data is remarkably good: \(R^2 = 0.98\) with a relative error of \(2.6\%\), which is significantly better than one would expect with zero fitting parameters in the leading order of the \(1/N_c^2\) expansion in real QCD.

A deeper consequence of this topological framework, which warrants further investigation, is the identification of an internal ``winding parity'' \(W = (-1)^{2a} = \pm 1\) that distinguishes the two closed-string sectors. As indicated by the evaluated functional determinants, the physical anti-periodic (\(a=1/2\), \(W=-1\)) sector is characterized by a non-trivial conformal metric and Liouville curvature, which strongly suppresses its transition amplitudes (\(\sim 10^{-5}\)). This geometric penalty perfectly aligns with the interpretation of these states as localized, \(s\)-channel physical glueballs that are notoriously difficult to produce experimentally. By contrast, the periodic (\(a=1\), \(W=+1\)) sector features a completely flat metric and macroscopically unsuppressed amplitudes (\(\sim 10^1\)). Lacking the intrinsic curvature required to form localized energy excitations, these states are likely forbidden from manifesting as observable asymptotic particles, strongly suggesting they exist purely as delocalized \(t\)-channel Regge poles.

This winding parity presents compelling candidates for resolving long-standing puzzles in high-energy scattering, dating back to the earliest formulations of Regge theory and the Pomeranchuk pole \cite{Pomeranchuk:1958aa, Gribov:1968fc}. The even-parity \(a=1\) sector yields a trajectory with an exact intercept of \(\alpha(0)=1\), capable of coupling directly to the vacuum channel. This provides a mathematically stable, non-perturbative candidate for the true bare Pomeron. This exact geometric result stands in stark contrast to perturbative QCD approximations involving multiple gluon emissions (such as the BFKL equation \cite{Kuraev:1977fs, Balitsky:1978ic}), which generate divergent ``hard'' Pomeron singularities with intercepts strictly greater than 1 that violate the Froissart unitarity bound \cite{Froissart:1961ux} and do not survive in the true non-perturbative continuum string limit.

Conversely, the \(n=-1\) daughter of the odd-parity \(a=1/2\) sector possesses a higher intercept of \(\alpha(0) = 1 + 1/12 = 13/12 \approx 1.083\). Because its odd winding parity fundamentally forbids it from mediating forward elastic vacuum scattering, this \(>1\) intercept poses no contradiction to unitarity. We hypothesize that this odd trajectory might provide the theoretical origin for the phenomenological ``soft Pomeron'' empirically fitted to world scattering data \cite{Donnachie:1992ny}. Exploring the exact role of these topological sectors in governing total hadronic cross-sections remains an exciting direction for future research. 

Taken together, these results determine the on-shell spectrum and the one-loop fluctuation determinant of the rigid twistor string in closed form, establish the existence of a positive pure-gauge mass gap, and strongly support this mathematical construction as a phenomenologically viable continuum realization of planar QCD.
\section*{Acknowledgements}
The author thanks N. Arkani-Hamed for the invitation to present this work at the IAS
Particle Physics (Pizza) Seminar, and the participants of the seminar for useful
discussions.

The author is also grateful to Nathan Seiberg for discussions concerning the relation
between this twistor-geometric construction and Seiberg--Witten theory.

Recent seminar in ENS, Paris, and subsequent long discussions with my old friend and colleague Volodya Kazakov helped me understand the correspondence of my theory with recent large $N_C$ computations in lattice gauge theories, including the TEK model. The results match for both glueballs and mesons (except $\pi$, for the reasons we discussed in Sections 8 and 9).

\paragraph{Declaration of generative AI in the writing process.}

During the preparation of this manuscript, the author used the generative AI systems
Gemini 3 Flash and ChatGPT 5.4 for assistance with editing and presentation. These tools
were used for correcting typographical and grammatical errors, improving the clarity of the
exposition, and standardizing notation and \LaTeX{} formatting.

All scientific content, derivations, and conclusions were developed and verified by the
author, who takes full responsibility for the manuscript.

\paragraph{Data availability.}
No data were created in this work. The computations of the meson and glueball masses were performed in \Mathematica{} notebooks \cite{MBTwistorAsymmetricMasses,MBTwistorStringDeterminant, MBGluonSpectrum}.


\bibliographystyle{JHEP}
\bibliography{bibliography.bib}

@article{MMEq79,
  author  = {Makeenko, Yu. M. and Migdal, A. A.},
  title   = {Exact equation for the loop average in multicolor QCD},
  journal = {Physics Letters B},
  volume  = {88},
  number  = {1},
  pages   = {135--137},
  year    = {1979},
  issn    = {0370-2693},
  doi     = {10.1016/0370-2693(79)90131-3},
  url     = {https://doi.org/10.1016/0370-2693(79)90131-3}
}

@article{Mig83,
  author  = {A. A. Migdal},
  title   = {Loop equations and $1/N$ expansion},
  journal = {Physics Reports},
  volume  = {102},
  number  = {4},
  pages   = {199--290},
  year    = {1983},
  doi     = {10.1016/0370-1573(83)90076-5}
}

@article{M81QCDFST,
title = {QCD=Fermi string theory},
journal = {Nuclear Physics B},
volume = {189},
number = {2},
pages = {253-294},
year = {1981},
issn = {0550-3213},
doi = {https://doi.org/10.1016/0550-3213(81)90381-3},
url = {https://www.sciencedirect.com/science/article/pii/0550321381903813},
author = {A.A. Migdal}
}

@misc{FP,
   author = "Wikipedia",
   title = "{Fundamental polygon} --- {W}ikipedia{,} The Free Encyclopedia",
   year = "2019",
   howpublished = {\url{http://en.wikipedia.org/w/index.php?title=Fundamental_polygon}},
   note = "[Online; accessed 29-November-2019]"
}

@article{Luscher1981,
  title = {How thick are chromoelectric flux tubes?},
  author = {L{\"u}scher, Martin and M{\"u}nster, Gernot and Weisz, Peter},
  journal = {Nuclear Physics B},
  volume = {180},
  number = {1},
  pages = {1--12},
  year = {1981},
  publisher = {Elsevier}
}

@article{Migdal2026GeometricQCDII,
title = {Geometric QCD II: The confining twistor string and meson spectrum},
author={Alexander Migdal},
journal = {Nuclear Physics B},
volume = {1026},
pages = {117424},
year = {2026},
issn = {0550-3213},
doi = {https://doi.org/10.1016/j.nuclphysb.2026.117424},
url = {https://www.sciencedirect.com/science/article/pii/S055032132600132X}
}

@misc{migdal2025geometric,
title = {Geometric QCD I: the Hodge-dual surface and quark confinement},
author = {Alexander Migdal},
journal = {Nuclear Physics B},
volume = {1025},
pages = {117380},
year = {2026},
issn = {0550-3213},
doi = {https://doi.org/10.1016/j.nuclphysb.2026.117380},
url = {https://www.sciencedirect.com/science/article/pii/S055032132600088X}
}

@inproceedings{Witten:1980ez,
    author = "Witten, Edward",
    title = "{The $1/N$ Expansion in Atomic and Particle Physics}",
    booktitle = "{Recent Developments in Gauge Theories. Proceedings, Nato Advanced Study Institute, Cargese, France, August 26 - September 8, 1979}",
    editor = "'t Hooft, G. and others",
    publisher = "Plenum Press",
    address = "New York",
    year = "1980",
    pages = "403--419",
    doi = "10.1007/978-1-4684-7571-5_21"
}

@article{Konopelchenko1999,
  author = {Konopelchenko, B. G. and Landolfi, G.},
  title = {Induced surfaces and their integrable dynamics {II}. {G}eneralized {W}eierstrass representations in {4D} spaces and string geometry},
  journal = {Journal of Geometry and Physics},
  volume = {29},
  number = {4},
  pages = {319--338},
  year = {1999},
  doi = {10.1016/S0393-0440(98)00045-3},
  eprint = {math/9810138},
  archivePrefix = {arXiv}
}

@misc{MBTwistorStringDeterminant,
    howpublished={\url{https://www.wolframcloud.com/obj/sasha.migdal/Published/TwistorStringDeterminant.nb}},
    author = {Alexander Migdal},
    title= {"Twistor String Determinant"},
    year = {2026},
    month = {04}
}

@misc{MBTwistorAsymmetricMasses,
    howpublished={\url{https://www.wolframcloud.com/obj/sasha.migdal/Published/TwistorAsymmetricMasses.nb}},
    author = {Alexander Migdal},
    title= {"QCD Twistor Asymmetric Masses"},
    year = {2026},
    month = {04}
}

@misc{MBGluonSpectrum,
    howpublished={\url{https://www.wolframcloud.com/obj/sasha.migdal/Published/GlueballSpectrumLinear.nb}},
    author = {Alexander Migdal},
    title= {"QCD Gluon Spectrum"},
    year = {2026},
    month = {05}
}

@article{Gribov:1968fc,
    author = "Gribov, V. N. and Migdal, A. A.",
    title = "{Strong Coupling in the Pomeranchuk Pole Problem}",
    journal = "Sov. Phys. JETP",
    volume = "28",
    pages = "784--795",
    year = "1969",
    note = "[Zh. Eksp. Teor. Fiz. 55, 1498 (1968)]"
}

@article{Pomeranchuk:1958aa,
    author = "Pomeranchuk, I. Ya.",
    title = "{Equality of the nucleon and antinucleon total cross sections at high energies}",
    journal = "Sov. Phys. JETP",
    volume = "7",
    pages = "499--501",
    year = "1958",
    note = "[Zh. Eksp. Teor. Fiz. 34, 725 (1958)]"
}

@article{Froissart:1961ux,
    author = "Froissart, M.",
    title = "{Asymptotic behavior and subtractions in the Mandelstam representation}",
    journal = "Phys. Rev.",
    volume = "123",
    pages = "1053--1057",
    year = "1961",
    doi = "10.1103/PhysRev.123.1053"
}

@article{Kuraev:1977fs,
    author = "Kuraev, E. A. and Lipatov, L. N. and Fadin, V. S.",
    title = "{The Pomeranchuk Singularity in Nonabelian Gauge Theories}",
    journal = "Sov. Phys. JETP",
    volume = "45",
    pages = "199--204",
    year = "1977",
    note = "[Zh. Eksp. Teor. Fiz. 72, 377 (1977)]"
}

@article{Balitsky:1978ic,
    author = "Balitsky, I. I. and Lipatov, L. N.",
    title = "{The Pomeranchuk Singularity in Quantum Chromodynamics}",
    journal = "Sov. J. Nucl. Phys.",
    volume = "28",
    pages = "822--829",
    year = "1978",
    note = "[Yad. Fiz. 28, 1597 (1978)]"
}

@article{Donnachie:1992ny,
    author = "Donnachie, A. and Landshoff, P. V.",
    title = "{Total cross-sections}",
    journal = "Phys. Lett. B",
    volume = "296",
    pages = "227--232",
    year = "1992",
    doi = "10.1016/0370-2693(92)90832-O"
}

@article{AthenodorouTeper2021,
    author = "Athenodorou, Andreas and Teper, Michael",
    title = "{SU(N) gauge theories in 3+1 dimensions: glueball spectrum, string tensions and topology}",
    eprint = "2106.00364",
    archivePrefix = "arXiv",
    primaryClass = "hep-lat",
    doi = "10.1007/JHEP12(2021)082",
    journal = "JHEP",
    volume = "12",
    pages = "082",
    year = "2021"
}

@article{Bonanno2026,
    author = "Bonanno, Claudio and Garc\'\i{}a P\'erez, Margarita and Gonz\'alez-Arroyo, Antonio and Ishikawa, Ken-Ichi and Okawa, Masanori",
    title = "{Meson spectrum and low-energy constants in large-$N$ QCD}",
    eprint = "2603.01583",
    archivePrefix = "arXiv",
    primaryClass = "hep-lat",
    journal = "PoS",
    volume = "LATTICE2025",
    pages = "447",
    year = "2026"
}
\appendix
\section{Analytical and numerical study of the Twistor String equation}
\label{sec:twistorSpectral}

The physical states of the twistor string are determined by the total action monodromy
\(\Delta S\) accumulated over one period of rotation. The Bohr--Sommerfeld quantization
condition requires \(\Delta S=2\pi n_{BS}\). As discussed in the main text, after factoring out
the overall twistor scale parameter \(a\), one can write \(\Delta S=a\Phi\), where the phase
functional sums the independent boundary contributions:
\EQ{
\Phi(K, \beta_u, \beta_d) &= 2\pi E K - 4\pi \left(J + \frac{h_u + h_d}{2}\right) - \frac{\pi \sigma K^2}{2} \left(\beta_u + \frac{\sin 2\beta_u}{2} + \beta_d + \frac{\sin 2\beta_d}{2}\right) \br
&- 2\pi K (m_u \cos\beta_u + m_d \cos\beta_d) - \frac{1}{3}(\tan\beta_u - \beta_u + \tan\beta_d - \beta_d).
}
Here \(K=R/a\) is the kinematic radius, and \(h_u, h_d \in \{0, \pm 1\}\) are the independent topological vector indices corresponding to the spin insertions at the respective boundaries.

\subsection{Reduction of the spectrum: only the \(n_{BS}=0\) level remains}

The classical string trajectory is determined by the saddle-point equations with respect to
the dynamical variables \(a\), \(K\), \(\beta_u\), and \(\beta_d\). Setting \(\partial_a \Delta S=0\), one obtains
\EQ{
\label{Phi0}
\Phi(K, \beta_u, \beta_d) = 0.
}
This implies that the Bohr--Sommerfeld condition
\EQ{
\Delta S = 2 \pi n_{BS}, \quad n_{BS} = 0, \pm 1, \pm 2, \dots
}
admits only the level \(n_{BS}=0\).

This follows from the fact that \(\Delta S\) depends linearly on \(a\) at fixed boundary phases and
\(K=R/a\). Passing to the variables \(a\), \(K\), and \(\beta_{u,d}\) removes \(a\) from the remaining
saddle-point equations and yields the additional condition \eqref{Phi0} for \(\beta_{u,d}\) and
\(K\).

Next, setting \(\partial_K \Phi=0\) yields the unified energy equation:
\EQ{
E = \frac{\sigma K}{2} \left(\beta_u + \frac{\sin 2\beta_u}{2} + \beta_d + \frac{\sin 2\beta_d}{2}\right) + m_u \cos\beta_u + m_d \cos\beta_d.
\label{eq:E_param}
}
Setting the independent boundary variations \(\partial_{\beta_u} \Phi=0\) and \(\partial_{\beta_d} \Phi=0\) gives two decoupled quadratic equations for \(K\):
\EQ{
\sigma \cos^2\beta_{u} \, K^2 - 2 m_{u} \sin\beta_{u} \, K + \frac{1}{3\pi} \tan^2\beta_{u} &= 0, \br
\sigma \cos^2\beta_{d} \, K^2 - 2 m_{d} \sin\beta_{d} \, K + \frac{1}{3\pi} \tan^2\beta_{d} &= 0.
}
Thus the saddle-point system is perfectly constrained. The single physical radius \(K\) forces the two asymmetric boundary phases to lock together dynamically.

\subsection{Algebraic simplifications of the spectral equation}

For the analytical and numerical analysis, it is convenient to work with dimensionless
variables. We introduce the dimensionless renormalized boundary mass parameters
\(x_{u,d}=m_{u,d}/\sqrt{\sigma}\). Solving the boundary quadratic equations for the radius, the global kinematic match requires:
\EQ{
\mathcal K(\beta_u, x_u) = \mathcal K(\beta_d, x_d) \equiv \sqrt{\sigma} K,
}
where
\EQ{
\mathcal K(\beta, x) = \frac{\sin \beta}{\cos^2 \beta} \left( x + \sqrt{x^2 - \frac{1}{3\pi}} \right).
}
For a physical string to exist, the classical radius \(K\) must be real, so the discriminants
must be non-negative. This gives the  chiral bound at both boundaries:
\EQ{
x_{u,d} \ge \frac{1}{\sqrt{3\pi}}.
}

Substituting \(E\) from eq.~\eqref{eq:E_param} back into \(\Phi=0\), the boundary-mass
terms cancel, and one obtains the total parametric angular momentum:
\EQ{
J(\beta_u, \beta_d) = \frac{\sigma K^2}{8} \left(\beta_u + \frac{\sin 2\beta_u}{2} + \beta_d + \frac{\sin 2\beta_d}{2}\right) - \frac{1}{12\pi} (\tan\beta_u - \beta_u + \tan\beta_d - \beta_d) - \frac{h_u + h_d}{2}.
\label{eq:J_param}
}

To fit the fundamental parameters---the dimensionless constituent quark masses
\((x_u,x_s,x_c,x_b)\) and the universal string tension---to the experimental Particle Data
Group data, we use the action monodromy condition \(\Phi=0\). Rather than fitting \(M^2\)
or \(J\) separately, one may minimize the physical constraint of the model itself.

For a given observed resonance with experimental mass \(M\) and spin \(J\), one first
inverts the coupled geometric system eq.~\eqref{eq:J_param} to determine the corresponding angle pair \(\beta_u, \beta_d\) such that
\(J_{\rm param}(\beta_u, \beta_d)=J\). Substituting these angles back into the phase equation, the angular
momentum terms cancel. Since the theoretical parameters satisfy \(\Phi=0\), the area,
boundary-mass, and Liouville terms combine to leave the geometric phase mismatch
\EQ{
\Phi(M, J) = 2\pi K \left( M - E_{\rm param}(\beta_u, \beta_d) \right).
}
This target function has a natural interpretation. Since \(K\propto 1/\sqrt{\sigma}\), the
quantity \(\Phi\) is dimensionless. At high energies, where \(K\propto E/\sigma\), the action
monodromy scales as
\[
\Phi \propto \frac{M\,\Delta M}{\sigma} \propto \frac{\Delta(M^2)}{2\sigma},
\]
which reproduces the expected mass-squared weighting of Regge trajectories. 

\section{Quadratic expansion of the geometric Hessian}
\label{app:GeometricHessian}

In this Appendix we compute the local quadratic form of the geometric part of the
effective action around the classical Minkowski helicoid background. We use the
norm-preserving multiplicative parametrization
\EQ{
\tilde\phi(\xi)=\exp{f_1(\xi)}\phi_0(\xi),
\qquad
\tilde\psi(\eta)=\exp{f_2(\eta)}\psi_0(\eta),
\qquad
f_1^\dagger=-f_1,
\qquad
f_2^\dagger=-f_2,
\label{AppGeomExpAnsatz}
}
where \(\xi=\tau+\theta\), \(\eta=\tau-\theta\), and \(\phi_0,\psi_0\) are the classical
twistors of the rotating Minkowski helicoid. In contrast with the linear ansatz
\((I+f_{1,2})\), the exponential form preserves the equal norms .

We expand the anti-Hermitian fluctuation matrices in the Pauli basis,
\EQ{
f_1=\I\left(\alpha_1\mathbf 1+u_i\sigma_i\right),
\qquad
f_2=\I\left(\alpha_2\mathbf 1+v_i\sigma_i\right),
\label{AppPauliDecomp}
}
with real coefficients \(\alpha_{1,2},u_i,v_i\). It is convenient to introduce
\EQ{
s\equiv \alpha_1+\alpha_2,
\qquad
\Delta_i\equiv v_i-u_i,
\qquad
C_i\equiv (\vec u\times \vec v)_i.
\label{AppDeltaCDefs}
}

The basic geometric quantity is the twistor bracket
\EQ{
B(\tau,\theta)\equiv
\tilde\phi_1\tilde\psi_2-\tilde\phi_2\tilde\psi_1
=
\tilde\phi^T\varepsilon\,\tilde\psi,
\qquad
\varepsilon=
\begin{pmatrix}
0&1\\
-1&0
\end{pmatrix},
\label{AppBDef}
}
because \(\Omega^2=|B|^2\) and \(\rho=\log|B|\). For the classical solution of section~6.2 one has
\EQ{
\phi_0(\xi)=\sqrt{\frac R2}
\begin{pmatrix}
\exp{\I(a\xi+\pi/4)}\\
\exp{-\I(a\xi+\pi/4)}
\end{pmatrix},
\qquad
\psi_0(\eta)=\sqrt{\frac R2}
\begin{pmatrix}
\exp{\I(a\eta-\pi/4)}\\
\exp{-\I(a\eta-\pi/4)}
\end{pmatrix},
\label{AppClassicalTwistors}
}
and therefore, with \(x\equiv 2a\theta\), \(t\equiv 2a\tau\), and \(c\equiv \cos x\), the classical bracket is
\EQ{
B_0=\phi_0^T\varepsilon\psi_0=\I R\cos x.
\label{AppB0Def}
}
We also define the auxiliary bilinears
\EQ{
D_i\equiv \phi_0^T\varepsilon\sigma_i\psi_0,
\label{AppDiDef}
}
which evaluate to
\EQ{
D_1=\I R\sin t,
\qquad
D_2=\I R\cos t,
\qquad
D_3=R\sin x.
\label{AppDiExplicit}
}

Using the identity \(A^T\varepsilon+\varepsilon A=(\tr A)\varepsilon\) for \(2\times2\) matrices, the bracket expands as \(B=B_0+\delta B+B^{(2)}+\mathcal{O}(f^3)\).
The linear term is
\EQ{
\delta B=\I\,s\,B_0+\I\,\Delta_iD_i.
\label{AppdBLinear}
}
Equivalently, \(\frac{\delta B}{B_0}=Y+\I Z\), with
\EQ{
Y=\Delta_3\tan x,
\qquad
Z=s+\frac{\Delta_1\sin t+\Delta_2\cos t}{\cos x}.
\label{AppYZdefs}
}
The quadratic correction is
\EQ{
\frac{B^{(2)}}{B_0}
=
-\frac12\left(s^2+\Delta^2\right)
-s\,\frac{\Delta_1\sin t+\Delta_2\cos t}{\cos x}
+\I\,s\,\Delta_3\tan x
+C_3\tan x
+\I\,\frac{C_1\sin t+C_2\cos t}{\cos x}.
\label{AppB2overB0}
}

\subsection{Area density and Liouville field}

Since \(\Omega^2=|B|^2\), we write \(\Omega^2=\Omega_0^2+\delta\Omega^2+\Omega_{(2)}^2+\mathcal{O}(f^3)\), where \(\Omega_0^2=R^2\cos^2 x\).
The linear term is \(\delta\Omega^2=2\Omega_0^2\,Y\). The quadratic term simplifies to
\EQ{
\frac{\Omega_{(2)}^2}{\Omega_0^2}
=
X^2+Y^2-\Delta^2+2C_3\tan x,
\qquad
X\equiv \frac{\Delta_1\sin t+\Delta_2\cos t}{\cos x}.
\label{AppOmega2Quadratic}
}
Equivalently,
\EQ{
\Omega_{(2)}^2
=
R^2\Big[
(\Delta_1\sin t+\Delta_2\cos t)^2
+\Delta_3^2\sin^2 x
-\cos^2 x\,\Delta^2
+2\sin x\cos x\,C_3
\Big].
\label{AppOmega2QuadraticExplicit}
}

For the Liouville field we write \(\rho=\rho_0+\rho_1+\rho_2+\mathcal{O}(f^3)\). Then
\EQ{
\rho_1=Y=\Delta_3\tan x,
\qquad
\rho_2
=
-\frac12\Delta^2
+\frac12X^2
-\frac12Y^2
+C_3\tan x.
\label{AppRho2}
}
Two consequences follow. First, the common phase mode
\(s=\alpha_1+\alpha_2\) drops out of both \(\Omega^2\) and \(\rho\) through quadratic order.
Second, only the relative fluctuation \(\Delta_3=v_3-u_3\) enters the Liouville field at
linear order.

\subsection{Quadratic variations of the action}

The target-time term of section~6.5 has no quadratic fluctuation. Indeed, the anti-Hermitian
exponential ansatz preserves \(\tilde\phi^\dagger\tilde\phi\) and
\(\tilde\psi^\dagger\tilde\psi\) pointwise, so \(\delta^2(E\Delta X^0)=0\).

For the minimal area term,
\EQ{
S_{\rm area}=\sigma \int_0^{2\pi} d\tau \int_{-\theta_d}^{\theta_u} d\theta\,\Omega^2,
\label{AppAreaTermDef}
}
the quadratic variation over the asymmetric domain is
\EQ{
\delta^2 S_{\rm area}
=
\sigma \int_0^{2\pi} d\tau \int_{-\theta_d}^{\theta_u} d\theta\,\Omega_{(2)}^2.
\label{AppAreaQuadratic}
}
Using \eqref{AppOmega2QuadraticExplicit}, we obtain
\EQ{
\delta^2 S_{\rm area}
=
\sigma R^2\int_0^{2\pi} d\tau \int_{-\theta_d}^{\theta_u} d\theta\,
\Big[
(\Delta_1\sin t+\Delta_2\cos t)^2
+\Delta_3^2\sin^2 x
-\cos^2 x\,\Delta^2
+2\sin x\cos x\,C_3
\Big].
\label{AppAreaQuadraticExplicit}
}

The boundary proper-mass term evaluated at the distinct constituent boundaries is
\EQ{
S_{\rm mass}
=
\int_0^{2\pi} d\tau\, \Big[ m_u \Omega(\tau,\theta_u) + m_d \Omega(\tau,-\theta_d) \Big].
\label{AppMassTermDef}
}
Expanding \(\Omega=\Omega_0\left(1+\rho_1+\rho_2+\frac12\rho_1^2\right)+\mathcal{O}(f^3)\),
the quadratic contribution becomes
\EQ{
\delta^2 S_{\rm mass}
=
\int_0^{2\pi} d\tau\, R \Bigg[ &m_u \cos\beta_u \left( -\frac12\Delta_u^2 + \frac12 X_u^2 + C_{3,u}\tan\beta_u \right) \br
&+ m_d \cos\beta_d \left( -\frac12\Delta_d^2 + \frac12 X_d^2 - C_{3,d}\tan\beta_d \right) \Bigg],
\label{AppMassQuadratic}
}
where the subscripts \(u, d\) denote evaluation at \(x=\beta_u\) and \(x=-\beta_d\) respectively.

Finally, the quadratic variation for the Liouville term integrated over the exact asymmetric bulk is
\EQ{
\delta^2 S_L
=
\frac{1}{12\pi}
\int_0^{2\pi} d\tau \int_{-\theta_d}^{\theta_u} d\theta\,
\Big[
(\partial_\theta\rho_1)^2-(\partial_\tau\rho_1)^2
+2\,\partial_\theta\rho_0\,\partial_\theta\rho_2
\Big].
\label{AppLiouvilleQuadratic}
}
Since \(\partial_\theta\rho_0=-2a\tan x\), this may also be written as
\EQ{
\delta^2 S_L
=
\frac{1}{12\pi}
\int_0^{2\pi} d\tau \int_{-\theta_d}^{\theta_u} d\theta\,
\Big[
(\partial_\theta\rho_1)^2-(\partial_\tau\rho_1)^2
-4a\tan x\,\partial_\theta\rho_2
\Big].
\label{AppLiouvilleQuadraticExplicit}
}

These formulas provide the complete, parity-broken local quadratic form of the geometric Hessian. The next step is to expand the coefficient functions \(u_j(\xi)\) and \(v_j(\eta)\) in Fourier modes and assemble the reduced one-dimensional mode matrices.

\section{Fourier reduction of the geometric Hessian}
\label{app:FourierGeometricHessian}

Having established the local quadratic variations, we now reduce the corresponding functional determinant to Fourier modes. The analyticity of the twistor
construction leads instead to a Fourier reduction to finite-dimensional matrix blocks.

Because we no longer assume symmetric quark masses, we integrate the quadratic effective action over the asymmetric physical string width \(x \in [-\beta_d, \beta_u]\). This breaks the parity reflection symmetry on the spatial strip, meaning the left-moving and right-moving twistor fluctuations no longer cleanly decouple.

\subsection{Fourier expansion in the co-rotating frame}

We begin by defining the complex co-rotating transverse fields and the real longitudinal
fields from the Pauli expansion:
\EQ{
    p(\xi) &\equiv \exp{2\I a\xi}(u_1 + \I u_2), \quad
    q(\eta) \equiv \exp{2\I a\eta}(v_1 + \I v_2), \br
    r(\xi) &\equiv u_3(\xi), \quad s(\eta) \equiv v_3(\eta).
}
Using the chiral light-cone coordinates, the dimensionless spatial variable \(x=2a\theta\), and the frequency parameter
\(\nu_n=n/(2a)\), we expand these fields in Fourier series:
\EQ{
    p(\xi) &= \sum_{n\in\mathbb{Z}} p_n \exp{\I n \tau} \exp{\I \nu_n x}, \quad
    q(\eta) = \sum_{n\in\mathbb{Z}} q_n \exp{\I n \tau} \exp{-\I \nu_n x}, \br
    r(\xi) &= \sum_{n\in\mathbb{Z}} r_n \exp{\I n \tau} \exp{\I \nu_n x}, \quad
    s(\eta) = \sum_{n\in\mathbb{Z}} s_n \exp{\I n \tau} \exp{-\I \nu_n x}.
}
Because \(u_j\) and \(v_j\) are real fields, the negative-frequency modes are related by
complex conjugation: \(\bar p_n=p_{-n}^*\), \(\bar q_n=q_{-n}^*\), \(r_{-n}=r_n^*\), and \(s_{-n}=s_n^*\).

Substituting these into the combinations \(D=s(\eta)-r(\xi)\) and \(W=\exp{\I x}q(\eta)-\exp{-\I x}p(\xi)\) gives:
\EQ{
    D(\tau, \theta) &= \sum_{n \in \mathbb{Z}} \exp{\I n \tau} \left( s_n \exp{-\I \nu_n x} - r_n \exp{\I \nu_n x} \right) \equiv \sum_{n \in \mathbb{Z}} \exp{\I n \tau} D_n(x), \br
    W(\tau, \theta) &= \sum_{n \in \mathbb{Z}} \exp{\I n \tau} \left( q_n \exp{\I (1 - \nu_n) x} - p_n \exp{-\I (1 - \nu_n) x} \right) \equiv \sum_{n \in \mathbb{Z}} \exp{\I n \tau} W_n(x).
}
Integrating the quadratic action over the temporal cycle \(\tau\in[0,2\pi]\) orthogonalizes the Fourier modes, decoupling the transverse and longitudinal sectors.

\subsection{The transverse Hessian (\(4\times4\) blocks)}

To compute the transverse fluctuation determinant for unequal masses, we can no longer impose the parity-adapted strip relation \(q_n=p_n\), as the limits of integration \([-\beta_d, \beta_u]\) explicitly break the spatial parity symmetry. Furthermore, because the quadratic action contains the term \((\Im W)^2\), the positive- and negative-frequency modes naturally mix.

For each \(n>0\), the transverse fluctuations therefore couple into an isolated \(4\times4\) block connecting the full conjugate multiplet. Defining the extended state vector \(V_n=\begin{pmatrix} p_n & q_n & p_{-n}^* & q_{-n}^* \end{pmatrix}^T\), the spatial integration over the physical interval gives
\EQ{
    \delta^2 S_{\perp}^{(n)} = \frac{\pi}{a} V_n^\dagger \mathcal{H}_n^{(\perp)} V_n, \quad \mathcal{H}_n^{(\perp)} = \int_{-\beta_d}^{\beta_u} dx \, \mathbf{M}^{(\perp)}(x, \nu_n).
}
The \(4\times4\) matrix integrand \(\mathbf{M}^{(\perp)}(x, \nu_n)\) is obtained by expanding \(|W|^2\), \((\Im W)^2\), and \(C_3\). Because the integration bounds are asymmetric, the definite integrals of the odd trigonometric cross-terms no longer vanish, directly populating the fully coupled matrix blocks. The boundary proper-mass terms at \(\beta_u\) and \(-\beta_d\) add algebraic contributions directly to these integrated matrix elements.

\subsection{The longitudinal Hessian (\(2\times2\) blocks over complex fields)}

Similarly, because the parity constraint \(s_n=\pm r_n\) is broken by the unequal integration limits, the longitudinal fluctuations \(r_n\) and \(s_n\) form a coupled \(2\times2\) block over independent complex variables for each \(n>0\). Defining the extended vector \(U_n=\begin{pmatrix} r_n & s_n \end{pmatrix}^T\), the longitudinal action reduces to
\EQ{
    \delta^2 S_{\parallel}^{(n)} = \frac{\pi}{a} U_n^\dagger \mathcal{H}_n^{(\parallel)} U_n, \quad \mathcal{H}_n^{(\parallel)} = \int_{-\beta_d}^{\beta_u} dx \begin{pmatrix} L_{11}(x, \nu_n) & L_{12}(x, \nu_n) \\ L_{12}^*(x, \nu_n) & L_{22}(x, \nu_n) \end{pmatrix}.
}
Because \(\rho_1=D\tan x\) is built from Fourier exponentials, the kinetic Liouville term
\((\partial_\theta\rho_1)^2-(\partial_\tau\rho_1)^2\) simply produces identical algebraic trigonometric coefficients:
\EQ{
    \partial_x \rho_{1,n} = \left(-\I \nu_n s_n \exp{-\I \nu_n x} - \I \nu_n r_n \exp{\I \nu_n x}\right) \tan x + D_n(x) \sec^2 x.
}
Substituting this expression into the geometric action yields the \(2\times2\) algebraic integrands \(L_{ij}(x, \nu_n)\).

\subsection{Factorization and \(n\leftrightarrow -n\) symmetry}

Projecting the functional integral onto this Fourier basis replaces the boundary-value
problem by an infinite product of exact algebraic matrices:
\EQ{
    \det \mathcal{H}_{geom} = \prod_{n=1}^\infty \det \mathcal{H}_n^{(\parallel)} \det \mathcal{H}_n^{(\perp)}.
}
All matrix integrands are exact rational trigonometric functions on the finite interval
\([-\beta_d,\beta_u]\), so each block reduces to a precise spatial quadrature.

While spatial parity is broken, the Fourier representation perfectly preserves the local temporal \(n\leftrightarrow -n\) symmetry. Substituting \(\nu_n\to-\nu_n\) is equivalent to complex conjugation and exchanging the positive/negative frequency components, which leaves the determinant invariant:
\EQ{
    \det \mathcal{H}_n^{(\perp)} = \det \mathcal{H}_{-n}^{(\perp)}, \quad \det \mathcal{H}_n^{(\parallel)} = \det \mathcal{H}_{-n}^{(\parallel)}.
}

\subsection{The zero-frequency sector (\(n=0\))}

To complete the determinant, one must evaluate the \(n=0\) Fourier sector. Because \(\nu_0=0\), the zero-mode state vectors fuse into real and imaginary combinations \(V_0 = \begin{pmatrix} p_R & q_R & p_I & q_I \end{pmatrix}^T\). The diagonalized geometric densities evaluate to:
\EQ{
    \lambda_+(x) &= \sigma R^2 \left( 4\sin^4(x) + 2\sin^2(x) \cos(x) \right) + \frac{2a^2}{3\pi} \sec^2(x) \left( 2\tan^2(x) + \tan(x)\sin(x) \right), \br
    \lambda_-(x) &= \sigma R^2 \left( -4\sin^2(x) \cos^2(x) + 2\sin^2(x) \cos(x) \right) + \frac{2a^2}{3\pi} \sec^2(x) \tan(x)\sin(x).
}
Integrating these functions over the asymmetric interval \(x\in[-\beta_d,\beta_u]\) requires preserving the odd components, coupling the block directly to the respective algebraic boundary-mass terms.

\subsection{Complete determinant and cancellation of the overall scale}

The total geometric fluctuation determinant is obtained by taking the regularized product over all Fourier modes. Changing variables to \(x=2a\theta\) gives \(d\theta=dx/(2a)\), while \(R=aK\). Hence, every term in the geometric action extracts a common overall factor \(a^2/a=a\):
\EQ{
    \mathcal{H}_n^{(k,a)} = (ka) \widetilde{\mathcal{H}}_n(\beta_d, \beta_u, K; a).
}
The power of \(ka\) extracted from the full determinant is fixed by the total number of independent real degrees of freedom. Breaking the geometric parity expands the isolated Fourier blocks (from \(2\times 2\) to \(4\times 4\)), thus squaring the individual matrix contributions. However, because the fundamental identity \(\tr_{\mathbb{Z}}\mathbf{1}=1+2\zeta(0)=0\) applies  to each independent real twistor field, the total zero-mode and nonzero-mode scale factors precisely cancel:
\EQ{
    \det \mathcal{H}_{geom}^{(k, a)} = (ka)^0 \det \widetilde{\mathcal{H}}_{geom}(\beta_d, \beta_u, K; a) = \det \widetilde{\mathcal{H}}_{geom}(\beta_d, \beta_u, K; a).
}
Thus the generalized geometric fluctuation determinant remains entirely independent of \(k\), and its overall scaling with \(a\) cancels .

\subsection{Analytical integration of the zero-mode sector}

The \(n=0\) sector quadratures can be evaluated directly over the asymmetric limits. Extracting the \(a^2\) scaling from the local integrands \(\lambda_\pm(x)\), the bulk contributions require evaluating the exact trigonometric primitives via the fundamental theorem of calculus:
\EQ{
    \int \left( 4\sin^4 x + 2\sin^2 x \cos x \right) dx &= \frac{3x}{2} - \sin(2x) + \frac{1}{8}\sin(4x) + \frac{4}{3}\sin^3 x, \br
    \int \left( -4\sin^2 x \cos^2 x + 2\sin^2 x \cos x \right) dx &= -\frac{x}{2} + \frac{1}{8}\sin(4x) + \frac{4}{3}\sin^3 x,
}
and for the Liouville components:
\EQ{
    \int \sec^2 x \left( 2\tan^2 x + \tan x \sin x \right) dx &= \frac{2}{3}\tan^3 x + \frac{1}{2}\sec x \tan x - \frac{1}{2}\ln|\sec x + \tan x|, \br
    \int \sec^2 x \left(\tan x \sin x \right) dx &= \frac{1}{2}\sec x \tan x - \frac{1}{2}\ln|\sec x + \tan x|.
}
Evaluating these exact primitive functions \([ F(x) ]_{-\beta_d}^{\beta_u}\) over the asymmetric boundary limits yields the bulk zero-mode contributions:
\EQ{
    I_\pm^{bulk}(\beta_d, \beta_u, K) &= \frac{1}{2a}\int_{-\beta_d}^{\beta_u} \lambda_\pm(x) dx \equiv a \;\widetilde{I}_\pm^{bulk}(\beta_d, \beta_u, K).
}
Combining these integrals with the asymmetric boundary matrices \(\widetilde{B}_\pm\) evaluated at \(x=\beta_u\) and \(x=-\beta_d\) provides the exact zero-mode matrix determinant.

\subsection{Matrix elements, exact analytic integration, and Abel regularization}

The Fourier reduction expresses the functional determinant as an infinite product of exact algebraic matrices. To strictly isolate the physical degrees of freedom, the functional determinant must be restricted to the non-singular $4\times 4$ transverse matrix block. The longitudinal determinant corresponds to pure gauge modes of the worldsheet reparametrization invariance, whose zero-mode vanishes but is  canceled by the corresponding conformal Faddeev-Popov ghost determinant. 

\EQ{
\log \det \widetilde{\mathcal{H}}_{geom} \approx \sum_{n=1}^\infty \log \mathcal{D}_\infty + \sum_{n=1}^\infty \log\lrb{\frac{\big|\det \widetilde{\mathcal{H}}^{(\perp)}_n\big|}{\mathcal{D}_\infty}}.
}
The approximate identity in this sum is understood in a sense of averaging over the period of oscillation of the $\log\det H^{\perp}_n$.
In the transverse sector, the frequency \(\nu_n=n/(2a)\) appears inside oscillatory functions. By the Riemann--Lebesgue lemma, the oscillatory terms vanish after integration over fixed \([-\beta_d,\beta_u]\), while the squared terms cleanly average to \(1/2\). Evaluating the remaining asymptotic baseline primitives across the asymmetric boundaries yields:
\EQ{
I_{11}^\infty &= \left[ -\frac{\sigma K^2}{2} \sin(2x) + \frac{1}{3\pi}\left( \tan x + \frac{1}{3}\tan^3 x \right) \right]_{-\beta_d}^{\beta_u} \equiv I_{22}^\infty, \br
I_{12}^\infty &= \left[ -\frac{\sigma K^2}{2} \sin(2x) + \frac{1}{3\pi}\left( -\tan x + \frac{1}{3}\tan^3 x \right) \right]_{-\beta_d}^{\beta_u}.\\
\mathcal{D}_\infty^{(\perp)} &= (I_{11}^\infty)^2-(I_{12}^\infty)^2;
}
Because \(\sum_{n=1}^\infty 1\to\zeta(0)=-1/2\), the \(\zeta\)-regularized baseline \(\mathcal{D}_\infty=\mathcal{D}_\infty^{(\perp)}\) evaluates cleanly to:
\EQ{
\det \widetilde{\mathcal{H}}_{geom} = \frac{1}{\sqrt{\mathcal{D}_\infty}} \det \widetilde{\mathcal{H}}^{(\perp)}_0 \exp{ \sum_{n=1}^\infty \log\lrb{\frac{\big|\det \widetilde{\mathcal{H}}^{(\perp)}_n\big|}{\mathcal{D}_\infty}} }.
}

At finite \(n\), the matrix elements reduce to exact oscillatory spatial quadratures of the form
\EQ{
E(w, p; \beta_d, \beta_u) \equiv \int_{-\beta_d}^{\beta_u} \frac{\cos(wx)}{\cos^p(x)} dx, \quad \text{for integers } p \ge 0.
}
To completely bypass numerical integration and its associated memory overhead and instabilities for highly oscillatory phases, the negative powers of the cosine are expanded into a geometric series of complex exponentials. Integrating this series analytically term-by-term yields an exact, closed-form elementary primitive in terms of the Gauss hypergeometric function \({}_2F_1\):
\EQ{
\int \frac{e^{iwx}}{\cos^p(x)} dx = -i \frac{2^p}{w+p} e^{i(w+p)x} \,_2F_1\left(p, \frac{w+p}{2}; \frac{w+p}{2}+1; -e^{2ix}\right).
}
Because the physical turning points lie strictly within the classical interval \(-\pi/2 < -\beta_d \le \beta_u < \pi/2\), the argument \(-e^{2ix}\) resides exclusively in the left-half complex plane and never crosses the branch cut of the hypergeometric function. Consequently, the matrix integrals converge unconditionally and are given  by the real and imaginary parts of this primitive evaluated at the boundaries.

For exceedingly large frequencies (\(w \ge 50\)), one guarantees numerical precision and prevents underflow in the hypergeometric series by seamlessly matching to the stable asymptotic expansion obtained by repeated integration by parts:
\EQ{
E(w, p; \beta_d, \beta_u) &= \sum_{k=0}^{M} \frac{1}{w^{k+1}} \left( \sin\left(w\beta_u + \frac{k\pi}{2}\right) \left. \frac{d^k}{dx^k} \sec^p(x) \right|_{x=\beta_u} \right. \br
&\quad \left. - \sin\left(-w\beta_d + \frac{k\pi}{2}\right) \left. \frac{d^k}{dx^k} \sec^p(x) \right|_{x=-\beta_d} \right) + \mathcal{O}\left(w^{-(M+2)}\right).
}
Finally, to evaluate the conditionally convergent residual logarithmic series, one introduces the Abel regulator
\(\exp{-\epsilon n}\), computes the truncated sum at finite \(\epsilon\), and extracts the exact physical
limit \(\epsilon\to0^+\) using an exact ordinary least-squares matrix regression over a stable domain of regularizers:
\EQ{
S(a) = \lim_{\epsilon \to 0^+} \left[ -\frac{1}{2} \sum_{n=1}^{N_{\text{max}}} \exp{-\epsilon n} \log\lrb{\frac{\big|\det \widetilde{\mathcal{H}}^{(\perp)}_n(\beta_d, \beta_u, K; a)\big|}{\mathcal{D}_\infty}} \right].
}
This  analytical point-splitting procedure perfectly isolates the finite transverse quantum fluctuation contribution of the asymmetric meson string.

\section{Quadratic expansion of the geometric Hessian at $q=0$}
\label{app:TrigGeometricHessian}

In this Appendix we compute the local quadratic form of the geometric part of the effective action around the pure Yang--Mills closed twistor string background. Because the macroscopic glueball string dynamically equilibrates precisely at the stable minimum $q=0$, the complex elliptic geometry of the worldsheet analytically collapses into elementary trigonometric functions.

We use the norm-preserving multiplicative parametrization
\EQ{
\tilde\phi(\xi)=\exp{f_1(\xi)}\phi_0(\xi),
\qquad
\tilde\psi(\eta)=\exp{f_2(\eta)}\psi_0(\eta),
\qquad
f_1^\dagger=-f_1,
\qquad
f_2^\dagger=-f_2,
}
where \(\xi=\tau+\sigma\), \(\eta=\tau-\sigma\), and \(\phi_0,\psi_0\) are the classical twistors. We expand the anti-Hermitian fluctuation matrices strictly in the Pauli basis:
\EQ{
f_1=\I\left(\alpha_1\mathbf 1+u_i\sigma_i\right),
\qquad
f_2=\I\left(\alpha_2\mathbf 1+v_i\sigma_i\right),
}
defining $s \equiv \alpha_1+\alpha_2$, $\Delta_i \equiv v_i-u_i$, and $C_i \equiv (\vec u\times \vec v)_i$.

\subsection{1. The Trigonometric Classical Background}
For a closed string with topologically quantized monodromy $a \in \{1/2, 1\}$, the classical twistors are governed by Riemann Theta functions. At $q=0$, these strictly reduce to pure phases $\vartheta_{\pm a}(z) \to \exp{\pm \I a z}$. The four exact combinations that parameterize the background metric natively evaluate to standard circular trigonometric functions:
\EQ{
\Theta_0(\tau,\sigma) &= \frac{1}{2} \left[ e^{\I a\xi}e^{-\I a\eta} + e^{-\I a\xi}e^{\I a\eta} \right] = \cos(2a\sigma), \br
\hat\Theta_1(\tau,\sigma) &= \frac{1}{2\I} \left[ e^{\I a\xi}e^{\I a\eta} - e^{-\I a\xi}e^{-\I a\eta} \right] = \sin(2a\tau), \br
\hat\Theta_2(\tau,\sigma) &= \frac{1}{2} \left[ e^{\I a\xi}e^{\I a\eta} + e^{-\I a\xi}e^{-\I a\eta} \right] = \cos(2a\tau), \br
\hat\Theta_3(\tau,\sigma) &= \frac{1}{2\I} \left[ e^{\I a\xi}e^{-\I a\eta} - e^{-\I a\xi}e^{\I a\eta} \right] = \sin(2a\sigma).
}
The classical bracket evaluates to \(B_0 = \I R\cos(2a\sigma)\). 

Dividing the linear variation $\delta B$ by $B_0$, we isolate the exact real and imaginary parts $\frac{\delta B}{B_0}=Y+\I Z$, with
\EQ{
Y &= \Delta_3\frac{\sin(2a\sigma)}{\cos(2a\sigma)} = \Delta_3\tan(2a\sigma), \br
Z &= s+\frac{\Delta_1\sin(2a\tau)+\Delta_2\cos(2a\tau)}{\cos(2a\sigma)}.
}

\subsection{2. Area density and Liouville field}
Since $\Omega^2=|B|^2$, the classical metric evaluates natively to $\Omega_0^2 = R^2\cos^2(2a\sigma)$. The quadratic area term  simplifies to:
\EQ{
\frac{\Omega_{(2)}^2}{R^2}
=
\big(\Delta_1\sin(2a\tau)+\Delta_2\cos(2a\tau)\big)^2
+\Delta_3^2\sin^2(2a\sigma)
-\cos^2(2a\sigma)\,\Delta^2
+2\cos(2a\sigma)\sin(2a\sigma)\,C_3.
}

For the Liouville field we write $\rho=\ln|B|=\rho_0+\rho_1+\rho_2$. The linear anomaly field is  $\rho_1 = Y = \Delta_3\tan(2a\sigma)$. The quadratic Liouville variation identically matches the transverse area norm:
\EQ{
\rho_2
=
-\frac12\Delta^2
+\frac12\left(\frac{\Delta_1\sin(2a\tau)+\Delta_2\cos(2a\tau)}{\cos(2a\sigma)}\right)^2
-\frac12\Delta_3^2\tan^2(2a\sigma)
+C_3\tan(2a\sigma).
}
The continuous translation phase $s$ strictly drops out of both $\Omega^2$ and $\rho$ through quadratic order. Only the relative fluctuation $\Delta_3$ enters the Liouville field at linear order.

\subsection{3. Total Geometric Action}
When summing over macroscopic configurations, the target-time component has  zero quadratic fluctuation, $\delta^2(w \I q_0\Delta X_0)=0$. The boundary proper-mass variation is identically zero, $\delta^2 S_{\rm mass} = 0$.

For the minimal area term integrated over the spatial body $\sigma \in [0, 2\pi]$, the quadratic variation evaluates to:
\EQ{
\delta^2 S_{\rm area}
=
w \sigma R^2\int_0^{2\pi} d\tau \int_0^{2\pi} d\sigma\, \frac{\Omega_{(2)}^2}{R^2}.
}

We define the spatial conformal derivative $H(\sigma) \equiv \frac{\partial_\sigma\Theta_0}{\Theta_0} = -2a\tan(2a\sigma)$. The exact Liouville geometric Hessian can be identically written as:
\EQ{
\delta^2 S_L
=
\frac{w}{12\pi}
\int_0^{2\pi} d\tau \int_0^{2\pi} d\sigma\,
\Big[
(\partial_\sigma\rho_1)^2-(\partial_\tau\rho_1)^2
+2\,H(\sigma)\,\partial_\sigma\rho_2
\Big].
}

\section{Explicit Computation of the Trigonometric Hessian}
\label{app:TrigFourierHessian}

To evaluate the functional determinant of the pure Yang--Mills twistor string, the geometric action is  reduced to algebraic matrix elements. By analytically performing integration-by-parts on the periodic worldsheet, we  transfer the Liouville curvature onto the classical spatial background, which cleanly decouples the space-time components.

Using the translation-invariant gauge $A_{i,0} = B_{i,0} = 0$, the Pauli coefficients are expanded in pure Fourier modes:
\EQ{
u_i(\xi) = \frac{1}{\sqrt{2\pi}} \sum_{n \neq 0} A_{i,n} \exp{\I n \xi}, \qquad
v_i(\eta) = \frac{1}{\sqrt{2\pi}} \sum_{m \neq 0} B_{i,m} \exp{-\I m \eta}.
}

\subsection{1. The Transverse Matrix Integrands and Banded Sparsification}
The transverse sector isolates $(u_1, u_2, v_1, v_2)$, meaning $u_3=v_3=0$, which yields $\Delta_3 = 0$ and $C_3 = u_1 v_2 - u_2 v_1$. The geometric area density expands algebraically as a quadratic form over $\Psi = (u_1, u_2, v_1, v_2)^T$:
\EQ{
\frac{\Omega_{(2)\perp}^2}{R^2} = \sum_{i,j=1}^4 Q_{ij}(\tau,\sigma) \Psi_i \Psi_j.
}
Expanding the brackets natively separates the variables:
\EQ{
Q_{11} &= Q_{33} = \sin^2(2a\tau) - \cos^2(2a\sigma), \br
Q_{22} &= Q_{44} = \cos^2(2a\tau) - \cos^2(2a\sigma), \br
Q_{12} &= Q_{21} = Q_{34} = Q_{43} = \sin(2a\tau)\cos(2a\tau) = \frac{1}{2}\sin(4a\tau), \br
Q_{13} &= Q_{31} = \cos^2(2a\sigma) - \sin^2(2a\tau), \br
Q_{24} &= Q_{42} = \cos^2(2a\sigma) - \cos^2(2a\tau), \br
Q_{14} &= Q_{41} = -\frac{1}{2}\sin(4a\tau) + \frac{1}{2}\sin(4a\sigma), \br
Q_{23} &= Q_{32} = -\frac{1}{2}\sin(4a\tau) - \frac{1}{2}\sin(4a\sigma).
}

Because $\Delta_3=0$, the linear anomaly vanishes ($\rho_1=0$). The scalar variation is  proportional to the area density, $\rho_2^{(\perp)} = \frac{1}{2\cos^2(2a\sigma)} (\Omega_{(2)\perp}^2 / R^2)$. Integrating by parts over the periodic $\sigma$ coordinate strictly removes the derivative on the fluctuations:
\EQ{
\INT{0}{2\pi} d\sigma \, 2 H(\sigma) \partial_\sigma \rho_2^{(\perp)} = \INT{0}{2\pi} d\sigma \left( -\frac{\partial_\sigma H}{\cos^2(2a\sigma)} \right) \left( \frac{\Omega_{(2)\perp}^2}{R^2} \right).
}
Using $H(\sigma) = -2a \tan(2a\sigma)$, the gradient evaluates to $\partial_\sigma H = -4a^2 \sec^2(2a\sigma)$. The entire transverse Lagrangian factorizes strictly into a purely spatial conformal scale multiplying the geometric matrix $\mathbf{Q}$:
\EQ{
\mathcal{L}_\perp(\tau,\sigma) = \Lambda_\perp(\sigma) \sum_{i,j=1}^4 Q_{ij}(\tau,\sigma) \Psi_i \Psi_j, \quad \text{where} \quad \Lambda_\perp(\sigma) = w \sigma R^2 + \frac{w a^2}{3\pi} \sec^4(2a\sigma).
}
Because $\Lambda_\perp(\sigma)$ is completely independent of worldsheet time $\tau$, the temporal integration $\int_0^{2\pi} d\tau$ acts directly on $Q_{ij}(\tau, \sigma)$. The trigonometric components $1, \cos(4a\tau)$, and $\sin(4a\tau)$ generate exact, rigid Kronecker deltas $\delta_{n+n', 0}$ and $\delta_{n+n', \pm 4a}$. The previously dense infinite integral therefore sparsifies mathematically into a tightly banded Toeplitz operator.

\subsection{2. The Longitudinal Matrix Integrands and Exact Decoupling}
The longitudinal sector isolates $u_3$ and $v_3$. Defining $\Delta_3 = v_3 - u_3$, the area variation evaluates algebraically to:
\EQ{
\mathcal{L}_{\text{area}}^\parallel = w \sigma R^2 \Delta_3^2 \left( \sin^2(2a\sigma) - \cos^2(2a\sigma) \right) = -w \sigma R^2 \cos(4a\sigma) \Delta_3^2.
}

The Liouville functions are $\rho_1 = \Delta_3 Y$ and $\rho_2^{(\parallel)} = -\frac{1}{2}\Delta_3^2 (1 + Y^2)$, with $Y(\sigma) = \tan(2a\sigma)$. We  expand the kinetic Liouville term $(\partial_\sigma \rho_1)^2 - (\partial_\tau \rho_1)^2$. Integrating the cross-terms by parts across the torus algebraically transfers all derivatives strictly onto the spatial background:
\EQ{
V_{\text{kin}}(\sigma) &= (\partial_\sigma Y)^2 - (\partial_\tau Y)^2 - \frac{1}{2}\partial_\sigma^2(Y^2) + \frac{1}{2}\partial_\tau^2(Y^2) = -Y(\sigma) \partial_\sigma^2 Y(\sigma) \br
&= -\tan(2a\sigma) \partial_\sigma \big( 2a\sec^2(2a\sigma) \big) = -8a^2 \tan^2(2a\sigma) \sec^2(2a\sigma).
}
For the potential Liouville term, integration by parts yields:
\EQ{
\INT{0}{2\pi} d\sigma \, 2 H \partial_\sigma \rho_2^{(\parallel)} = \INT{0}{2\pi} d\sigma \left( \partial_\sigma H \right) (1 - Y^2) \Delta_3^2.
}
Substituting $\partial_\sigma H = -4a^2 \sec^2(2a\sigma)$, the total Liouville curvature evaluates algebraically to:
\EQ{
-8a^2 \tan^2(2a\sigma) \sec^2(2a\sigma) - 4a^2 \sec^2(2a\sigma) \left( 1 - \tan^2(2a\sigma) \right) = -4a^2 \sec^4(2a\sigma).
}
Summing the Area and total Liouville components, the longitudinal Lagrangian definitively partitions into a purely spatial mass potential and a decoupled kinetic interaction:
\EQ{
\mathcal{L}_\parallel(\tau,\sigma) = \frac{w}{3\pi} \tan^2(2a\sigma) (\partial_\tau u_3)(\partial_\tau v_3) + V_{\text{eff}}^{\parallel}(\sigma) (u_3^2 + v_3^2 - 2 u_3 v_3),
}
where the exact effective longitudinal mass potential is strictly given by:
\EQ{
V_{\text{eff}}^{\parallel}(\sigma) = -w \sigma R^2 \cos(4a\sigma) - \frac{w a^2}{3\pi} \sec^4(2a\sigma).
}
Because $V_{\text{eff}}^{\parallel}(\sigma)$ is completely free of $\tau$-dependence, substituting the Fourier modes $(\partial_\tau u_3)(\partial_\tau v_3) \to n m u_3 v_3$ yields orthogonal integrals. The temporal integral $\int_0^{2\pi} d\tau$  imposes $\delta_{n,-n'}$ and $\delta_{n,m}$. 

The infinite, dense functional path integral is thus perfectly resolved. The fully defined double integrals for the $2 \times 2$ longitudinal matrix blocks algebraically collapse into  mathematically isolated 1D spatial quadratures for each independent frequency $n$:
\EQ{
\mathcal{M}^{AA,\parallel}_{n,n'} &= \delta_{n,-n'} \INT{0}{2\pi} d\sigma \, V_{\text{eff}}^{\parallel}(\sigma), \br
\mathcal{M}^{BB,\parallel}_{m,m'} &= \delta_{m,-m'} \INT{0}{2\pi} d\sigma \, V_{\text{eff}}^{\parallel}(\sigma), \br
\mathcal{M}^{AB,\parallel}_{n,m} &= \delta_{n,m} \INT{0}{2\pi} d\sigma \left[ -2 V_{\text{eff}}^{\parallel}(\sigma) + \frac{w n^2}{3\pi} \tan^2(2a\sigma) \right] \exp{2\I n\sigma}.
}
These  closed forms directly evaluate the discrete string spectrum,  validating the complete structural stability of the $q=0$ ground state.
\subsection{Szeg\H{o} Factorization and the Exact Zero-Mode Annihilation}
\label{app:SzegoZeta}

To  evaluate the functional determinant of the dense spatial mixing matrix, we must isolate its principal extensive macroscopic scaling. The uncoupled spatial matrix elements are derived from the Fourier projection of the background effective potential $V_{\rm eff}^{\parallel}(\sigma)$, giving $\mathcal{D}_{n,m} = 2\pi \tilde{V}_{n+m}$. 

For a purely multiplicative background operator, the asymptotic volume growth of the functional determinant is mathematically governed by Szeg\H{o}'s Strong Limit Theorem. The effective uniform baseline per Fourier mode is defined by the continuous geometric mean of the absolute effective potential over the periodic domain:
\EQ{
G \equiv \exp{ \frac{1}{2\pi} \INT{0}{2\pi} d\sigma \ln \big| 2\pi V_{\rm eff}^{\parallel}(\sigma) \big| } = \lim_{N \to \infty} \exp{ \frac{1}{N} \sum_{n=1}^N \ln \mathcal{D}_{n,n} }.
}
By explicitly factoring out this geometric baseline, the normalized fluctuation matrix $\bar{\mathcal{D}}_{n,m} = \mathcal{D}_{n,m} / G$ yields a regularized determinant $|\det \bar{\mathcal{D}}|$ that unconditionally converges to the finite Widom anomaly.

In the continuous functional Hilbert space, the overall extensive volume prefactor is defined via zeta-function regularization over the discrete mode spectrum. At the exact $q=0$ trigonometric limit, this geometric scaling dynamically separates based strictly on the topological spin structures:

\begin{itemize}
    \item \textbf{The Periodic Sector ($a=1$):} The classical twistors evaluate to pure phases, yielding a flat metric $A(\sigma) = 1$ and zero Liouville curvature $P(\sigma) = 0$. The potential evaluates to a strict constant $V_{\rm eff}^{\parallel} = -\sigma R^2$, meaning the geometric mean is  $G = 2\pi\sigma R^2$. The Fourier spectrum consists of integer modes ($n \in \Z \setminus \{0\}$). The effective dimensionality  regularizes to $\sum_{n \neq 0} 1 = 2\zeta(0) = -1$. Because the string dynamically couples both left- and right-moving chiral modes, the macroscopic geometric baseline of the non-zero modes evaluates to an exact functional prefactor of $G^{2\zeta(0)} = G^{-2}$ overall. 
    
    \item \textbf{The Anti-Periodic Sector ($a=1/2$):} The intrinsic topological twist yields a conformal metric $A(\sigma) = \cos^2(\sigma)$. The boundary conditions shift the Fourier spectrum strictly to half-integers ($r \in \Z + 1/2$). The effective dimensionality analytically regularizes via the Hurwitz zeta function to $\sum_{r} 1 = 2\zeta(0, 1/2) = 0$. The geometric phase-space volume  collapses ($G^0 = 1$), isolating the determinant strictly to the scale-invariant anomaly.
\end{itemize}

Crucially, this analytic factorization dynamically resolves the zero-mode scalar ghost divergence natively inherent to the translational degrees of freedom in the $a=1$ sector. The complete macroscopic S-matrix transition amplitude $\mathcal{A}(J)$ requires assembling all functional measure factors:
\EQ{
\mathcal{A}(J) \propto \mathcal{P}_{\text{WKB}} \times \mathcal{J}_{\text{FP}} \times \left[ \det \mathcal{M}_{n \neq 0} \right]^{-1/2}.
}
For the $a=1$ sector, the continuous WKB propagator diverges perfectly upon intercepting the classical target-space geometric origin, yielding a kinematic pole $\mathcal{P}_{\text{WKB}} = 1 / M^2$. The Faddeev--Popov integration Jacobian $\mathcal{J}_{\rm FP}$ strictly equates to the proper spatial area of the unperturbed minimal surface, $\mathcal{J}_{\rm FP} = 4\pi^2 R^2$. Because the string dynamically obeys $R^2 = M^2 / (16\pi^2\sigma^2)$, the translational measure scales precisely as $\mathcal{J}_{\rm FP} \propto M^2$. 

Simultaneously, the regularized functional determinant of the non-zero modes evaluates to $G^{-2} |\det \bar{\mathcal{D}}|^2$. Extracting the inverse square root contributes a continuous numerator factor of  $G^1 = 2\pi\sigma R^2 \propto M^2$. 

Multiplying all three fundamental path-integral components together yields an exact, analytical algebraic collapse for the periodic $a=1$ amplitude:
\EQ{
\mathcal{A}_{a=1}(J) \propto \left(\frac{1}{M^2}\right)_{\text{pole}} \times \left(M^2\right)_{\text{FP}} \times \left(M^2\right)_{G^{1}} \propto M^2.
}
As $M \to 0$ (at  $J=0$), the geometric phase-space suppression dynamically forces the complete transition amplitude to rigidly evaluate to zero ($\mathcal{A}_{J=0} = 0$). The fake scalar pole is not merely canceled; it is completely annihilated. This strictly proves that the $a=1$ topological string definitively mathematically decouples from the physical target space at $J=0$, leaving a fully stable string partition function.

Conversely, for the physical anti-periodic $a=1/2$ glueballs, the absence of continuous topological zero-modes ($\mathcal{J}_{\rm FP} = 1$) and the nullified geometric dimension ($G^0 = 1$) leave the exact transition amplitude scaling natively as $\mathcal{A}_{a=1/2} \propto 1/M^2 \times 1 \times 1$. This preserves the $1/M^2$ kinematic pole, correctly propagating genuine physical states with finite quantum amplitudes. To avoid the unphysical tachyonic ground state, the physical scalar spectrum at $J=0$ natively resides on the first daughter trajectory ($n=1$), yielding a strictly real, positive squared mass $M^2 = 4\pi\sigma(11/12) \approx 2.07 \text{ GeV}^2$. This corresponds perfectly to the observed $f_0(1500)$ scalar glueball, keeping the physical spectrum strictly free of tachyons.
The \Mathematica{} code computing the amplitudes for various values of J and both cases ($a =1/2, 1$) is published in the notebook \cite{MBGluonSpectrum}.
\end{document}